\documentclass{aa}

\usepackage{graphicx}
\usepackage{txfonts}
\usepackage{lipsum}
\usepackage{subcaption}         % necessary for continued figures, example in section 3
\usepackage{lscape}             % to rotate a single page table, example in appendix.
\usepackage{placeins}           % useful with \FloatBarrier, to keep 
\usepackage{xcolor}
\usepackage{wasysym}

\def\chisqr{\hbox{$\chi^2_{\rm r}$}}
\def\msun{\hbox{${\rm M}_{\odot}$}}
\def\mjup{\hbox{${\rm M}_{\jupiter}$}}

\def\mspy{\hbox{${\rm M}_{\odot}$\,yr$^{-1}$}}

\def\rsun{\hbox{${\rm R}_{\odot}$}}
\def\lsun{\hbox{${\rm L}_{\odot}$}}
\def\rcor{\hbox{$r_{\rm cor}$}}
\def\rmag{\hbox{$r_{\rm mag}$}}
\def\mstar{\hbox{$M_{\star}$}}
\def\rstar{\hbox{$R_{\star}$}}
\def\lstar{\hbox{$L_{\star}$}}
\def\teff{\hbox{$T_{\rm eff}$}}
\def\logg{\hbox{$\log g$}}

\def\vD{\hbox{$v_{\rm D}$}}

\def\kms{\hbox{km\,s$^{-1}$}}
\def\cmss{\hbox{cm\,s$^{-2}$}}

\def\vsini{\hbox{$v \sin i$}}

\def\mic{\hbox{$\mu$m}}

\def\emr{}
\def\Bl{\hbox{$B_{\rm \ell}$}}
\def\Bd{\hbox{$B_{\rm d}$}}

\def\degr{\hbox{$^\circ$}}

\def\Mdot{\hbox{$\dot{M}$}}

\def\Prot{\hbox{$P_{\rm rot}$}}

\newcommand{\feiif}{\hbox{[Fe$\;${\sc ii}]}}
\newcommand{\htw}{\hbox{H$_2$}}
\newcommand{\hei}{\hbox{He$\;${\sc i}}}

\newcommand{\pab}{\hbox{Pa${\beta}$}}
\newcommand{\brg}{\hbox{Br${\gamma}$}}

\begin{document}

%%%%%%%%%%%%%%%%%%% TITLE PAGE %%%%%%%%%%%%%%%%%%%

\title{Monitoring the magnetospheric accretion of the classical T~Tauri star DO~Tau with SPIRou} 

   \author{J.-F.~Donati\inst{1}
      \and P.I.~Cristofari\inst{2}
      \and A.~Carmona\inst{1}
      \and A.~Lavail\inst{1}
      \and C.~Moutou\inst{1}
      \and J.~Bouvier\inst{3}
      \and K.~Perraut\inst{3}
      \and S.H.P.~Alencar\inst{4} 
      \and F.~M\'enard\inst{3} 
      \and M.~Audard\inst{5}
      \and P.~Petit\inst{1}
      \and E.~Alecian\inst{3}
      \and T.~Ray\inst{6}
      \and the SPIRou science team
          }
   \institute{Univ.\ de Toulouse, CNRS, IRAP, 14 avenue Belin, 31400 Toulouse, France (\email{jean-francois.donati@cnrs.fr})
        \and Leiden Observatory, Leiden University, Niels Bohrweg 2, 2333 CA Leiden, the Netherlands
	\and Univ.\ Grenoble Alpes, CNRS, IPAG, 38000 Grenoble, France 
        \and Departamento de F\'{\i}sica -- ICEx -- UFMG, Av. Ant\^onio Carlos, 6627, 30270-901 Belo Horizonte, MG, Brazil
        \and Department of Astronomy, University of Geneva, Chemin Pegasi, 51, Versoix CH-1290, Switzerland 
        \and Dublin Institute for Advanced Studies, Astronomy \& Astrophysics Section, 31 Fitzwilliam Place, Dublin D02 XF86, Ireland
             }

% These dates will be filled out by the publisher
\date{Submitted 2025 Dec.~19 -- Accepted 2026 Feb.~22} 

% Abstract of the paper
\abstract{We present observations of the classical T~Tauri star DO~Tau collected with the near-infrared SPIRou spectropolarimeter and precision 
velocimeter at the Canada-France-Hawaii Telescope from early 2020 to late 2025.  Circularly polarized Zeeman signatures were clearly detected at most epochs in the atomic 
spectral lines of DO~Tau, yielding longitudinal magnetic fields of up to 280~G modulated with a period of $5.128\pm0.002$~d which we identified as the rotation period of DO~Tau.  
Applying Zeeman-Doppler imaging to the SPIRou data recorded in 2021, 2024 and 2025, we found that DO~Tau hosts an unusual large-scale magnetic field that is weaker, less poloidal, more 
inclined to the rotation axis, and varies more rapidly with time than those of previously studied T~Tauri stars, possibly as a result of intense accretion 
between the inner disk and the stellar surface.  The dipole component of this large-scale field of about 0.2-0.3~kG even flipped polarity toward the end of our 
observing campaign, making DO~Tau the first T~Tauri star for which a magnetic polarity reversal is reported.  The magnetospheric gap surrounding the central star was quite compact,  
extending to $\simeq$1.6~\rstar\ (0.014~au) as a result of the strong accretion rate ($\log\Mdot=-7.7$~\mspy), with the inner accretion disk being warped by 
the tilted stellar magnetic field.  Radial velocity variations suggest the presence of a close-in planet of a few \mjup\ or a density structure in the inner accretion 
disk at an orbital period of 21~d (corresponding to 0.12~au), which might be linked to the wiggle in the jet axis of DO~Tau. } 


\maketitle

%%%%%%%%%%%%%%%%%%%%%%%%%%%%%%%%%%%%%%%%%%%%%%%%%%

%%%%%%%%%%%%%%%%% BODY OF PAPER %%%%%%%%%%%%%%%%%%

\section{Introduction}
\label{sec:int}

With an age of only a few Myr, pre-main-sequence (PMS) low-mass T~Tauri stars, and in particular the classical T~Tauri stars (cTTSs) that still accrete material from their 
accretion disk, are key targets for investigating the late phases of star formation and the multiple physical processes that take place in the close stellar environment, 
including magnetospheric accretion, outflows and jets \citep[e.g.,][]{Romanova21}.  They are also ideal objects for constraining models of planetary formation and migration 
at a stage for which very little observational material is yet available.  

A few dozen weakly accreting cTTSs have been observed at multiple epochs to address these points.  We used spectropolarimetry to characterize 
the magnetic field of the central star and its interactions with the inner accretion disk, and velocimetry to investigate the potential presence of close-in massive 
companions at the outer edge of the stellar magnetospheres \citep[e.g.,][]{Zaire24,Donati24,Donati24d}, in particular those transiting their host stars \citep{Barber24,Donati25c}.  
These results, some from data obtained with the near-infrared (nIR) SPIRou spectropolarimeter and velocimeter at the Canada-France-Hawaii Telescope (CFHT), used Zeeman-Doppler 
imaging (ZDI) to reconstruct the large-scale magnetic topologies of the observed cTTSs.  These data also clarify how magnetospheric accretion proceeds, emphasizing the 
role of magnetic fields in controlling the rotation of cTTSs \citep{Romanova25b}.  

Very few such studies exist for cTTSs that strongly accrete material from their disks.  Characterizing such an evolution phase, through which all PMS stars must go at some point 
\citep[e.g.,][]{Fischer23}, is essential for a thorough understanding 
of star and planet formation.  The effect of accretion flows on the formation of low-mass stars \citep[e.g.,][]{Geroux16}, and more specifically, on the internal structure, 
convection patterns, dynamo processes, and large-scale magnetic fields, must be investigated to determine how these fields interact with accretion, 
outflows, and jets from the inner disks \citep{Blinova16}, and how they affect stellar rotation \citep[e.g.,][]{Amard23}.  In addition, detecting close-in massive planets 
and characterizing how they migrate within the disk and trigger enhanced accretion \citep{Romanova25} is essential for better constraining theoretical 
models at early stages of planet formation.  

To address these points we ran a long-term observing program on DO~Tau, a low-mass cTTS with a high accretion rate, using SPIRou at the CFHT and collected 
spectra over 5.7 years.  After recalling the main parameters of DO~Tau and detailing the SPIRou observations we collected in Secs.~\ref{sec:par} and \ref{sec:obs}, 
we model the temporal evolution of its magnetic field with Gaussian process regression (GPR) and ZDI in Secs.~\ref{sec:mag} and \ref{sec:zdi}, and 
we investigate in Sec.~\ref{sec:rvs} whether its radial velocity (RV) variations reveal the presence of a close-in massive planet.  In Sec.~\ref{sec:act} we study 
what the emission lines tell us about magnetospheric accretion taking place in DO~Tau, and we summarize the results in Sec.~\ref{sec:dis}.

\section{The T~Tauri star DO~Tau} 
\label{sec:par}

Located in the Taurus star-forming region \citep[in the C7-L1527 spatial subgroup, aged $2.59\pm0.75$~Myr;][]{Krolikowski21} at a distance of $138.52\pm0.68$~pc from 
the Earth \citep{Gaia20}, DO~Tau is a low-mass M1 cTTS undergoing intense accretion as signaled by strong emission lines across its spectrum \citep{Alcala21,Sousa23}.  
It is surrounded by a compact accretion disk of several dozen au that is inclined at $27.6\pm0.3$\degr\ to the line of sight \citep{Long19}, whose inner region is the 
source of a CO outflow and a bipolar jet \citep[e.g.,][]{Fernandez20,Erkal21}.  Its complex circumstellar environment, featuring arms and streams, suggests an unusual 
evolution that might be caused by interactions with the nearby cTTS HV~Tau or with remnant material from the star formation process \citep{Winter18,Huang22}.  The morphology of the 
jet, and in particular the axis wiggle in both lobes, also suggests that the inner disk of DO~Tau, from which the jet originates, might host a massive planet on an 
inclined orbit or be warped by a variable magnetic field \citep{Erkal21}.  

As DO~Tau accretes material from the inner regions of its accretion disk at a relatively high rate, its spectrum is subject to veiling, i.e., a dilution of its photospheric 
lines that results from the addition of a mostly featureless spectral contribution from warm accretion features (strongest in the blue) and from the inner edge of 
the accretion disk (strongest in the red).  Photospheric spectral lines of DO~Tau can thus be weaker by two to seven times than expected from a weak-line TTS of the same spectral type 
\citep[][see also Sec.~\ref{sec:act}]{Alcala21}, rendering fundamental parameters hard to estimate precisely.  Several studies quoted values inferred from high-resolution spectra, 
e.g., \citet{Alcala21} who derived an effective temperature of $\teff=3694\pm104$~K from optical spectra 
and \citet{Lopez-Valdivia21} who reached similar results with larger error bars from nIR spectra ($\teff=3704\pm350$~K and $\logg=3.44\pm0.50$~dex).  
A study from low-resolution spectra aimed at explaining the $IYJ$ continuum excess emission of cTTSs suggested a much lower effective temperature 
\citep[$\teff=2947\pm25$~K;][]{Paolino25b}.  Our own analysis of the least veiled SPIRou spectra yielded an intermediate value ($\teff=3450\pm50$~K and 
$\logg=3.6\pm0.1$~dex, see Sec.~\ref{sec:obs}), which is more or less consistent with previous measurements from high-resolution spectroscopy.  

The velocity field in the disk derived from molecular line ALMA images indicated a dynamical mass of $\mstar=0.54\pm0.07$~\msun\ for DO~Tau \citep{Braun21}, 
which, when it is coupled to our estimate of \logg\, implied a stellar radius of $\rstar=1.9\pm0.2$~\rsun\ and a logarithmic luminosity of $\log \lstar/\lsun=-0.3\pm0.2$.  
The original evolutionary models of \citet{Feiden16} that include magnetic fields are able to reconcile our \teff\ and \logg\ estimates with this mass 
measurement, yielding an age of $1.7\pm0.5$~Myr (consistent with that of the C7-L1527 Taurus subgroup).  In their most recent version (where the surface magnetic field 
strength was updated at each model iteration to be equal to the pressure equipartition value; Feiden, private communication), the mass that best matches our measurements 
is $0.42\pm0.05$~\msun, slightly lower than (but still compatible with) the dynamical mass estimate of DO~Tau.  The nonmagnetic models of \citet{Feiden16} 
and those of \citet{baraffe15} yield an even lower mass of 0.30$-$0.32~\msun, which is inconsistent with the dynamical estimate.  
DO~Tau is still fully convective according to evolutionary models of \citet{Feiden16}, although convection might be affected by strong accretion \citep{Geroux16}.  

Our spectropolarimetric observations of DO~Tau with SPIRou (see Secs.~\ref{sec:obs} and \ref{sec:mag}) revealed that the stellar rotation period is $\Prot=5.128\pm0.002$~d 
and the rotational broadening of spectral lines is $\vsini=13.0\pm0.5$~\kms, implying an inclination angle $i$ of the stellar rotation axis to the line of sight of 
$i=45\pm8$\degr, consistent with the inclination of the inner disk from interferometric measurements \citep{Perraut25}.  
This is higher than the inclination of the outer disk and is an additional indication that the evolution history of DO~Tau is complex and that 
its inner disk is warped.  Table~\ref{tab:par} lists the parameters we used.  

\begin{table}[t!]
\caption{Parameters of DO~Tau} 
\centering
\resizebox{\linewidth}{!}{  
\begin{tabular}{ccc}
\hline
distance (pc)    & $138.52\pm0.68$ & \citet{Gaia20}    \\ 
\teff\ (K)       & $3450\pm50$     & this paper \\ 
\logg\ (\cmss)   & $3.60\pm0.10$   & this paper \\ 
\mstar\ (\msun)  & $0.54\pm0.07$   & \citet{Braun21} \\ 
\rstar\ (\rsun)  & $1.9\pm0.2$     & this paper \\ 
$\log \lstar (\lsun)$ & $-0.3\pm0.2$& this paper \\ 
age (Myr)        & $1.7\pm0.5$     & \citet{Feiden16} \\ 
\Prot\ (d)       & $5.128\pm0.002$ & this paper        \\ 
\vsini\ (\kms)   & $13.0\pm0.5$    & this paper \\ 
$i$ (\degr)      & $45\pm8$        & from \Prot, \vsini, and \rstar \\ 
$i_{\rm disk}$ (\degr) & $27.6\pm0.3$ & \citet{Long19} \\ 
<$B$> (kG)       & $2.5\pm0.3$     & this paper        \\ 
\rcor\ (\rstar, au) & 5.4, 0.047   & from \Prot, \mstar, and \rstar  \\ 
$\log \Mdot$ (\mspy) & $-7.70\pm0.19$   & this paper       \\ 
RV (\kms)        & $16.3\pm0.1$    & this paper    \\ 
\hline
\end{tabular}
}
\label{tab:par}
\end{table} 

DO~Tau is known to exhibit significant Zeeman broadening of photospheric lines, which is attributed to small-scale magnetic fields at the surface of the star with an average 
strength <$B$>~$=1.78\pm0.85$~kG \citep{Lopez-Valdivia21}.  Zeeman signatures were also detected in the 588~nm \hei\ line of DO~Tau, demonstrating a magnetic 
field of 0.45$-$0.79~kG at the footpoints of the magnetospheric accretion funnels linking the star to the inner disk \citep{Dodin13}.

Several studies estimated mass accretion at the surface of DO~Tau, in most cases by exploiting accretion fluxes derived from several emission lines using calibrated 
scaling laws \citep{Alcala17,Fiorellino25}.  Measuring accretion fluxes in 16 emission lines from the UV to the nIR, \citet{Alcala21} obtained that 
$\log \Mdot=-7.73\pm0.40$ (with \Mdot\ in \mspy).  From SPIRou spectra (some in common with our study), 
\citet{Sousa23} obtained a similar estimate ($\log \Mdot=-7.54\pm0.30$) using \pab\ and \brg\ alone.  Along with the strong veiling, it 
demonstrates that DO~Tau is one of the strongest cTTS accretors in its mass range \citep[see Fig.~9 of][]{Alcala21}. 

DO~Tau was monitored by TESS in 2021 (September~6 to November~6, sectors 43 and 44) and 2023 (September~20 to November~11, sectors 70 and 71) and exhibited a chaotic and 
aperiodic light curve of amplitude $\simeq$1~mag, consistent with the K2 light curve based on which DO~Tau was classified as a burster with 
photometric variations mostly attributed to accretion processes and not to rotational modulation \citep{Cody22}.

\section{SPIRou observations}
\label{sec:obs}

We observed DO~Tau between 2020 February~19 and 2025 November~13, with the SPIRou nIR spectropolarimeter \citep{Donati20} at the CFHT, within the SPIRou Legacy Survey and the 
SPICE and PLANETS large programs (RUNIDs 20AP40, 20BP40, 23AP45, 23BP45, 24AP45, and 25BP45, PI J.-F.~Donati).  SPIRou collects unpolarized and polarized stellar spectra, 
covering a range of 0.95--2.50~\mic\ at a resolving power of 70\,000 in a single exposure.  Polarimetric observations consist of a sequence of four 
subexposures, with each subexposure corresponding to a different azimuth of the Fresnel rhomb retarders of the SPIRou polarimetric unit.  With this procedure, we are able to 
remove systematics in polarized spectra to first order \citep[][]{Donati97b}.  For DO~Tau, we focused on circular polarization, each recorded sequence of four 
subexposures yielding one unpolarized (Stokes $I$) and one circularly polarized (Stokes $V$) spectrum, as well as one null polarization check (called $N$) used to diagnose 
potential instrumental or data reduction issues.  
We collected 77 polarization sequences over five consecutive observing seasons covering 2094~d, 3 in early 2020, 8 in late 2020 $-$ early 2021, 3 in 
early 2023, 20 in early 2024, and 43 in late 2025.  The total exposure times ranged from 1582 to 1627~s, while the signal-to-noise ratios (S/Ns) per 2.3~\kms\ pixel in the $H$ band 
ranged from 154 to 321 (median 260).  

All spectra were processed with \texttt{Libre ESpRIT}, the nominal reduction pipeline of ESPaDOnS at CFHT, optimized for spectropolarimetry 
and adapted for SPIRou \citep{Donati20}.  To all reduced spectra, we applied least-squares deconvolution \citep[LSD;][]{Donati97b} using a line mask computed with the VALD-3 
database \citep{Ryabchikova15} for atmospheric parameters matching those of DO~Tau ($\teff=3500$~K, $\logg=3.5$).  We only selected atomic lines deeper than 10\% of the 
continuum level $I_c$ (with negligible contribution from the disk) for a total of $\simeq$1500 lines with an average wavelength and Land\'e factor equal to 1750~nm 
and 1.2.  This yielded noise levels, $\sigma_V$, in the resulting Stokes $V$ LSD profiles ranging from 1.6 to 4.1 {\emr (median 2.1), in units of $10^{-4} I_c$}. 
Zeeman signatures were detected at most epochs, with typical peak-to-peak amplitudes of a few 0.1\%.  We computed the longitudinal field \Bl\ {\emr \citep[i.e., the 
line-of-sight component of the magnetic vector averaged over the visible stellar hemisphere, an ideal proxy for diagnosing rotational modulation; e.g.,][]{Rescigno24}}
from LSD Stokes $I$ and $V$ profiles following \citet{Donati97b}, integrating over an interval of $\pm$30~\kms\ adequate for DO~Tau.  We found that \Bl\ was well detected, 
and the reduced chi square assuming no field was $\chisqr=43$.  The same operation applied to the null polarization spectrum $N$ yielded $\chisqr=0.81$, 
which is consistent with no spurious signal and indicates no issues in the observation and reduction procedures.  
\Bl\ ranged from $-193$ to 276~G (median error bar 19~G, see Fig.~\ref{fig:gpb} and Sec.~\ref{sec:mag}) and was clearly rotationally modulated.  
We also computed Stokes $I$ LSD profiles with a line mask that only included the magnetically insensitive CO bandhead lines from 2.29 
to 2.40~\mic\ deeper than 10\% of $I_c$ ($\simeq$500 lines), seen in absorption and coming mostly from the star, to examine how they differed from the LSD profiles 
of atomic lines.  

We also reduced our SPIRou data with the latest version of \texttt{APERO} (v0.7.294), the nominal SPIRou reduction pipeline \citep{Cook22} optimized for RV 
precision and the correction of telluric lines.  We applied the line-by-line (LBL) technique \citep[v0.65;][]{Artigau22} to the reduced data, yielding RVs and differential 
temperatures $dT$ estimated from variations {\emr in the relative depths of veiling-corrected spectral lines} with respect to their median profiles \citep{Artigau24}.   However, LBL 
was designed for measuring subtle changes in stellar spectra and might not be well suited for highly veiled cTTSs like DO~Tau.  We found that while LBL RVs show a clear 
rotational modulation (see Sec.~\ref{sec:rvs}), $dT$ is only marginally modulated at \Prot\ with a periodogram peaking at 12.9~d and showing several 
peaks of similar strength between 5 and 30~d.  We also applied LSD to the \texttt{APERO} spectra with the masks mentioned above and derived alternative RVs, separately for 
atomic and CO bandhead lines, by describing each individual LSD profile as a simple first-order Taylor expansion constructed from the median of all LSD profiles 
\citep[as in previous studies; e.g.,][]{Donati24d,Donati24c}.  The LBL RVs were found to correlate tightly with the RVs of CO lines (Pearson's coefficient $R\simeq0.90$) and 
more loosely with those of atomic lines ($R\simeq0.80$), but showed a significantly weaker modulation amplitude than either (see Sec.~\ref{sec:rvs}).  

We used the median of the least veiled \texttt{APERO}-reduced spectra of DO~Tau to measure the stellar parameters with ZeeTurbo \citep[][]{Cristofari23} 
simultaneously with the small-scale magnetic field estimated from the Zeeman broadening of the spectral lines.  For the stellar parameters, we found $\teff=3450\pm50$~K and 
$\logg=3.60\pm0.10$ (assuming solar metallicity).  We also deduced that DO~Tau hosts a small-scale magnetic field of <$B$>~=~$2.5\pm0.3$~kG, with the 2 and 4~kG components 
covering 85\% and 8\% of the visible hemisphere on average, and the other (including the field-free component) contributing less than a few percent.  This is stronger  
than the value derived by \citet[][equal to $1.78\pm0.85$~kG]{Lopez-Valdivia21}, possibly due to long-term evolution.  Given the strong veiling, individual spectra 
were too noisy for ZeeTurbo to yield reliable results.  

The full log of our observations is provided in Table~\ref{tab:log}.  Phases and rotation cycles were derived assuming a rotation period of $\Prot=5.128$~d (see Table~\ref{tab:par}), 
counting from an arbitrary starting barycentric Julian date (BJD) of 2\,458\,898.7 (slightly before our first observation of DO~Tau).

\section{Magnetic field and temperature changes}
\label{sec:mag}
 
To quantify the temporal fluctuations of \Bl, arranged in a vector denoted $y$, we applied GPR \citep[][]{Haywood14,Rajpaul15} and employed 
the following quasi-periodic (QP) covariance function $c(t,t')$
\begin{eqnarray}
c(t,t') = \theta_1^2 \exp \left( -\frac{(t-t')^2}{2 \theta_3^2} -\frac{\sin^2 \left( \frac{\pi (t-t')}{\theta_2} \right)}{2 \theta_4^2} \right) 
\label{eq:covar}
\end{eqnarray}
where $\theta_1$ is the amplitude (in G) of the Gaussian process (GP), $\theta_2$ is its recurrence period (measuring \Prot), $\theta_3$ is the evolution timescale on which the \Bl\ curve 
changes shape (in d), and $\theta_4$ is a smoothing parameter controlling the amount of harmonic complexity.
A fifth hyperparameter, $\theta_5$, describes the excess of uncorrelated noise on \Bl.  This ensures that the QP GPR fit can diagnose cases where the standard error 
bars are underestimated, and reach the solution with highest likelihood, $\mathcal{L}$, defined by
\begin{eqnarray}
2 \log \mathcal{L} = -n \log(2\pi) - \log|C+\Sigma+S| - y^T (C+\Sigma+S)^{-1} y
\label{eq:llik}
\end{eqnarray}
where $C$ is the covariance matrix for all observing epochs, $\Sigma$ is the diagonal variance matrix associated with $y$, $S=\theta_5^2 J$ is the contribution of the additional
white noise, with $J$ the identity matrix, and $n$ is the number of data points.

\begin{table}[t!]
\caption{MCMC GPR modeling of the \Bl\ curve of DO~Tau}
\centering
\resizebox{\linewidth}{!}{
\begin{tabular}{cccc}
\hline
Parameter   & Symbol & Value & Prior   \\
\hline
GP amplitude (G)     & $\theta_1$  & $159\pm48$       & mod Jeffreys ($\sigma_{\Bl}$) \\
Rec.\ period (d)     & $\theta_2$  & $5.128\pm0.002$  & Gaussian (5.1, 1.0) \\
Evol.\ timescale (d) & $\theta_3$  & 600              & fixed \\ 
Smoothing            & $\theta_4$  & $0.87\pm0.21$    & Uniform  (0, 3) \\
White noise (G)      & $\theta_5$  & $22.6\pm3.6$     & mod Jeffreys ($\sigma_{\Bl}$) \\
Rms (G)              &             & 25.4             & \\ 
$\chisqr$            &             & 1.83             & \\
\hline
\end{tabular}}
\tablefoot{For each hyper parameter, we list the fitted value along with the corresponding error bar and the assumed prior.  The knee of the modified Jeffreys prior is set to
the median error bar of \Bl\ (19~G).  The evolution timescale $\theta_3$ was poorly constrained by the data and set to 600~d.  }
\label{tab:gpr}
\end{table}

\begin{figure}[ht!]
\centerline{\includegraphics[scale=0.38,angle=-90]{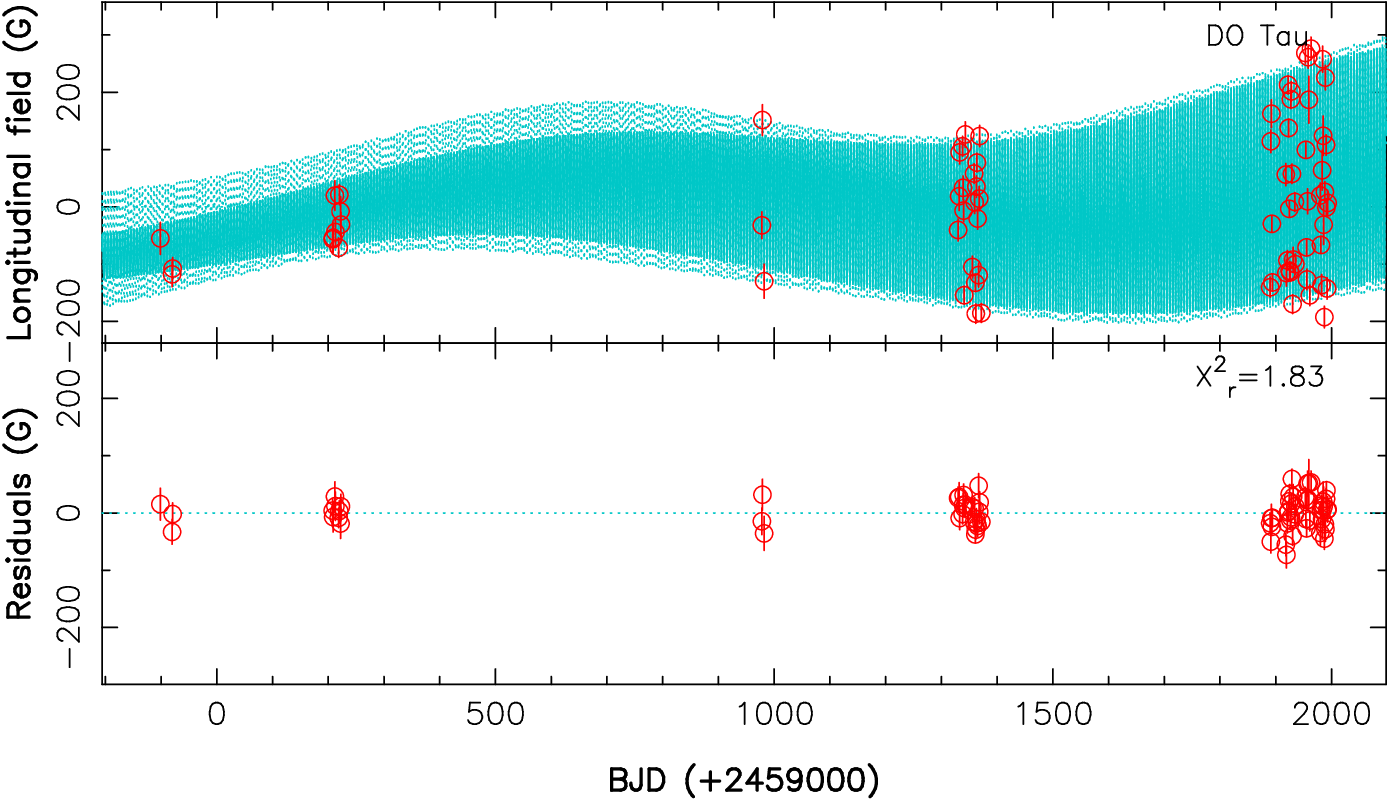}\vspace{2mm}}
\centerline{\includegraphics[scale=0.38,angle=-90]{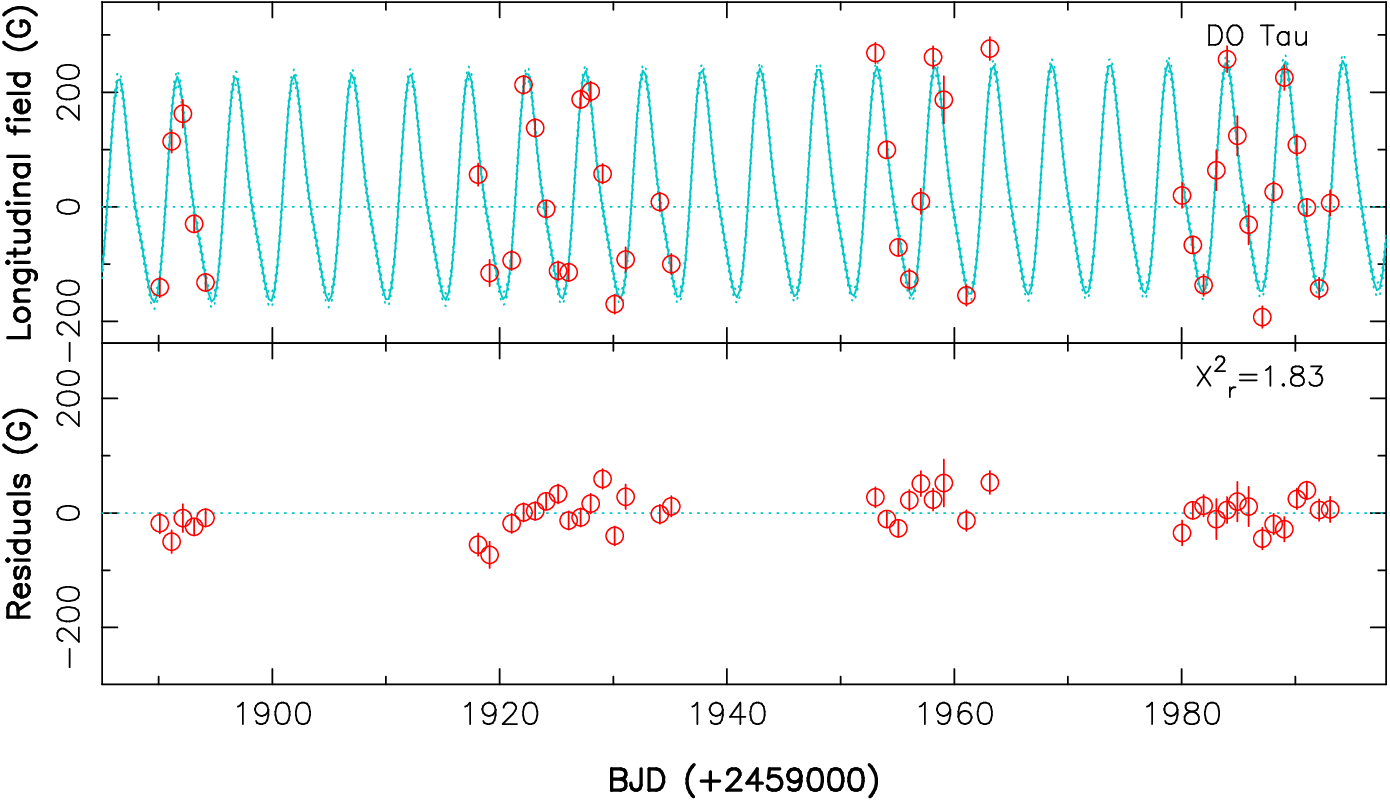}}
\caption[]{Longitudinal magnetic field \Bl\ (red dots), and QP GPR fit to the data (full cyan line) with corresponding 68\% confidence intervals (dotted cyan lines).  
The bottom panel zooms in on the 2025 data.  The residuals, shown in the bottom plot of each panel, have an rms of 25~G ($\chisqr=1.8$). }
\label{fig:gpb}
\end{figure}

We explored the hyperparameter domain with a Markov chain Monte Carlo (MCMC) process and retrieved posterior distributions and error bars for all parameters.  The MCMC and 
GPR modeling tools were those used in our previous studies \citep[e.g.,][]{Donati23}.  The MCMC process is a single-chain Metropolis-Hastings scheme
running over a few $10^5$ steps, including the first few $10^4$ steps as burn-in.  Convergence was checked with an autocorrelation analysis to verify that the burn-in and main
phase vastly exceeded the autocorrelation lengths of all parameters.  We computed the marginal logarithmic likelihood, $\log \mathcal{L}_M$, following \citet{Chib01}.

The GPR fit we obtained is shown in Fig.~\ref{fig:gpb}, with a zoom-in on the 2025 data.  The derived GPR hyperparameters are listed in Table~\ref{tab:gpr}.  {\emr The rotation 
period we infer ($5.128\pm0.002$~d) is shorter than that of prototypical mildly accreting cTTSs (e.g., AA~Tau or BP~Tau) and agrees better with that of lower-mass cTTSs 
\citep[e.g.,V2247~Tau,][]{Donati10} and strong accretors \citep[e.g., RU~Lup,][]{Armeni24}.}
The rotational modulation semi-amplitude of \Bl\ increased steadily, growing from $\simeq$50~G in 2020 to $\simeq$200~G in 2025.  
\Bl\ progressively evolved from being always or mostly negative in 2020 (average $-94$~G) and 2021 (average $-28$~G) to mostly positive in 2025 (average 24~G).   
Given the good sampling in three of our five seasons, this suggests a topological change in the large-scale field of DO~Tau over our observing campaign.  The poloidal field 
component (to which \Bl\ is mostly sensitive) was slightly tilted to the stellar rotation axis in 2020 (given the low-amplitude modulation with respect to the average $|\Bl|$; see 
Fig.~\ref{fig:gpb}), became more tilted in 2021 and 2024 (lower average $|\Bl|$ and stronger modulation), and was even more strongly tilted in 2025 following a polarity switch (maximum 
amplitude modulation and sign change of the average \Bl; see Fig.~\ref{fig:gpb}).  The achieved \chisqr\ (1.83) is significantly larger than 1, indicating that intrinsic variability, likely 
due to accretion, induced random variations in \Bl\ stronger than those from photon noise.  

We carried out a similar analysis on $dT$ and confirmed that rotational modulation was marginal.  The semi-amplitude of $dT$ fluctuations was only $\simeq$7~K (rms 14~K) 
and roughly constant with time, indicating that {\emr surface brightness features (brighter or darker than the photosphere)} were either too small or the contrast was too low 
to induce detectable rotational modulation.

\section{ZDI modeling}
\label{sec:zdi}

We applied ZDI to the LSD Stokes $I$ and $V$ profiles of DO~Tau to model the rotational modulation of the line profiles, assuming their shapes and distortions are 
caused by brightness and magnetic surface features.  
{\emr As in previous studies, LSD profiles were corrected for veiling to remove contributions from accretion funnels, flares and the inner disk to first order.} 
For highly veiled cTTSs, however, occultation by dense accretion funnels or strong warps in the inner 
accretion disk \citep[e.g., induced by highly tilted magnetic fields,][]{Romanova03} might also generate profile distortions resembling those induced by brightness features;  
we did not include this option in the modeling to avoid making it overly complex by adding still more free parameters, but we kept it in mind when we analyzed the ZDI results.  
Of the five seasons for which we collected data, only two (2024 and 2025) featured a dense phase coverage to allow reliable brightness and magnetic imaging with ZDI, with a third 
season (2021) for which sparser sampling was achieved but on which we still applied ZDI (we kept in mind this caveat when discussing the results).  
Using both LSD Stokes $I$ and $V$ (instead of only Stokes $V$) profiles in the ZDI modeling ensures that the Zeeman broadening of line profiles is taken into 
account and that the reconstructed field is not severely underestimated \citep[e.g.,][]{Donati25b}.  

In practice, ZDI operates as outlined in previous studies.  Beginning with empty magnetic maps and brightness distributions and assuming that the star rotates as a solid body, 
ZDI iteratively adds information on both images, exploring the parameter space with conjugate gradient techniques and comparing the synthetic Stokes profiles of the current images 
with observed ones at each iteration, until it reaches the requested level of agreement with the data (i.e., a given \chisqr).
The surface of DO~Tau was divided into 5000 grid cells, each associated with a value of the local surface brightness (relative to the quiet photosphere).  We proceeded in a 
different way for the magnetic field, described as a spherical harmonics (SH) expansion using the formalism of \citet{Donati06b} in its revised implementation \citep{Lehmann22, 
Finociety22}, where the poloidal and toroidal components of the vector field are expressed with three sets of complex SH coefficients, $\alpha_{\ell,m}$ and $\beta_{\ell,m}$ 
for the poloidal component, and $\gamma_{\ell,m}$ for the toroidal component ($\ell$ and $m$ denoting the degree and order of the corresponding SH term in the expansion).  
We limited the SH expansion to $\ell=10$, consistent with the measured \vsini\ (see Table~\ref{tab:gpr}).  The inversion problem is ill-posed and requires some regularization 
as the number of parameters is much larger than the number of independent data points.  ZDI chooses the simplest solution, i.e., the solution with minimum information or maximum 
entropy that matches the data at the requested \chisqr\ level, following \citet[][]{Skilling84}.

\begin{figure*}[ht!] 
\flushleft{\large\bf \hspace{3.7cm}2021\hspace{3.3cm}2024\hspace{4.7cm}2025\vspace{-4mm}}
\center{\includegraphics[scale=0.28,angle=-90]{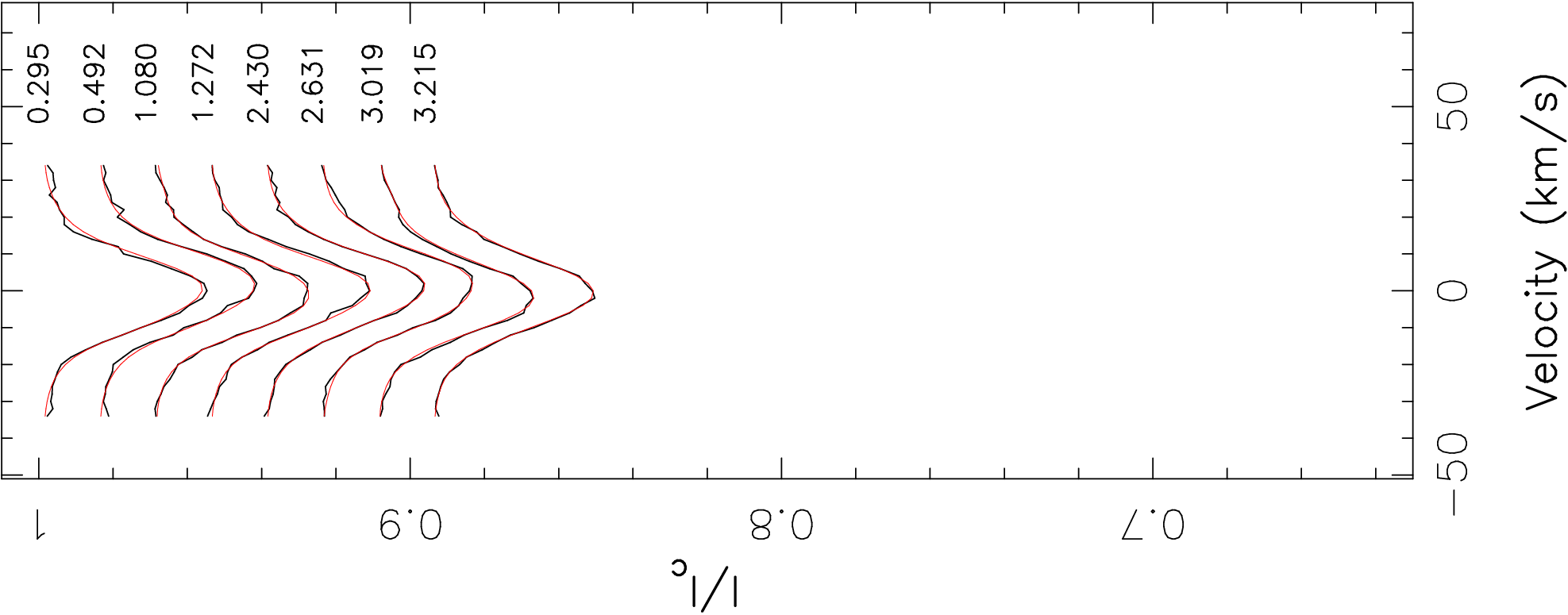}\hspace{2mm}\includegraphics[scale=0.28,angle=-90]{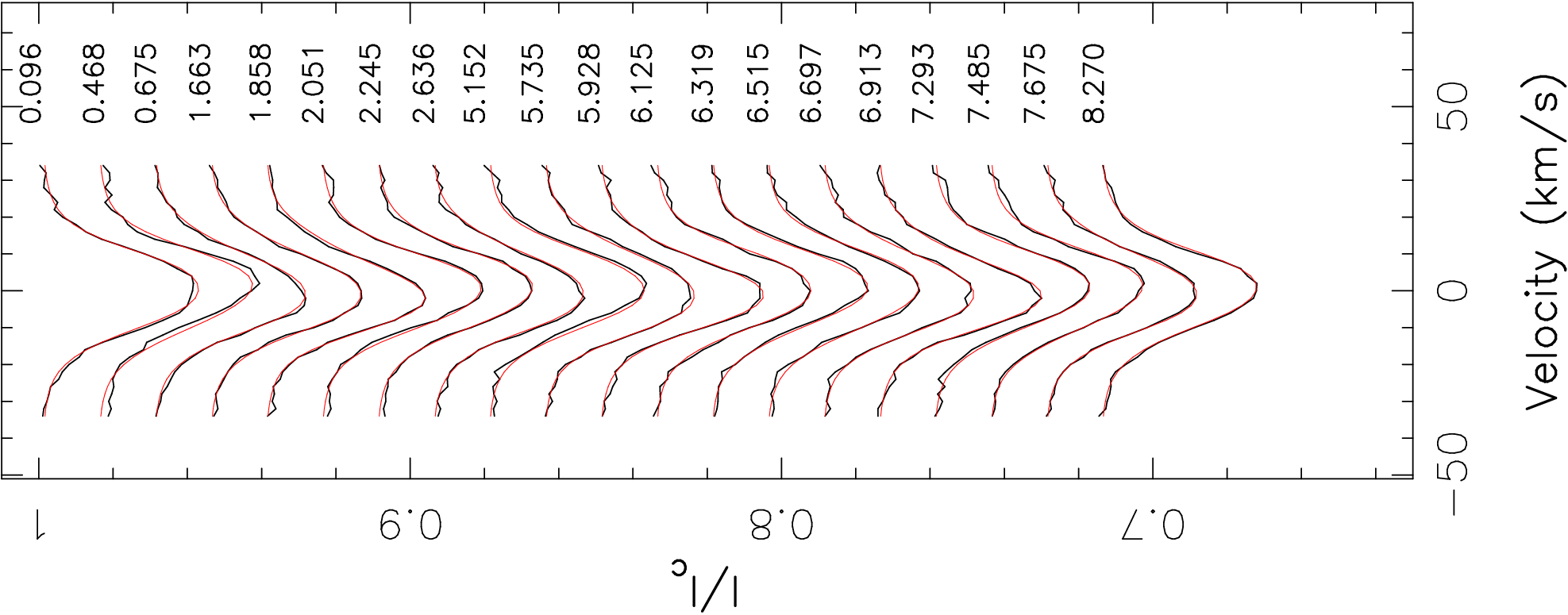}\hspace{2mm}\includegraphics[scale=0.28,angle=-90]{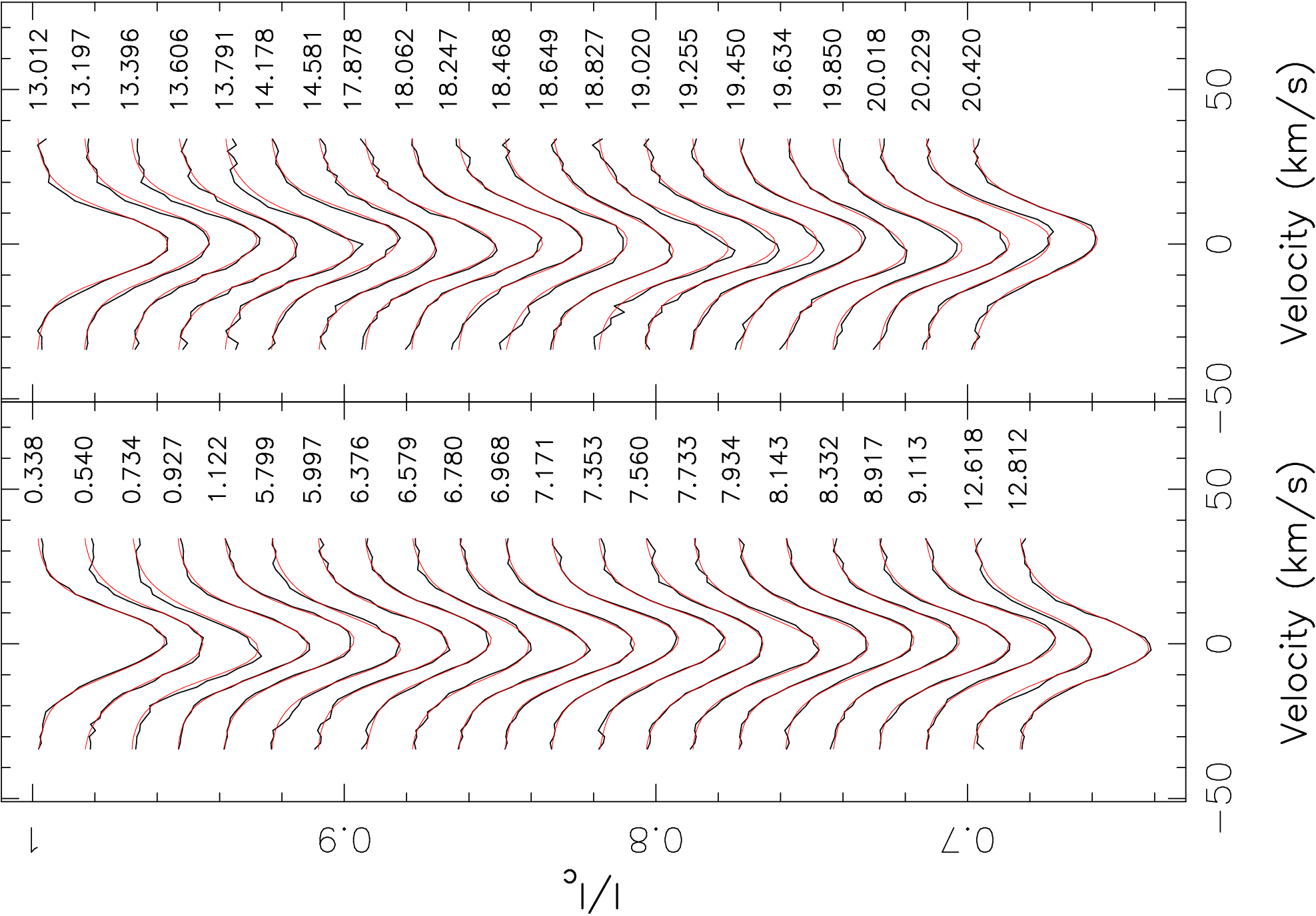}\vspace{1mm}}   
\center{\includegraphics[scale=0.28,angle=-90]{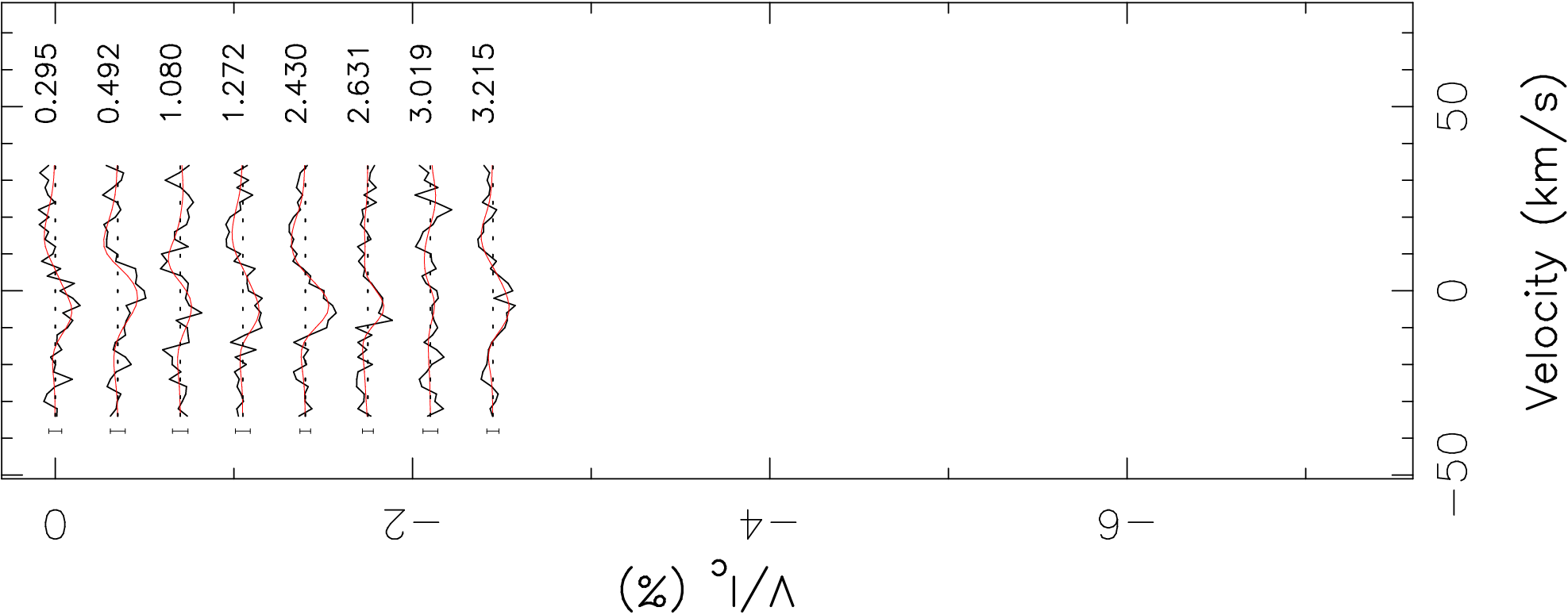}\hspace{2mm}\includegraphics[scale=0.28,angle=-90]{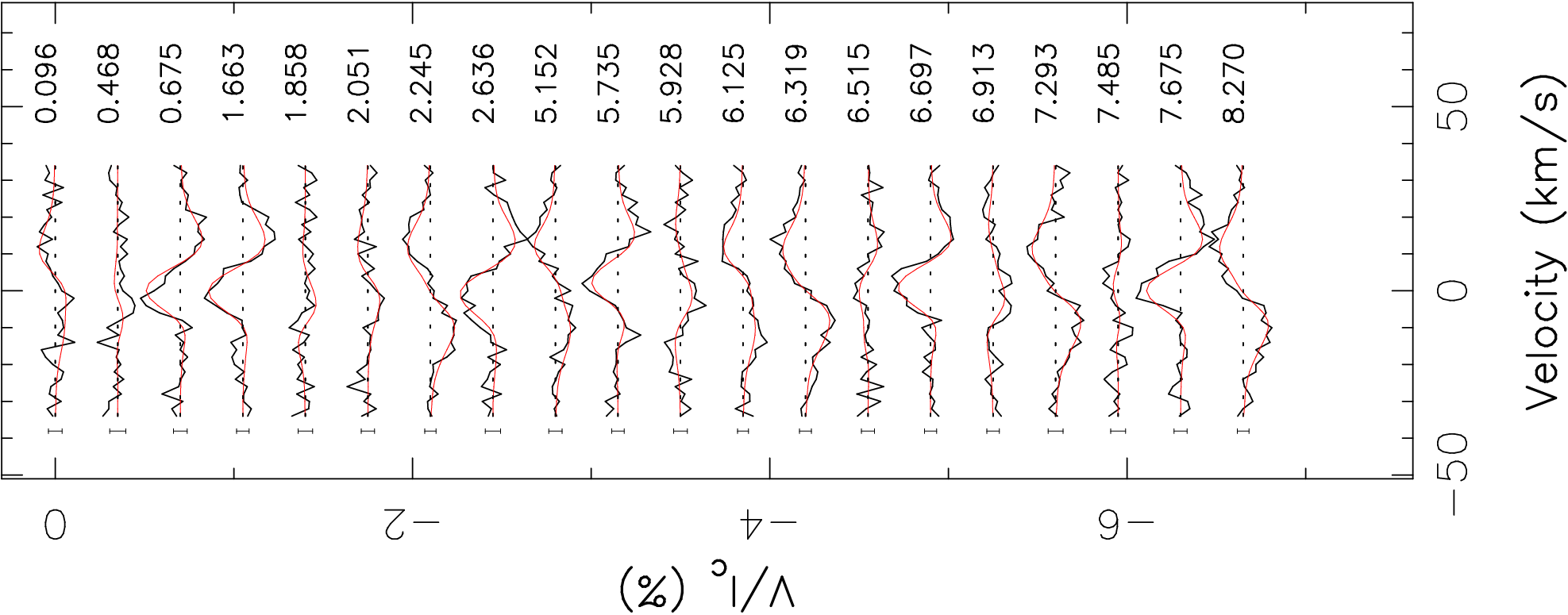}\hspace{2mm}\includegraphics[scale=0.28,angle=-90]{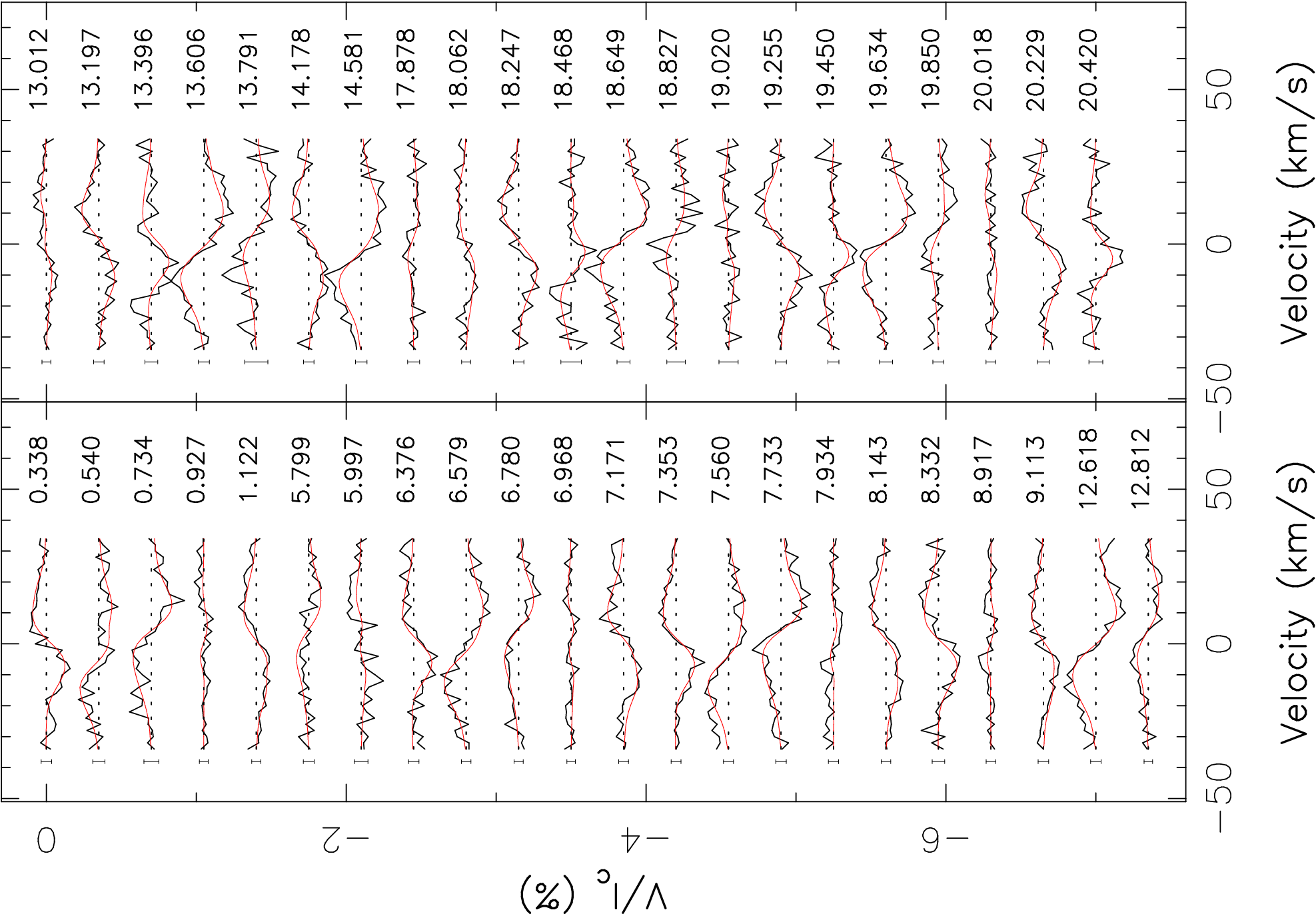}}   
\caption[]{Observed (thick black line) and modeled (thin red line) LSD Stokes $I$ (top row) and $V$ (bottom row) profiles of DO~Tau in 2021, 2024, and 2025 (from left to right).
Rotation cycles (counting from 60, 279, and 388 in 2021, 2024, and 2025, respectively, see Table~\ref{tab:log}) are indicated to the right of the LSD profiles, and $\pm$1$\sigma$ 
error bars are shown to the left of the Stokes $V$ profiles. }
\label{fig:fit}
\end{figure*}

\begin{figure*}[ht!] 
\centerline{\large\bf 2021\raisebox{0.30\totalheight}{\includegraphics[scale=0.40,angle=-90]{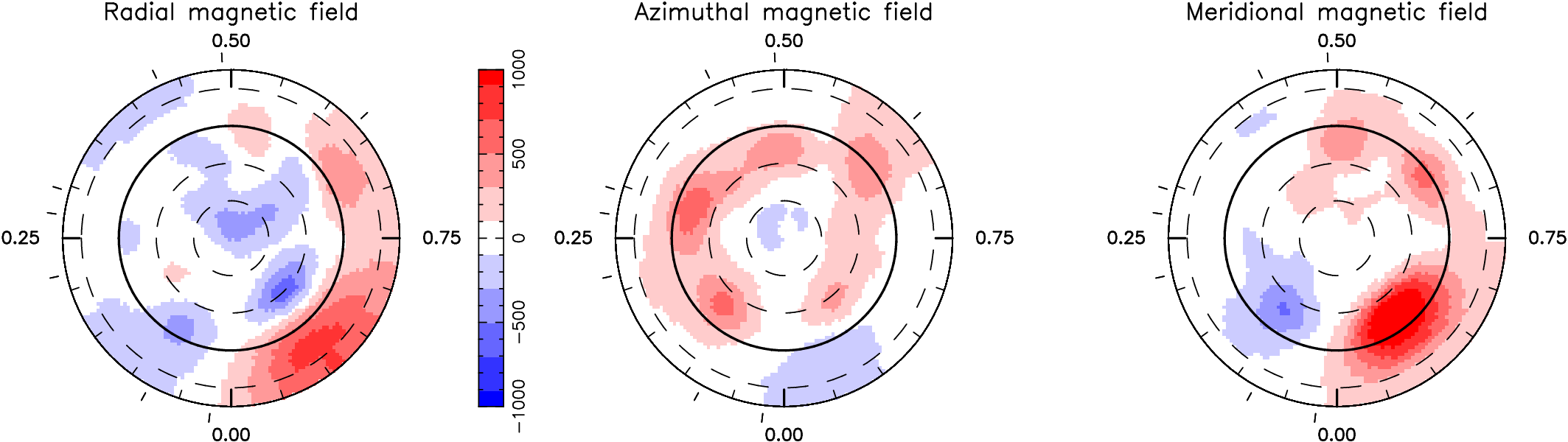}}\vspace{1mm}}
\centerline{\large\bf 2024\raisebox{0.30\totalheight}{\includegraphics[scale=0.40,angle=-90]{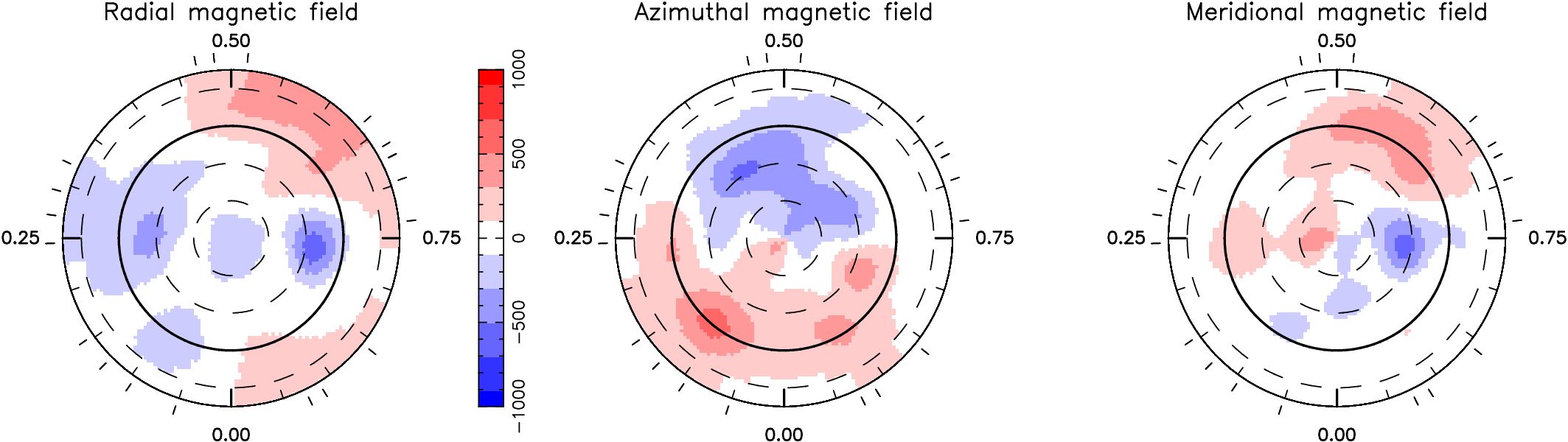}}\vspace{1mm}}
\centerline{\large\bf 2025\raisebox{0.30\totalheight}{\includegraphics[scale=0.40,angle=-90]{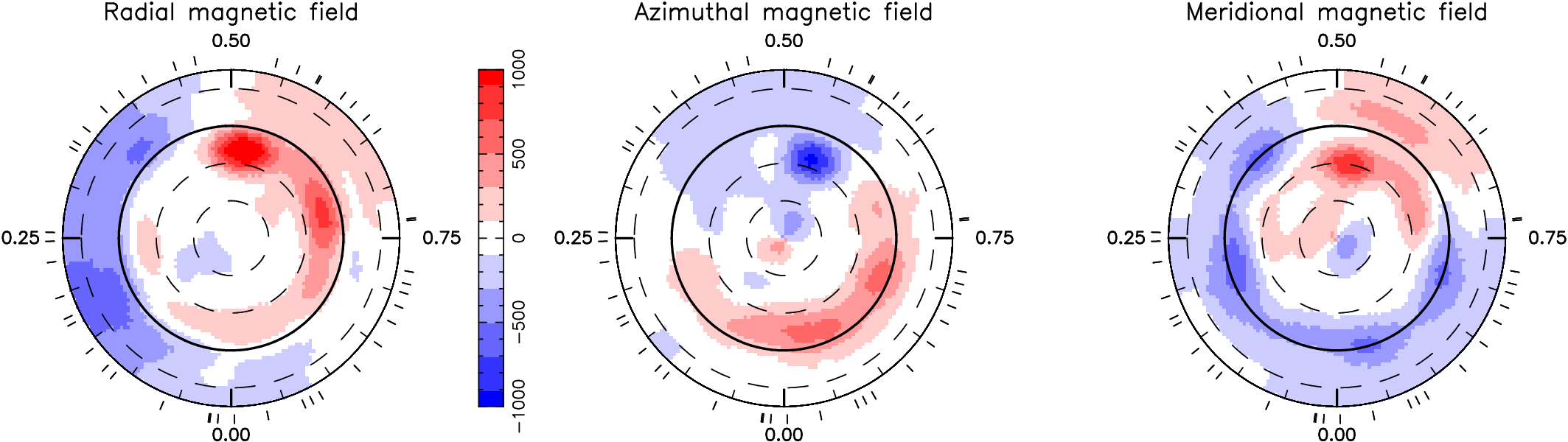}}} 
\caption[]{Reconstructed maps of the large-scale field of DO~Tau showing the radial, azimuthal and meridional components in spherical 
coordinates (left, middle and right columns, units in G), for seasons 2021, 2024, and 2025 (top to bottom rows, respectively).  
These maps, derived from the LSD Stokes $I$ and $V$ profiles of Fig.~\ref{fig:fit} using ZDI, are displayed in a flattened polar projection 
down to latitude $-45$\degr, with the north pole at the center and the equator depicted as a bold line.  Outer ticks mark the phases of
observations.  Positive radial, azimuthal, and meridional fields point outward, counterclockwise, and poleward, respectively. }
\label{fig:map}
\end{figure*} 

\begin{table}[t!]
\caption{Large-scale and small-scale fields measured from the magnetic topologies derived with ZDI in 2021, 2024, and 2025}
\centering
\resizebox{\linewidth}{!}{
\begin{tabular}{ccccccc}
\hline
         & \multicolumn{6}{c}{Stokes $I$ \& $V$ analysis ($f_I=0.9$, $f_V=0.2$, $\vD=3.0$~\kms)}    \\
\hline
Season   & <$B_V$> & <$B_I$> & <$B_s$> & \Bd  & tilt / phase & pol/axi   \\
         &  (kG)   & (kG)    & (kG)   & (kG) & (\degr / ) & (\%) \\
\hline
2021     & 0.39   & 1.8   & 1.4   & $-$0.19 &  50 / 0.31 & 63 / 27 \\
2024     & 0.30   & 1.4   & 1.4   & $-$0.19 &  45 / 0.16 & 72 / 17 \\
2025     & 0.40   & 1.8   & 1.5   &    0.32 &  66 / 0.68 & 76 / 34 \\
\hline
\end{tabular}}
\tablefoot{Columns 2 and 3 list the quadratically averaged large-scale field (<$B_V$>) and small-scale field (<$B_I$>) over the stellar surface.  
Column~4 gives the time-averaged small-scale field integrated over the visible hemisphere (<$B_s$>).  
Columns~5 to 7 list the polar strengths of the dipole component \Bd, the tilt of the dipole component to the rotation axis and the phase toward 
which it is tilted, and the amount of magnetic energy reconstructed in the poloidal component of the field and in the axisymmetric modes of this component.  
Typical error bars on field values, percentages and dipole tilts are equal to 20\%, 10\%, and 10\degr\ in 2024 and 2025, and twice as 
much in 2021 due to the sparser sampling. }
\label{tab:mag}
\end{table}

Local synthetic Stokes $I$ and $V$ profiles for each grid cell were computed using Unno-Rachkovsky's solution of the polarized radiative transfer equation in a plane-parallel 
Milne-Eddington atmosphere \citep{Landi04}, and assuming a linear center-to-limb darkening law for the continuum (with a coefficient of 0.3).  At each observed rotation 
phase, the spectral contributions from all visible cells were summed to obtain the overall synthetic profiles, whose mean wavelength and Land\'e factor mirror 
those of our LSD profiles (1750~nm and 1.2).  We assumed a Doppler width of the local profile of $\vD=3.0$~\kms, typical for such stars.  As in previous studies, 
we introduced a filling factor for the large-scale field, $f_V$, and another for the small-scale field, $f_I$, both assumed constant over the star.  This implies that 
each cell with a reconstructed field $B$ hosts a field strength $B/f_V$ in a fraction of the cell {\emr equal to $f_V$ ($f_I$) for Stokes $V$ ($I$) LSD profiles}.  
For this study, we set $f_I=0.9$ (consistent with the results of ZeeTurbo; see Sec.~\ref{sec:obs}) and $f_V=0.2$, yielding optimal fits as for similar young stars 
\citep[e.g.,][]{Donati25c}.  Synthetic light curves were computed in the same way by summing the contributions of all visible cells, estimated from their limb angle and 
reconstructed brightness at each rotation phase.  

Figure~\ref{fig:fit} shows the ZDI fits to the LSD Stokes $I$ and $V$ profiles for each season, and Fig.~\ref{fig:map} shows the corresponding reconstructed 
magnetic maps whose main characteristics are listed in Table~\ref{tab:mag}.  
As a byproduct of the ZDI fit to the LSD Stokes profiles, we inferred that the rotational line 
broadening of DO~Tau is equal to $\vsini=13.0\pm0.5$~\kms\ (in agreement with ZeeTurbo).  The derived magnetic topology was mainly poloidal over the three seasons, with the dipole 
component encompassing 25$-$55\% of the reconstructed poloidal field energy.  The field was unusually complex for a young fully convective star, featuring in particular 
a mostly nonaxisymmetric poloidal component and a significant toroidal component.  The only known cTTS with such a complex large-scale field is 
V2247~Oph \citep{Donati10}.  However, in contrast to DO~Tau, V2247~Oph exhibited rapid temporal evolution and strong differential rotation, suggesting a different 
origin for its magnetic complexity.  One option is that the strong accretion rate of DO~Tau affects its inner convection pattern \citep{Geroux16} and the associated 
dynamo processes, leading to more complex magnetic fields.  The large-scale field also evolved significantly from one season to the next, especially from 2024 
to 2025, when the dipole component switched sign (as anticipated from the \Bl\ variations; see Sec.~\ref{sec:mag} and Table~\ref{tab:mag}), a clear difference with 
the large-scale fields of weakly accreting cTTSs \citep{Zaire24,Donati24,Donati24b}.  
The small-scale field <$B_s$> of DO~Tau inferred with ZDI (1.5~kG) remained stable throughout the campaign as did $dT$ (see Sec.~\ref{sec:mag}), a good proxy 
for <$B$> \citep{Artigau24,Cristofari25}.  However, <$B_s$> is 40\% smaller than <$B$> derived with ZeeTurbo ($2.5\pm0.3$~kG) but still consistent with the 
estimate of \citet{Lopez-Valdivia21}.  The reconstructed field reached maximum strengths of 0.6 to 1.2~kG, consistent with the measurements of \citet{Dodin13} 
at the funnel footpoints.  

The brightness maps we simultaneously recovered with the magnetic maps all included a pair of bright and dark regions separated by about half a rotation cycle, directly 
reflecting the periodic distortions in the LSD Stokes $I$ profiles causing the observed RV changes (see Fig.~\ref{fig:fit} and Sec.~\ref{sec:rvs}).  Similar RV changes 
were also detected in the magnetically insensitive CO lines (see Sec.~\ref{sec:rvs}), demonstrating that these distortions are not caused by magnetic fields.  However, 
the photometric curves predicted by ZDI featured a semi-amplitude (of 6$-$8\%) larger than the observed fluctuations inferred from $dT$ (see Fig.~\ref{fig:pho}).  
A tentative interpretation is that the LSD Stokes $I$ profile distortions that ZDI interpreted as brightness spot signatures were actually caused by something else than 
temperature inhomogeneities, e.g., a corotating accretion funnel or a strong warp of the inner disk that regularly crossed and partially occulted the star, thereby 
generating no $dT$ variations (as opposed to brightness spots).

\section{RV modeling}
\label{sec:rvs}

As mentioned above, Stokes $I$ LSD profiles of the atomic lines of DO~Tau exhibited clear rotationally modulated distortions generating RV changes 
that ZDI interpreted as caused by brightness regions at the stellar surface.  The bisector span (BIS) of the LSD profiles of atomic lines 
varied by about $\pm$1~\kms\ (see Table~\ref{tab:log}) and showed clear rotational modulation anticorrelated with RVs ($R\simeq$$-0.68$), 
further indicating that RV changes were mostly due to {\emr activity at the stellar surface (or to corotating occulting features in the inner disk)}.  
A similar behavior was observed in the CO lines.  Modeling the RV activity jitter with QP GPR yielded semi-amplitudes $\theta_1$ of $\simeq$0.9~\kms\ for LSD profiles 
of atomic and CO lines, and $\simeq$0.5~\kms\ for LBL RVs, with all three GP models varying in phase with one another (as expected since the three RV sets are strongly correlated).  
The derived period of the RV modulation, equal to $5.129\pm0.003$~d in all three cases, is fully consistent with that derived from \Bl\ (see Sec.~\ref{sec:mag}).  

\begin{table}[t!]
\caption{MCMC results of the modeling of LBL RVs}
\centering
\resizebox{\linewidth}{!}{
\begin{tabular}{cccc}
\hline
Parameter          & No planet                 & Planet b        & Prior \\
                   &                           &                 &       \\
\hline
$\theta_1$ (\kms)  & $0.50^{+0.13}_{-0.10}$ & $0.53^{+0.13}_{-0.11}$ & mod Jeffreys ($\sigma_{\rm RV}$) \\
$\theta_2$ (d)     & $5.129\pm0.003$      & $5.130\pm0.003$      & Gaussian (5.1; 0.3) \\
$\theta_3$ (d)     & 600                  & 600                  & fixed \\
$\theta_4$         & 0.7                  & 0.7                  & fixed \\ 
$\theta_5$ (\kms)  & $0.21\pm0.04$        & $0.13\pm0.04$        & mod Jeffreys ($\sigma_{\rm RV}$) \\
\hline
$K_b$ (\kms)       &                      & $0.22\pm0.04$        & mod Jeffreys ($\sigma_{\rm RV}$) \\
$P_b$ (d)          &                      & $21.14\pm0.02$       & Gaussian (21.1; 0.2)        \\ 
$T_b$ (2459000+)   &                      & $1318.1\pm0.6$       & Gaussian (1318; 5)            \\ 
$M_b \sin i$ (\mjup) &                    & $2.0\pm0.4$          & from $K_b$, $P_b$, \mstar \\
$M_b$ (\mjup)      &                      & $2.8\pm0.6$          & assuming $i=45$\degr  \\
$a_b$ (au)         &                      & $0.12\pm0.01$        &  \\ 
\hline
\chisqr            & 1.71                 & 1.07                 &   \\
rms (\kms)         & 0.26                 & 0.21                 &   \\
$\log \mathcal{L}_M$ & $-28.1$            & $-14.8$              &   \\
$\Delta \log \mathcal{L}_M$ & 0.0         & 13.3                 &   \\
\hline
\end{tabular}}
\tablefoot{We list the recovered GP and candidate planet parameters with their error bars, as well as the priors used where relevant, for the model without planet (column 2) 
and the best model with a candidate planet at 21.14~d (column 3).  The last four rows report the \chisqr, the rms of the best fit to the RV data, the associated marginal 
logarithmic likelihood, $\log \mathcal{L}_M$, and the marginal logarithmic likelihood variation, $\Delta \log \mathcal{L}_M$, relative to the case without planet.}
\label{tab:rv}
\end{table}

\begin{figure}[ht!]
\centerline{\includegraphics[scale=0.38,angle=-90]{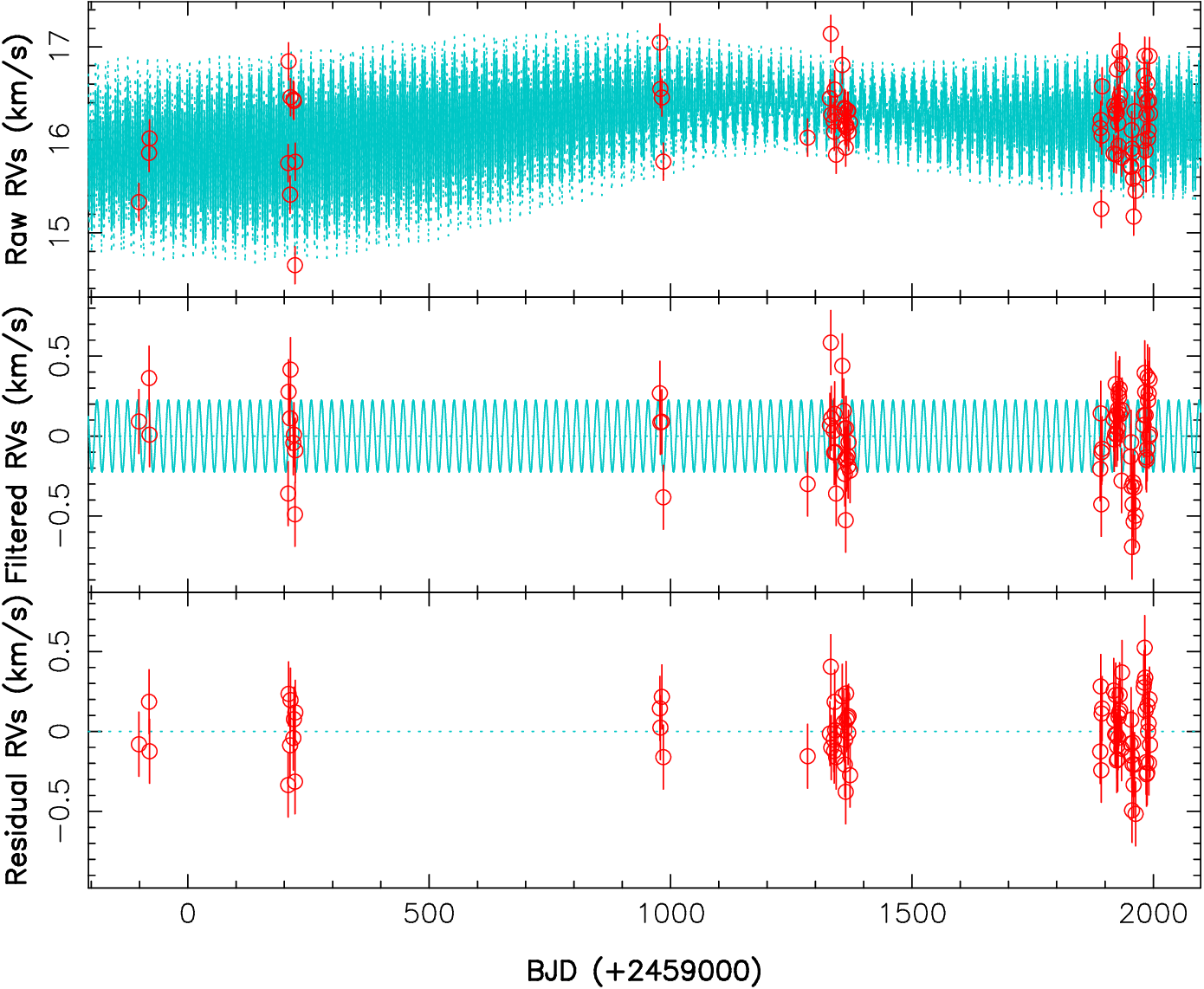}}
\caption[]{Raw (top), filtered (middle), and residual (bottom) LBL RVs of DO~Tau (red dots) over the observing period.  The top panel shows the MCMC fit to the data, including a 
QP GPR modeling of the activity and a candidate planet on a 21.14~d circular orbit (cyan).  The middle panel shows the planet RV signature (cyan) after activity was filtered out.  
The rms of the residuals is 0.21~\kms.  A zoom-in on the 2025 data is shown in Fig.~\ref{fig:rv2}.  }
\label{fig:rv}
\end{figure}

\begin{figure*}[ht!]
\centerline{\includegraphics[scale=0.37,angle=-90]{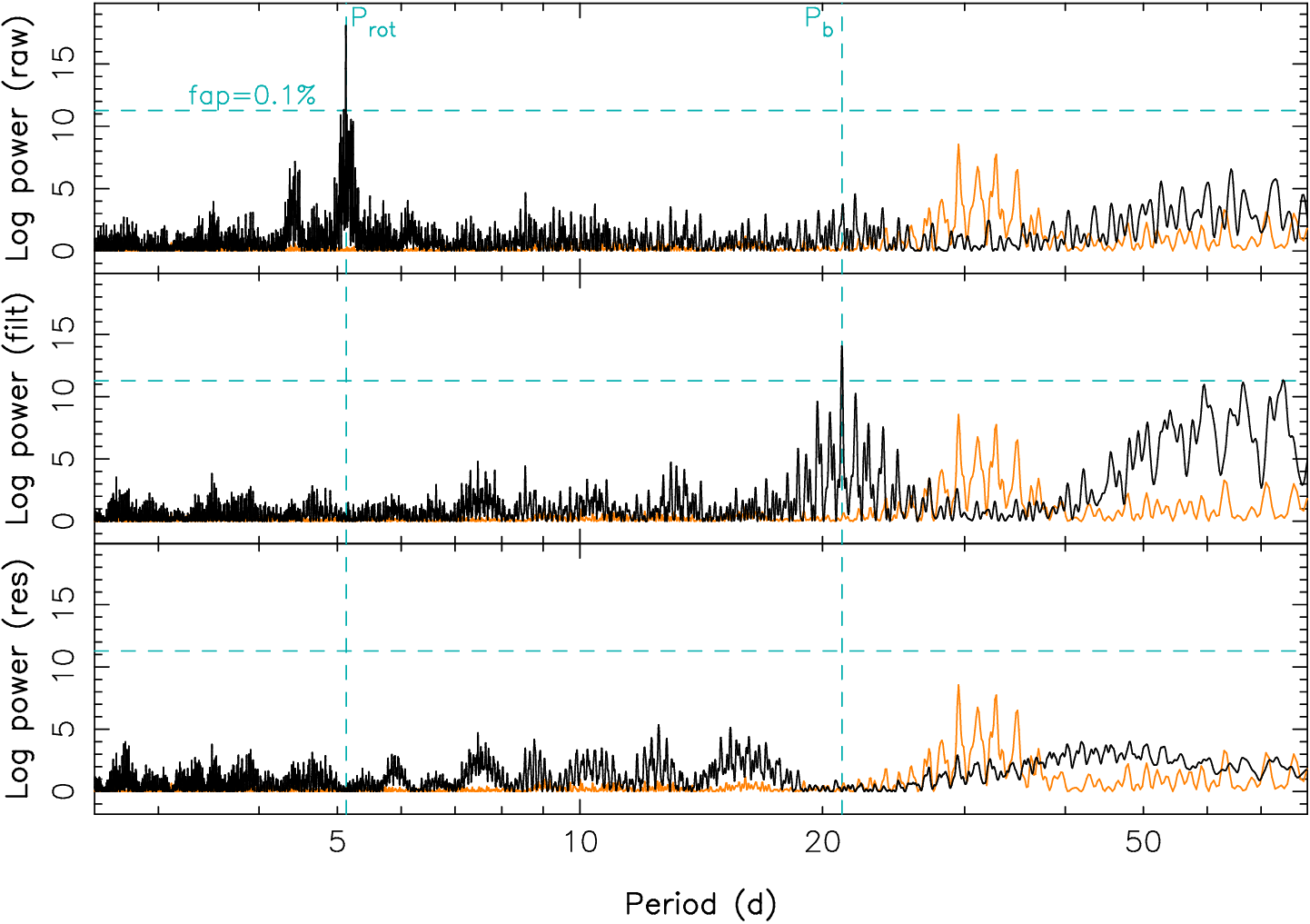}\hspace{3mm}\includegraphics[scale=0.32,angle=-90]{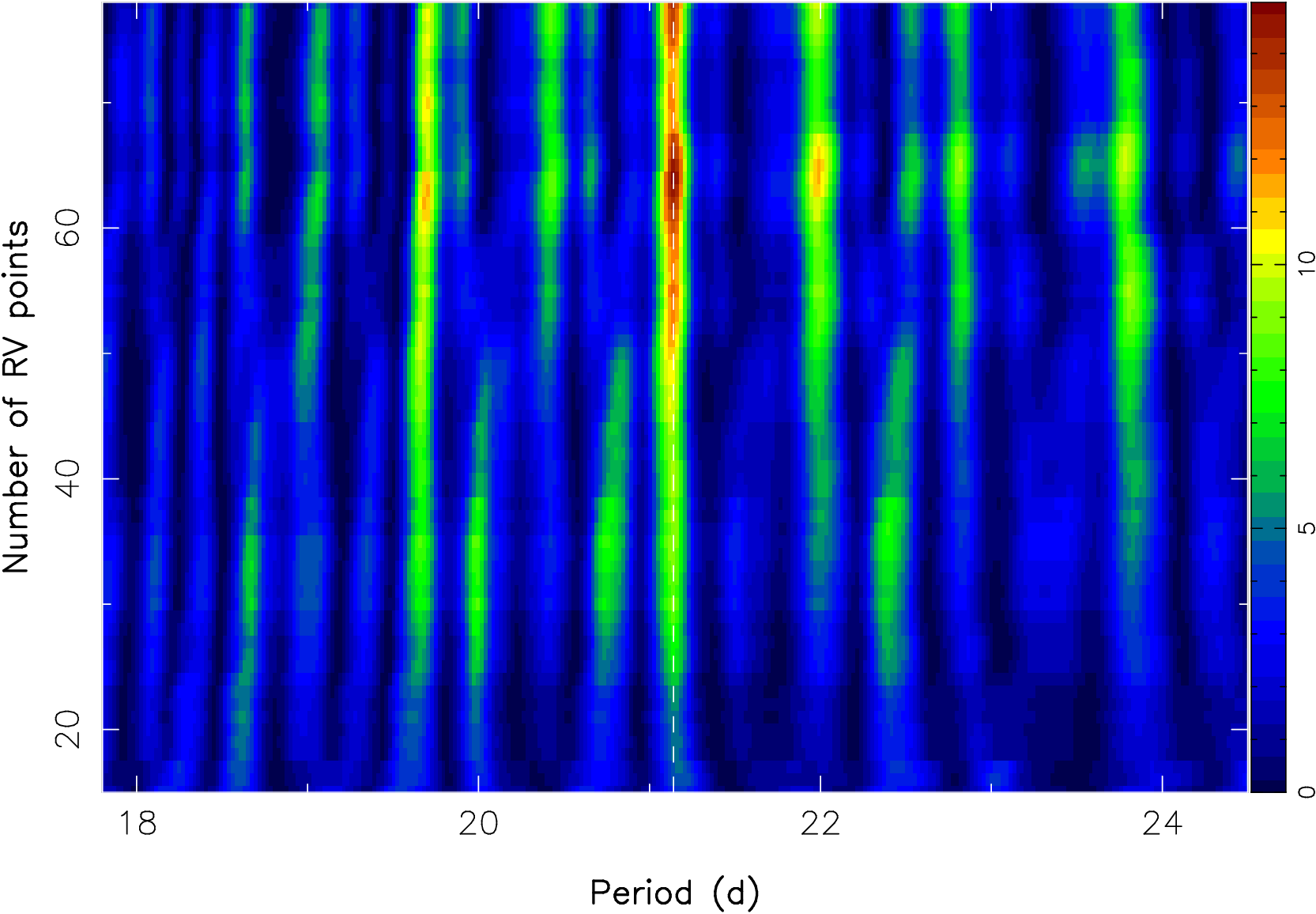}}  
\caption[]{Standard and stacked Lomb-Scargle periodograms of the RV data. 
Left panel: Periodogram of the raw (top), filtered (middle), and residual (bottom) LBL RV data, including a candidate planet on a 21.14~d circular orbit in the MCMC 
modeling.  The dashed vertical cyan lines trace the stellar rotation period and the candidate planet orbital period, and the dashed horizontal lines indicate a 0.1\% FAP level 
in the periodogram of the RV data.  
The orange curve depicts the periodogram of the window function. Right panel: Stacked periodograms of the filtered LBL RVs, as a function of the number of RV points included in 
the Fourier analysis.  The color-scale indicates the logarithmic power in the periodogram.  The vertical dashed line traces the candidate planet orbital period. } 
\label{fig:per} 
\end{figure*}

\begin{figure}[ht!]
\centerline{\includegraphics[scale=0.45,angle=-90]{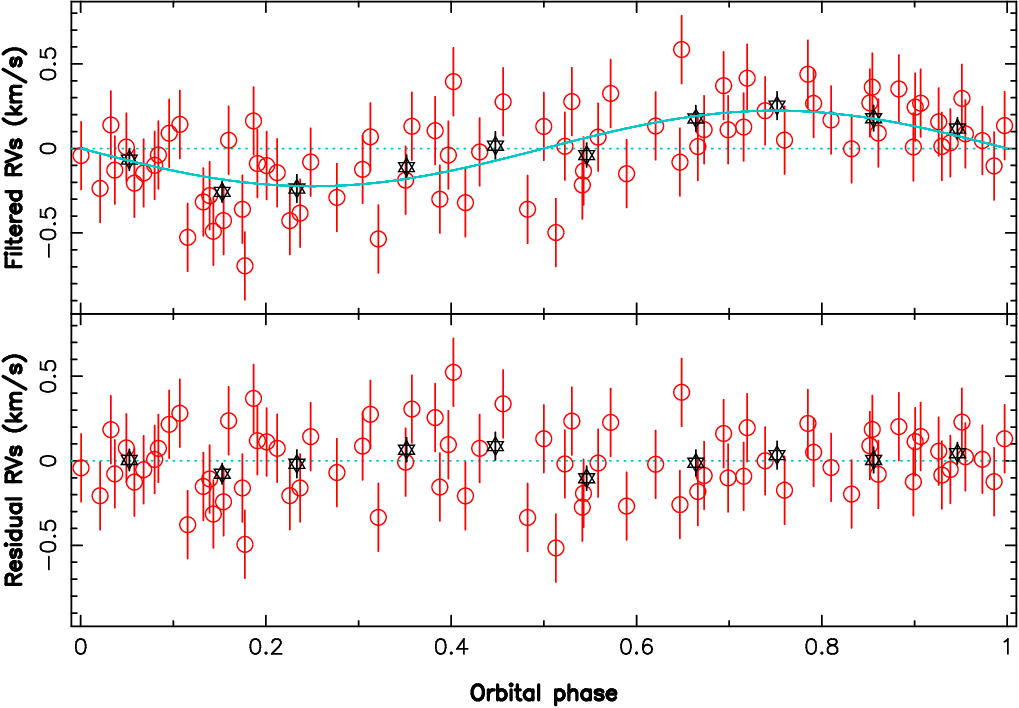}}
\caption[]{Phase-folded filtered (top) and residual (bottom) LBL RVs for the model with a candidate planet on a 21.14~d circular orbit.  
The red dots show the measured RVs, and the black stars indicate average RVs over ten even phase bins. } 
\label{fig:rvf}
\end{figure} 

To investigate whether DO~Tau might host a massive planet in its close circumstellar environment, we also modeled RVs by adding a sine wave to describe the RV 
effect of a putative massive planet on a circular orbit in addition to the QP GP term describing activity.  We studied the periodograms of the 
residuals of the activity-only GPR fit for the three sets of RVs and found that the LBL RVs exhibited the largest peaks at a period of 21.1~d and its most prominent one-month 
alias (66.5~d) with false-alarm probabilities (FAPs) of $\simeq$0.1\% (each peak surrounded by one-year aliases).  The 21.1~d peak also dominated the residual periodogram 
of the RVs of the CO lines (but at a weaker level, with an FAP of a few percent and a one-month alias at 68.2~d), accompanied by another peak at 7.5~d.  This is consistent with LBL 
RVs being dominated by information from molecular lines (including CO lines) in M stars.  In contrast, the residual periodogram of RVs of the atomic lines instead showed a 
main peak at 15.6~d (again with an FAP of a few percent) plus a number of weaker peaks (including one at 21.1~d).  That the RVs of atomic lines did not yield the same results as 
the other two RV sets suggests that these periodogram peaks might not be related to close-in planets (expected to affect all RV sets in an identical way) but 
rather to density structures in the inner accretion disk.  We nonetheless explored the possibility that the main period in LBL RVs (at 21.1~d), also detected in RVs of CO 
lines, probes a candidate planet, to estimate its minimum mass in case it is real.  

We achieved this by comparing the result of an MCMC fit to the LBL RVs with a model featuring a candidate planet and a QP GP describing activity, with that of a similar MCMC 
fit assuming no planet, the difference in $\log \mathcal{L}_M$ yielding the logarithmic Bayes factor in favor of the candidate planet.  We found that the optimal orbital 
period was $P_b=21.14\pm0.02$~d corresponding to an orbital distance of $0.12\pm0.01$~au, with a semi-amplitude of $K_p=0.22\pm0.04$~\kms\ indicating a minimum planet mass 
of $2.0\pm0.4$~\mjup\ (planet mass of $2.8\pm0.6$~\mjup\ assuming $i=45$\degr).  The detected signal is apparently significant, with a logarithmic Bayes factor 
of $\Delta \log \mathcal{L}_M=13.3$.  Assuming that the candidate planet is on a Keplerian orbit yielded only a marginal improvement in $\log \mathcal{L}_M$ over the case with a 
circular orbit.  The full result of the MCMC fit is outlined in Table~\ref{tab:rv} and the corresponding fit to the LBL RVs is shown in Fig.~\ref{fig:rv} 
(with a zoom-in on the 2025 data in Fig.~\ref{fig:rv2}).  The corresponding periodograms of the raw, filtered and residual LBL RVs are shown in the left panel of Fig.~\ref{fig:per} 
where the peak at $P_b$ features an FAP level of 0.006\%.  The stacked periodogram of the filtered RVs, displayed in the right panel of the same figure, exhibits 
a peak at $P_b$ that strengthened irregularly as the number of points increased.  This fluctuating peak strength might come from stochastic variability induced by accretion;  
alternatively, it might reflect that the detected RV signal comes from activity associated with a variable disk structure and not from a candidate planet.  The corresponding 
phase-folded filtered and residual RV curves are shown in Fig.~\ref{fig:rvf}.  

Finally, we ran injection recovery tests to investigate the semi-amplitude threshold above which detections can be considered as reliable, for orbital periods of 6 to 100~d 
(see Fig.~\ref{fig:det}). We found that, for periods in the range 8$-$25~d, $K_b\simeq0.23$~\kms\ (corresponding to minimal masses of the candidate planet of 1.5 to 
2.2~\mjup) is required on average to ensure a reliable detection with $\Delta \log \mathcal{L}_M=10$.  {\emr Beyond the drop in sensitivity
for periods close to the synodic period of the Moon}, the planet detectability weakens slightly, with $K_b\simeq0.25$~\kms\ being required for a secure detection at an orbital 
period of 100~d (corresponding to a minimum mass 3.8~\mjup).  This confirms that the candidate planet we identified is close to the detection threshold of our dataset.

\section{Accretion and activity}
\label{sec:act}

The spectrum of DO~Tau is subject to strong veiling, indicating high accretion rates at the surface of the star.  We measured 
this veiling by comparing the equivalent widths (EW) of the LSD Stokes $I$ profiles of DO~Tau with those of the weak-line T~Tauri stars TWA~25 and V819~Tau, whose spectral 
types are similar \citep[as in][]{Sousa23}.  We obtained a first estimate, $r_{JH}$, from atomic lines (concentrating mostly in the $JH$ band), and a second one, $r_K$, 
from the CO bandhead lines (redward of 2.28~\mic).  The values we obtained, listed in Table~\ref{tab:log}, range from 1.0 to 3.6 for $r_{JH}$ (median 1.5), and from 
1.6 to 6.1 for $r_K$ (median 2.7).  We found that $r_K$ correlated well with $r_{JH}$ ($R\simeq$$0.86$), with $r_K$ being twice as large as $r_{JH}$ on average.    
However, neither $r_{JH}$ nor $r_K$ showed rotational modulation, each exhibiting various peaks in the periodogram (including one at $\simeq$21~d).  

\begin{figure*}[ht!]
\resizebox{0.95\textwidth}{!}{
\centerline{\hspace{-2mm}\includegraphics[scale=0.32,angle=-90]{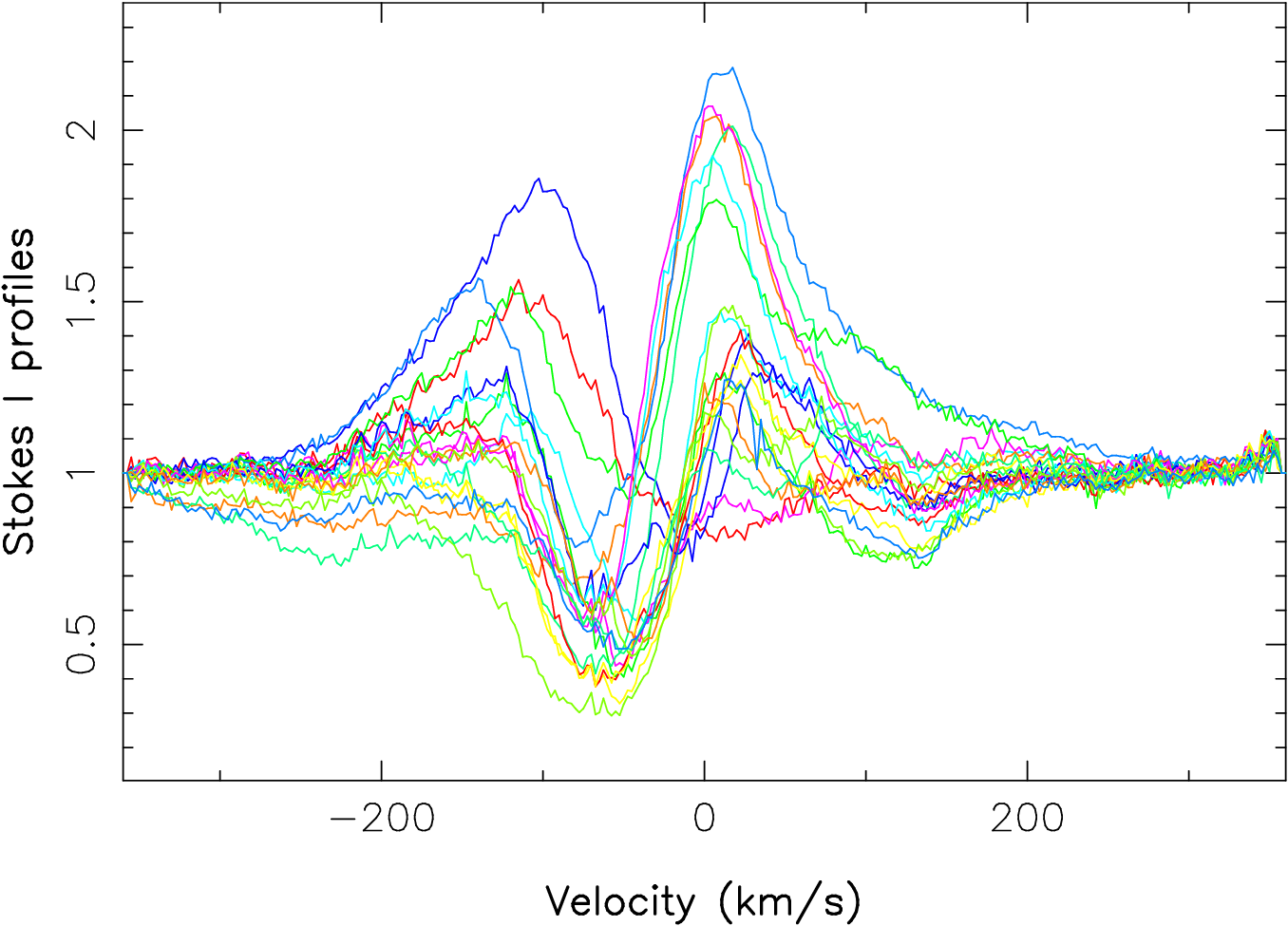}\hspace{14.5mm}\includegraphics[scale=0.32,angle=-90]{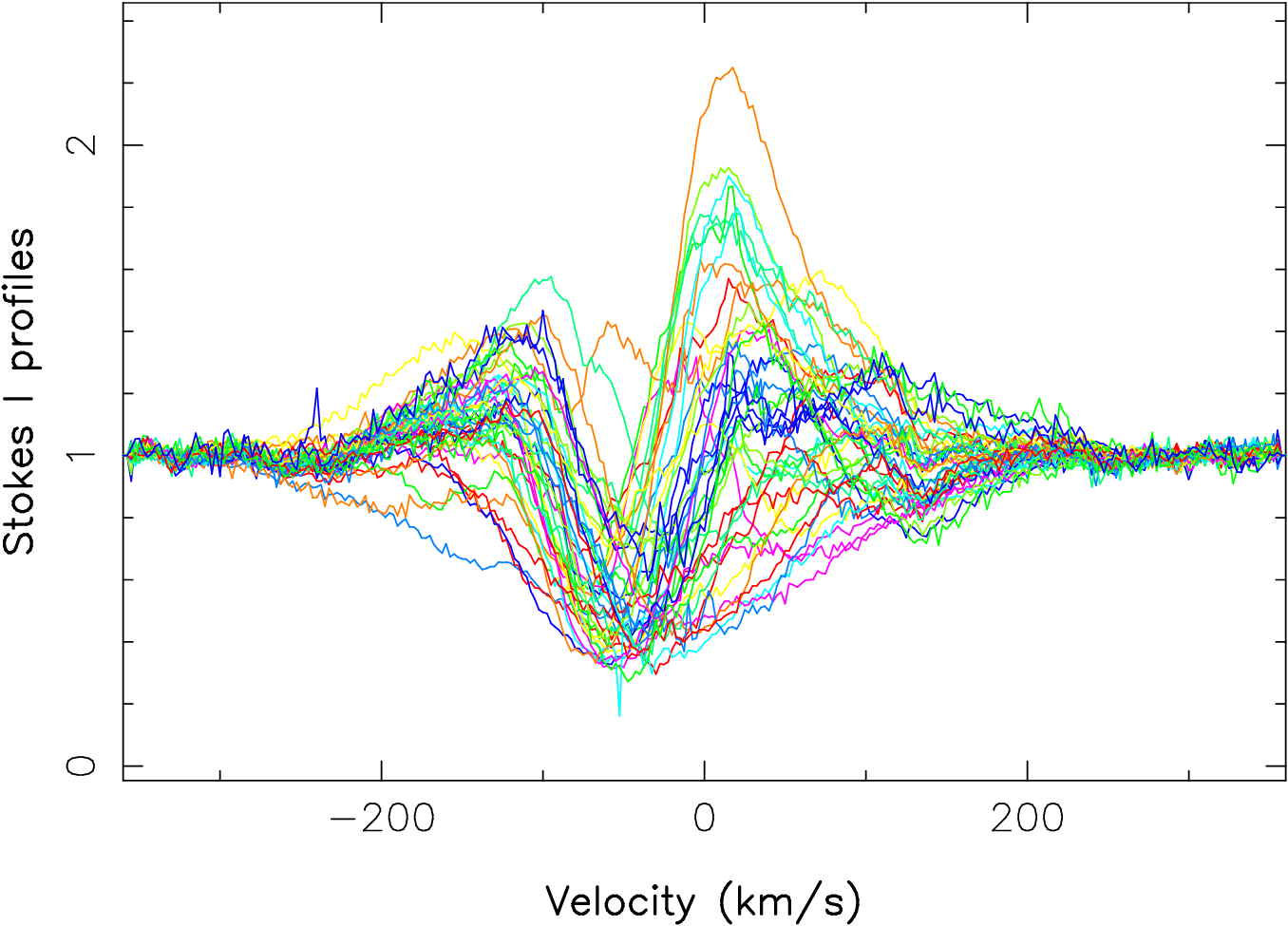}\vspace{2mm}}}
\resizebox{0.95\textwidth}{!}{
\centerline{\includegraphics[scale=0.55,angle=-90]{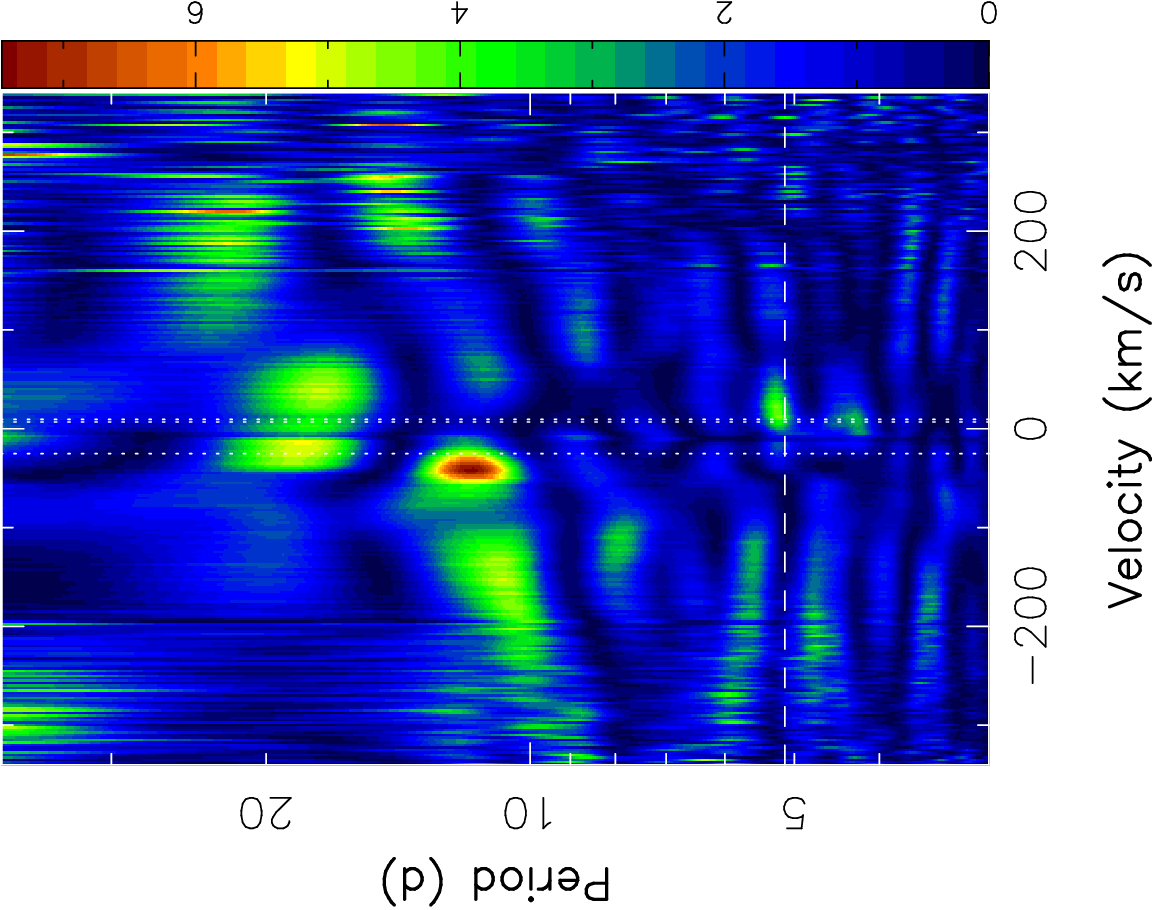}\hspace{3mm}\includegraphics[scale=0.55,angle=-90]{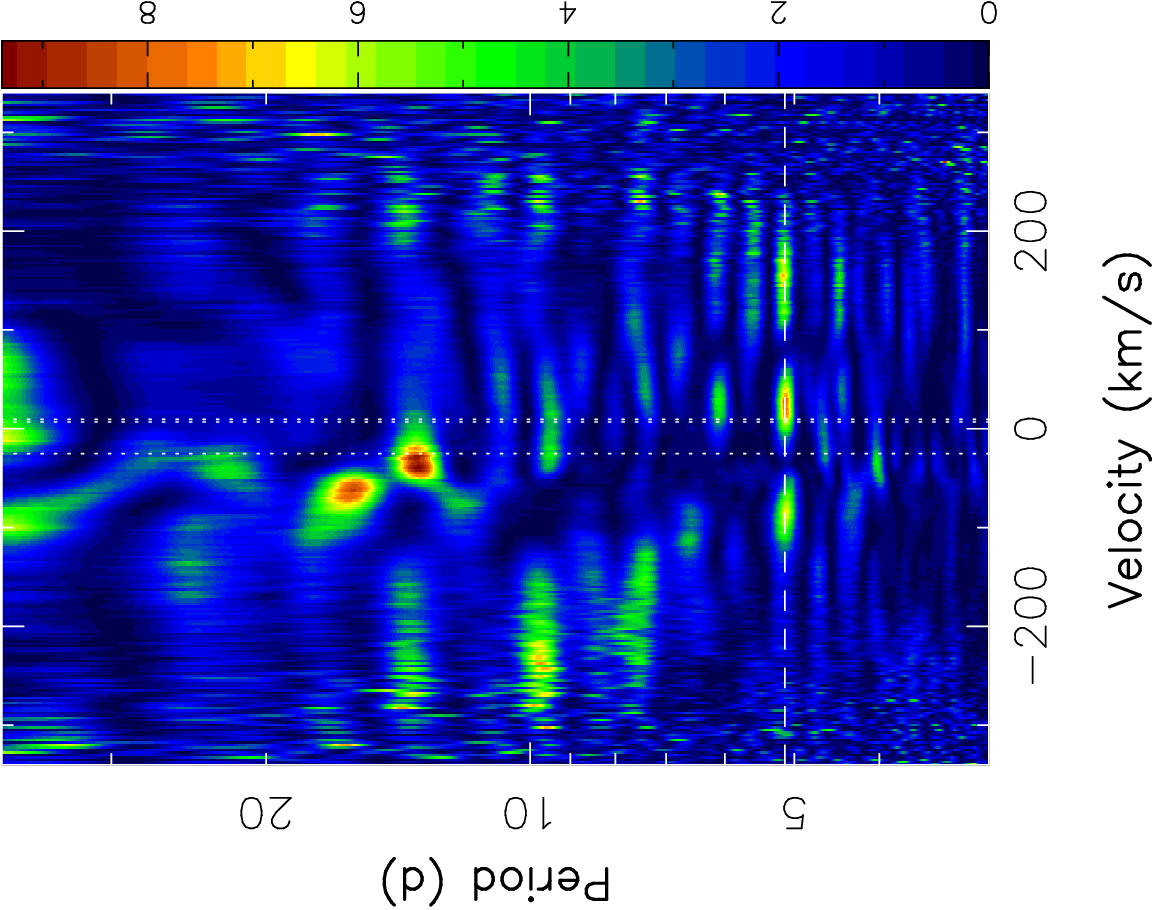}}}
\caption[]{
Stacked Stokes $I$ profiles (top plots) and 2D periodograms (bottom plots) of the 1083.3~nm \hei\ IRT in the stellar rest frame for seasons 2024 (left panels) and 2025 (right panels).  
The dashed horizontal line traces \Prot\ and the vertical dotted lines depict the velocities of the three components of the \hei\ triplet.  The color-scale traces the logarithmic 
power in the periodogram.  Only the main peaks (yellow to red and extending over at least several velocity bins) are likely to be significant. }
\label{fig:hei}
\end{figure*}

We also examined the 1083~nm \hei\ triplet and constructed the 2D periodograms of its profile for seasons 2024 and 2025 (see Fig.~\ref{fig:hei}).  The line profile featured
redshifted emission associated with accretion processes (extending to 250~\kms) as well as blueshifted absorption and emission probing the stellar and disk winds 
(extending to $-250$~\kms).  It also featured redshifted absorption (between 50 and 200~\kms) tracing accretion funnels or strong warps in the inner accretion disk regularly 
crossing the line of sight.  Rotational modulation was detected in the central emission peak, possibly related to chromospheric emission \citep[as in PDS~70,][]{Donati24c};  
in 2025, rotational modulation was also observed in the redshifted absorption.  
In the blueshifted component probing the stellar and disk winds, longer periods, ranging from 13~d (in 2024 and 2025) to $\simeq$16~d (in 2025 only), were also observed at 
low velocities, suggesting transient phenomena occurring at the base of the wind every few stellar rotation cycles.  In 2025, i.e., the season for which our dataset was 
most extended, the 2D periodogram rapidly evolved with time, with the blueshifted component showing a period of $\simeq$17~d in the first half of the 
dataset, then clear rotational modulation in the second half (see Fig.~\ref{fig:hei2}).  Rapid evolution was likely also at work in previous epochs.   

\begin{figure*}[ht!]
\resizebox{0.95\textwidth}{!}{
\centerline{\hspace{-2mm}\includegraphics[scale=0.32,angle=-90]{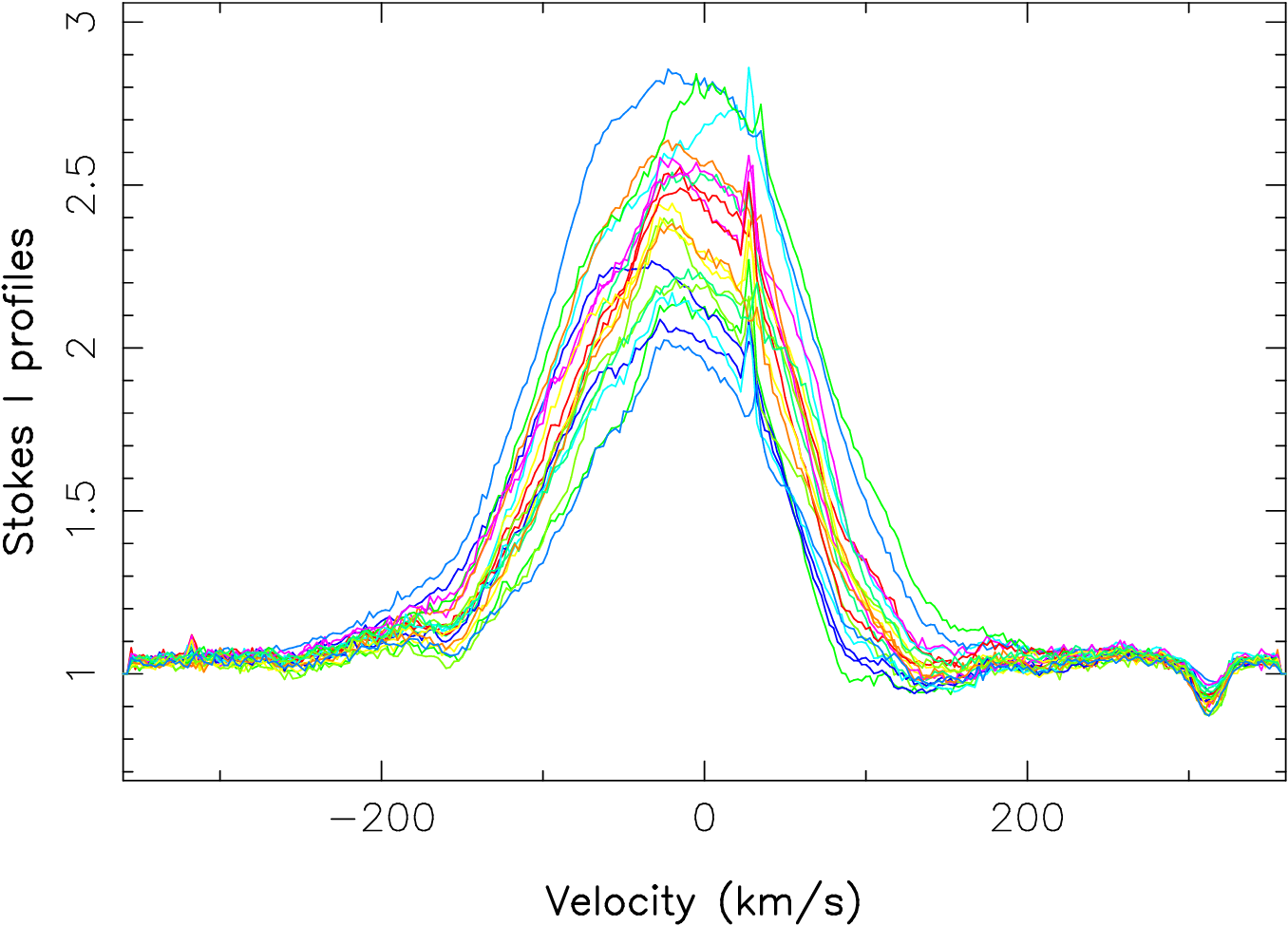}\hspace{14.5mm}\includegraphics[scale=0.32,angle=-90]{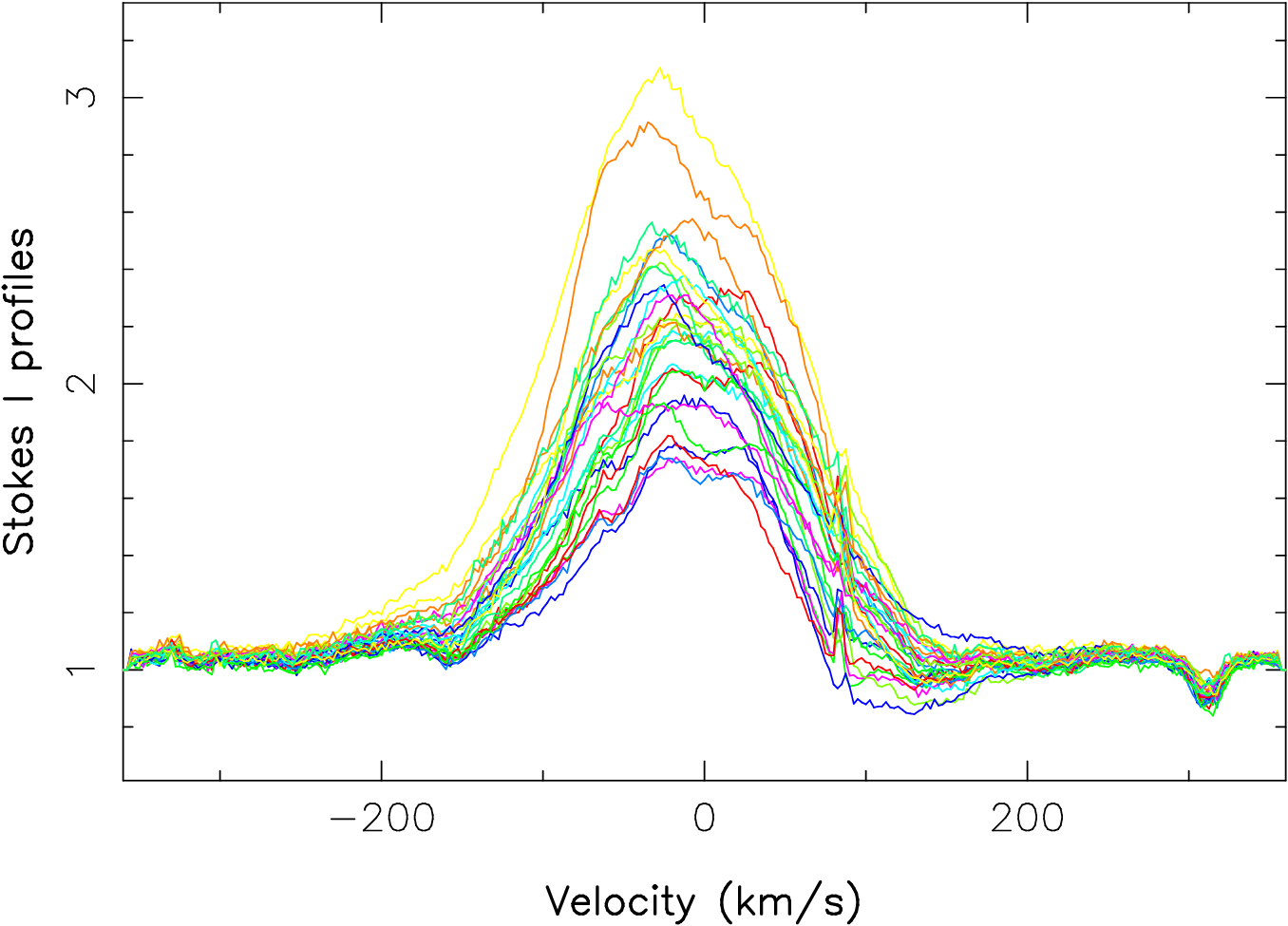}\vspace{2mm}}}
\resizebox{0.95\textwidth}{!}{
\centerline{\includegraphics[scale=0.55,angle=-90]{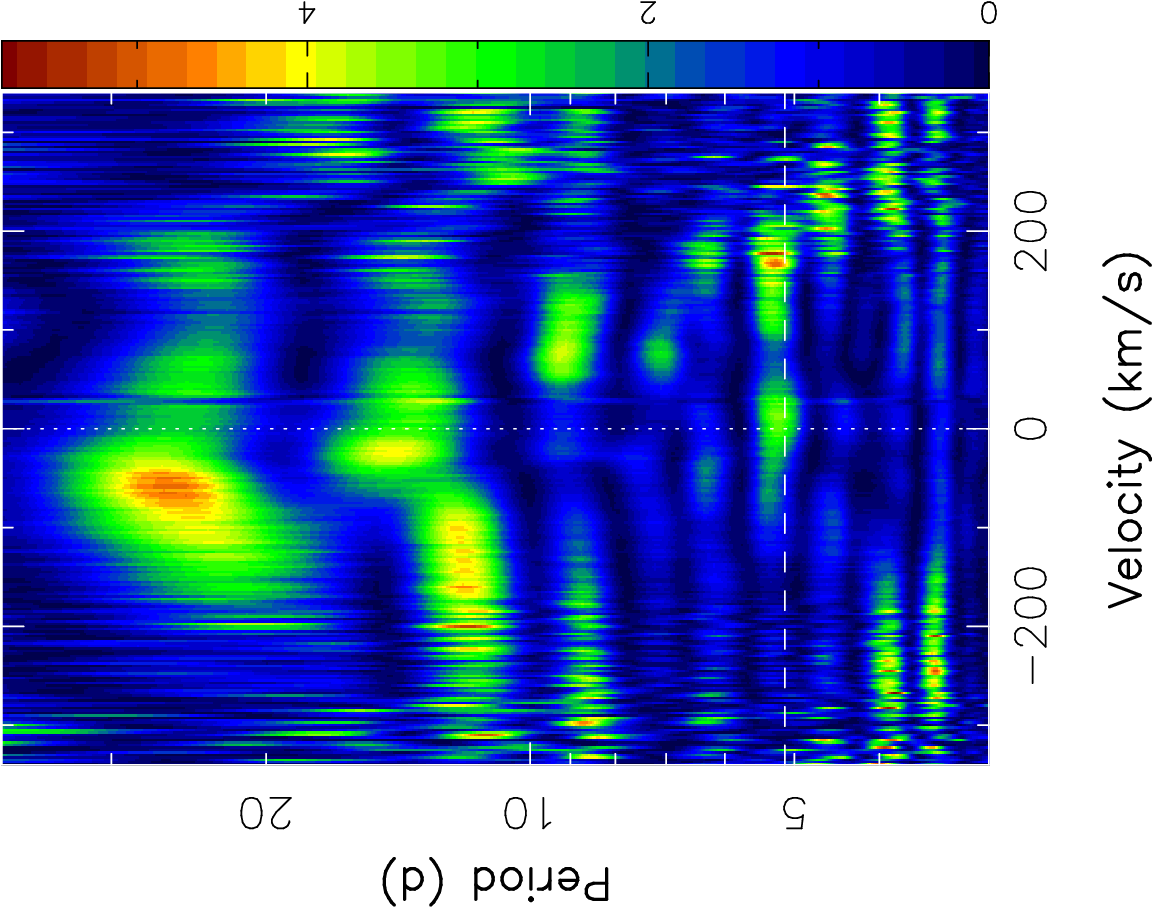}\hspace{3mm}\includegraphics[scale=0.55,angle=-90]{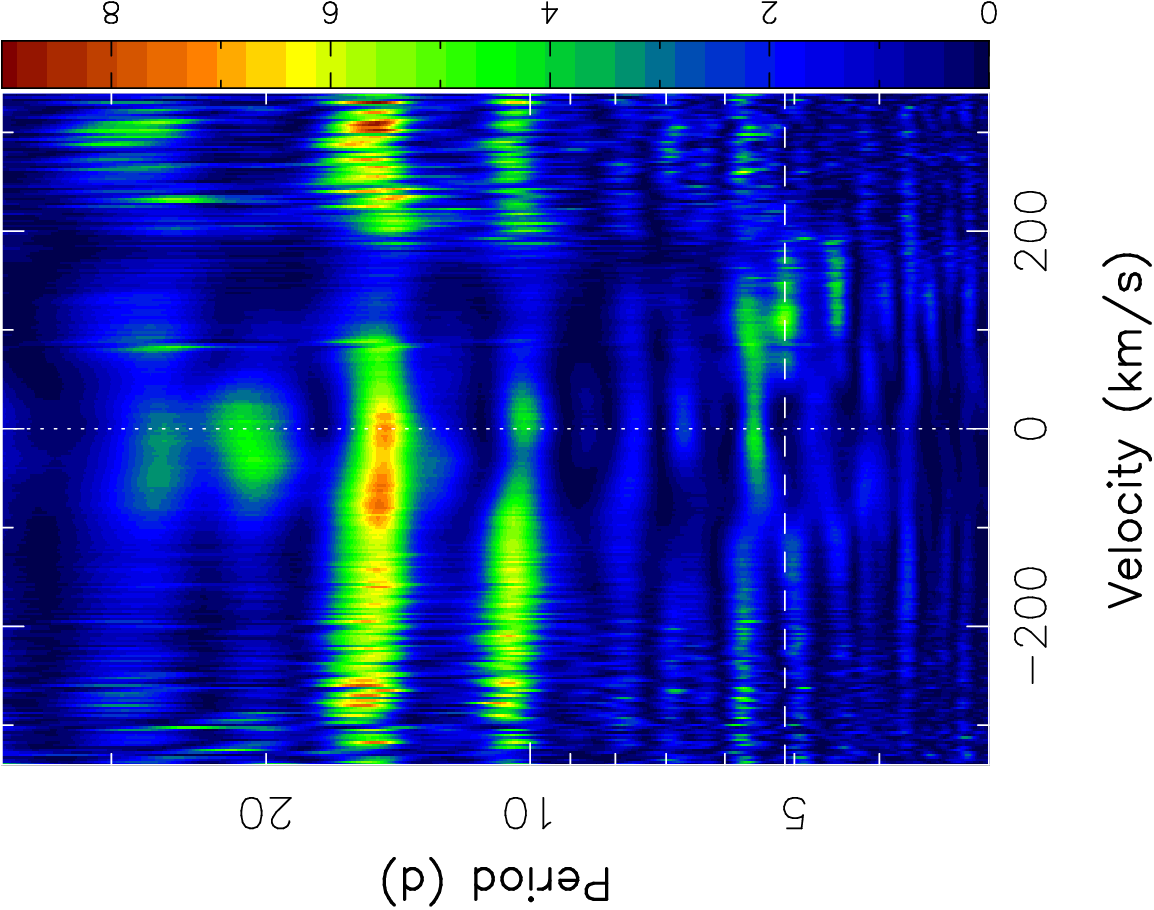}}}
\caption[]{
\emr Same as Fig.~\ref{fig:hei} for the 1282~nm \pab\ line (left: 2024, right: 2025).}  
\label{fig:pab}
\end{figure*}

We similarly investigated the 1282~nm \pab\ and 2166~nm \brg\ lines, whose profiles and 2D periodograms for seasons 2024 and 2025 are shown in Figs.~\ref{fig:pab} and \ref{fig:brg}. 
Both lines exhibited similar patterns, \brg\ being weaker and more noisy than \pab.  The blueshifted half of both lines showed periods similar to those seen in the blue wing of 
\hei, suggesting that it is also associated with winds from the star and the disk; 
the redshifted absorption of \pab\ and \brg, probing funnel or inner disk material crossing the line of sight, showed conspicuous 
rotational modulation from velocities 50 to 200~\kms\ in 2024, but only up to 150~\kms\ and occurring at lower velocities in \brg\ than in \pab\ in 2025.  As for \hei,  
the 2D periodograms of both lines evolved with time in 2025, showing both 11 and 16~d periods in the blueshifted half of the profile in the first half of 
the season, then no periodicity in the second half.  Similarly, rotational modulation of the redshifted absorption, obvious in the first half of the season, vanished almost 
completely in the second half (see Figs.~\ref{fig:pab2} and \ref{fig:brg2}).  This suggests that unstable accretion occurred at the stellar surface from the inner accretion 
disk.  

By summing the 50$-$200~\kms\ spectral bins of each line, we obtained fluxes that are rotationally modulated, more clearly for \brg\ than for \pab\ but nonetheless 
well correlated with one another ($R\simeq0.75$).  Though noisier than \pab, \brg\ was less subject to accretion-related stochastic variations, and exhibited
cleaner rotational modulation of the redshifted absorption.  A GPR fit to these fluxes (with $\theta_3$ and $\theta_4$ fixed to 600 and 0.7 as in Table~\ref{tab:rv}) yielded 
period estimates of $5.10\pm0.05$~d for both lines, consistent with our determination of \Prot\ within the error bars.    
Maximum redshifted absorption in \pab\ and \brg, signaling the phase at which the absorbing material crossed the line of sight, occurred at phases 0.80$-$0.90 in 2021, 
0.15$-$0.25 in 2024, and 0.40$-$0.50 in 2025.  In 2024, this phase roughly coincided with the phase toward which the reconstructed dipole component was tilted (0.16; 
see Table~\ref{tab:mag}) and with the bright region reconstructed with ZDI (see Fig.~\ref{fig:pho}), implying that the redshifted absorption was likely caused by the main 
accretion funnel crossing the line of sight.  The bright photospheric region might have reflected the hot chromospheric spot, although $dT$ did not show a maximum at this phase 
(see Fig.~\ref{fig:pho}).  In 2021, in contrast, the phase of maximum \pab\ and \brg\ redshifted absorption occurred about half a rotation cycle later than the phase toward 
which the inferred dipole component was tilted (0.31; see Table~\ref{tab:mag}), as if the accretion funnel crossing the line of sight was now connected to the positive 
(rather than negative) pole of the dipole.  It also coincided with the dark (rather than the bright) region reconstructed with ZDI (see Fig.~\ref{fig:pho}).  
In 2025, finally, maximum \pab\ and \brg\ redshifted absorption occurred roughly at mid distance between the crossing of the negative and positive poles (located 
close to the equator at phases 0.18 and 0.68, respectively; see Fig.~\ref{fig:map}), and matched the dark spot reconstructed with ZDI (see Fig.~\ref{fig:pho}), suggesting this 
time that it might probe a strong warp of the inner disk.  

We computed the emission EWs for \pab\ and \brg\ (see Table~\ref{tab:log}) and found that they correlated with veiling $r_{JH}$ and $r_K$ ($R\simeq0.7$ and 0.6 for \pab\ and 
\brg, respectively).  We also found that \pab\ and \brg\ EWs were strongly correlated with one another ($R\simeq0.95$), with \pab\ EWs being stronger than \brg\ ones by an average 
factor of 4.2 (2.7 after correcting for veiling).  We then used the scaling relations of \citet{Fiorellino25} and our stellar parameters (see Table~\ref{tab:par}) to turn the 
veiling-corrected EWs of \pab\ and \brg\ into accretion fluxes and the logarithmic mass-accretion rates $\log \Mdot$ for each visit.  
We derived time-averaged values of $-7.72\pm0.24$~dex (\Mdot\ in \mspy) for \pab\ and $-7.78\pm0.31$~dex for \brg, the error bar denoting the dispersion over 
all visits.  Taking the weighted average of both estimates yielded $\log \Mdot = -7.74\pm0.19$~dex, consistent with previous results \citep{Alcala21,Sousa23,Perraut25}.  
We found for individual seasons that DO~Tau accreted mass marginally faster in 2021 ($\log \Mdot = -7.47\pm0.15$~dex) and in 2024 ($-7.65\pm0.14$~dex) than 
in 2025 ($-7.87\pm0.16$~dex), without a clear variation between the first and second half of the 2025 season.   
From the average mass-accretion rate, the intensity of the reconstructed dipole component of the large-scale field (0.20$-$0.30~kG; see Table~\ref{tab:mag}) and the stellar parameters of 
Table~\ref{tab:par}, we derive following \citet{Bessolaz08}, that the magnetospheric gap of DO~Tau extended to a distance 
of $\rmag = 1.6\pm0.3$~\rstar\ or $0.014\pm0.002$~au (the error bar corresponding to variations in the dipole field strength and the mass accretion rate), implying that 
$\rmag/\rcor = 0.30\pm0.05$ where \rcor\ denotes the corotation radius (located at 5.4~\rstar, i.e., 0.047~au) at which the Keplerian angular velocity equals the rotation rate at the 
stellar surface. {\emr Our \rmag\ estimate is consistent with that inferred} with Gravity from the size of the \brg\ emitting region \citep{Perraut25}.  

In this regime and following \citet{Blinova16}, magnetospheric accretion is expected to be unstable but ordered, with one to two tongues of accreted material 
penetrating the magnetosphere in the equatorial plane.  For an inclined dipole, accretion at the surface of the star is expected to be rotationally modulated and 
to fluctuate with the Keplerian period of the inner disk (0.84~d), generating rapid intrinsic variability in addition to rotational modulation\footnote{\emr Periodograms 
of the TESS light curves also show peaks at periods shorter than \Prot\ down to about $\simeq$1~d, in line with model predictions.}.  
This is qualitatively consistent with what we observed, with maximum redshifted absorption occurring in conjunction with either poles of the magnetic dipole in 
2021 and 2024.  For dipoles that are strongly tilted to the rotation axis (as in 2025), magnetospheric accretion is predicted to proceed directly to the 
magnetic poles in the equatorial plane and to generate a strong warp in the inner disk at mid distance from magnetic poles \citep{Romanova03}.  
We speculate that this strong warp generated the periodic distortions in Stokes $I$ profiles and redshifted absorption in emission lines in 2025.  
The disk warp material, located well within the corotation radius and forced to corotate with the star, was unable to escape falling toward the star, albeit at 
reduced speed given the transverse orientation of the magnetic field.  This makes our observations 
qualitatively consistent with theoretical expectations.  Given the low \rmag, we also expect the higher SH terms of the magnetic topology to 
contribute to rendering the accretion pattern more complex and unstable.  The large-scale field of DO~Tau is likely too weak to prevent 
accretion of angular momentum from the disk, causing the star to speed up, in contrast to more slowly rotating cTTSs with stronger fields and lower accretion rates 
\citep[like CI~Tau,][]{Donati24}.  

We also studied the \feiif\ J and H and \htw\ lines (at 1257.02, 1644.00, and 2121.83~nm), which probe the jet and outflow of DO~Tau \citep{Erkal21} and were found to vary very weakly.  
We report a $\simeq$1~\kms\ shift of the \feiif\ lines from 2024 to 2025.

\section{Summary}
\label{sec:dis}

We carried out a spectropolarimetric study of DO~Tau based on 77 SPIRou Stokes $I$ and $V$ spectra collected from early 2020 to late 2025. From the 
detected Zeeman polarization signatures in LSD profiles of atomic lines, we infer that DO~Tau rotates in $\Prot=5.128\pm0.002$~d, with its rotation axis inclined at $45\pm8$\degr\ 
to the line of sight, slightly more than the outer accretion disk.  The rotational broadening of spectral lines, $\vsini=13.0\pm0.5$~\kms, implied that 
DO~Tau has a radius of $\rstar=1.9\pm0.2$~\rsun.  The atmospheric parameters derived with ZeeTurbo, $\teff=3450\pm50$~K and $\logg=3.6\pm0.1$, suggest that DO~Tau 
is still fully convective, with an age of $1.7\pm0.5$~Myr.  

\begin{figure}[ht!]
\centerline{\includegraphics[scale=0.35,angle=-90]{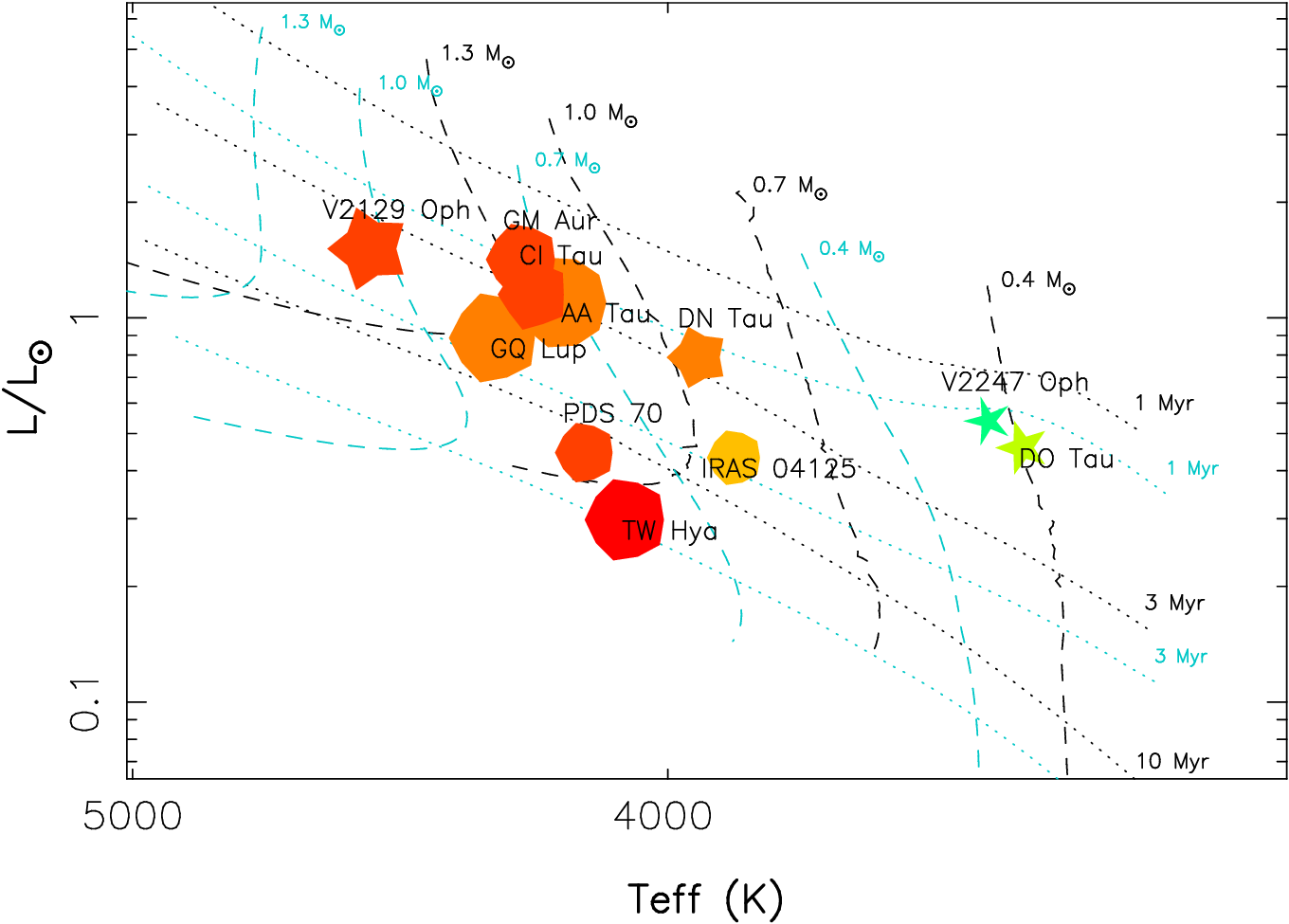}}
\caption[]{Basic properties of the large-scale magnetic topologies of cTTSs, as a function of their locations in the Hertzsprung-Russel diagram. The symbol size indicates 
relative magnetic intensities, the symbol colour illustrates the field configurations (red to blue for purely poloidal to purely toroidal fields), and the symbol shape depicts the
degree of axisymmetry of the poloidal field component (decagon and stars for purely axisymmetric and purely nonaxisymmetric poloidal fields, respectively).
We also show the most recent magnetic (black) and nonmagnetic (cyan) PMS evolutionary tracks (dashed lines) and isochrones (dotted lines) of \citet{Feiden16}.} 
\label{fig:hrd}
\end{figure} 

We found that the average large-scale and small-scale magnetic field strengths at the surface of DO~Tau were equal to 0.35 and 2.5~kG, respectively.  The large-scale field 
reconstructed with ZDI was mostly nonaxisymmetric and featured a significant toroidal component.  The dipole component of the poloidal field, ranging 
from 0.2 to 0.3~kG, became more inclined with respect to the rotation axis, with the negative pole eventually changing hemisphere between the last two seasons.  If this is 
confirmed, DO~Tau is the first cTTS for which a global polarity switch of the large-scale magnetic field is detected, bringing further 
evidence that large-scale magnetic fields of cTTSs are dynamo amplified rather than fossil fields.  The large-scale field of DO~Tau was unusually weak and complex for a cTTS 
(see Fig.~\ref{fig:hrd}).  DO~Tau even differs from its closest cTTS neighbor in our Hertzsprung-Russel diagram, V2247~Oph, whose complex field varied very rapidly under strong differential 
rotation \citep{Donati10}.  This difference might be caused by intense accretion from the inner disk that affects convection and dynamo action \citep{Geroux16}.  

Strong and variable veiling and fluxes in emission lines were detected at all times and were well correlated, but showed no rotational modulation, emphasizing the stochastic 
nature of accretion from the inner disk.  The average logarithmic mass-accretion rate derived from veiling-corrected \pab\ and \brg\ fluxes was equal to $\log \Mdot = -7.74\pm0.19$~dex,
marginally stronger in 2021 and 2024 than in 2025.  The magnetospheric radius we derived from the large-scale dipole, the mass accretion rate and the stellar parameters is equal 
to $1.6\pm0.3$~\rstar\ or $0.30\pm0.05\rcor$, suggesting that accretion is complex and unstable on DO~Tau, occurring via a few tongues penetrating the magnetosphere in the equatorial 
plane.  The redshifted absorption in \pab\ and \brg, most of the time modulated by rotation, was interpreted as due to accretion funnels linking the inner disk to one of the magnetic 
poles and regularly crossing the line of sight.  In 2025, however, this redshifted absorption was instead attributable to a warp in the inner accretion disk likely induced by 
the strongly tilted magnetosphere.  Stellar and disk wind signatures were also detected in the blue wings of the \hei\ triplet, \pab\ and \brg, with periods varying between 13 and 
16~d from one season to the next, presumably reflecting sporadic events induced by accretion bursts; rotational modulation of the \hei\ wind signatures was also observed in the second half 
of the 2025 season.  

The RVs of DO~Tau, either measured with LBL on all lines or on LSD profiles of atomic and CO lines, also varied, reflecting mainly rotational modulation caused by activity.  
We report a potential RV contribution from a candidate massive planet orbiting within the inner disk at a distance of $0.12\pm0.01$~au (orbital period $P_b=21.14\pm0.02$~d). 
The detected RV signature, best seen in LBL RVs and in CO lines but only marginally in atomic lines, featured a semi-amplitude of $K_b=0.22\pm0.04$~\kms, would imply a minimum 
mass of $2.0\pm0.4$~\mjup\ for the candidate planet, i.e., a mass of $2.8\pm0.6$~\mjup\ if the orbital motion takes place in the stellar equatorial plane.  The observed RV 
signal is close to the detection threshold of our dataset according to planet injection recovery tests, however.  Moreover, the three RV sets yielded differing periodograms,  
and the veiling also showed some power at 21~d, which might suggest that the reported RV signature comes from activity related to a density structure in the inner disk 
and not from a close-in planet.  

Our spectropolarimetric and velocimetric study of DO~Tau showed that collecting both types of data is essential for characterizing the physical processes taking 
place in the close environments of young cTTSs, especially those featuring strong accretion rates.  For instance, work like this can 
test the tentative scenarios invoked to explain the wiggle of the jet axis \citep{Erkal21}, i.e., a warp induced by a magnetic disk wind or a slightly 
inclined close-in massive planet.  If this is confirmed, the candidate planet is less massive than required (6$-$12~\mjup), but in the correct range of orbital distances 
(0.10$-$0.15~au) and on a possibly inclined orbit given the misalignment of the inner and outer disks. 
In addition, the variable stellar magnetic field switching polarity might be an alternative option for generating warps in the inner disk.  Follow-up monitoring of DO~Tau 
{\emr simultaneously with SPIRou and ESPaDOnS at optical and nIR wavelengths (i.e., with VISION, to be commissioned at the CFHT in 2026) and complementary instruments} 
should reveal which of these options explains the observations best.  It will also further document the temporal evolution of the large-scale magnetic field of DO~Tau and 
its possible cyclic variations, clarify the effect of strong accretion on the dynamo processes and the dynamics of convective zones in young low-mass stars, and ultimately 
improve physical realism in models of star and planet formation.

\section*{Data availability}  SPIRou data used in this study are, or will soon be, publicly available at the Canadian Astronomy Data Center (\url{https://www.cadc-ccda.hia-iha.nrc-cnrc.gc.ca}).       

\begin{acknowledgements}
We thank an anonymous referee for constructive comments that improved the paper.  
This work benefited from the SIMBAD CDS database at URL {\tt http://simbad.u-strasbg.fr/simbad} and the ADS system at URL {\tt https://ui.adsabs.harvard.edu}.  
We thank Dr Greg Feiden for providing us the tracks and isochrones of his latest magnetic and non-magnetic stellar evolution models.  
Our study is based on data obtained at the CFHT, operated by the CNRC (Canada), INSU/CNRS (France) and the University of Hawaii.  
The authors wish to recognise and acknowledge the very significant cultural role and reverence that the summit of Maunakea has always had
within the indigenous Hawaiian community.  SHPA acknowledges financial support from CNPq, CAPES and Fapemig. FM received funding from the European Research 
Council (ERC) under the European Union's Horizon Europe research and innovation program (grant agreement \#101053020 Dust2Planets).
\end{acknowledgements}

\bibliographystyle{aa}
\bibliography{aa58694-25}
\clearpage

\begin{appendix}

\section{SPIRou observations: additional material}
\label{sec:appA}

Table~\ref{tab:log} provides the observation log for the SPIRou spectra of DO~Tau, and the measurements derived from them at each epoch.

\begin{table*}[ht!]
\caption[]{Observing log of our SPIRou observations of DO~Tau}
\centering 
\resizebox{0.88\linewidth}{!}{   
\begin{tabular}{cccccccccccc} 
\hline
BJD        & UT date & BERV   & c / $\phi$ & t$_{\rm exp}$ & S/N   & $\sigma_P$           &  LBL RV  & \Bl\  &  BIS    &  $r_{JH}$ / $r_K$ & EW \pab\ / \brg \\
(2459000+) &         & (\kms) &            &   (s)         & ($H$) & ($10^{-4} I_c$)      &   (\kms) &(G)    & (\kms)  &                   & (\kms)          \\
\hline
-101.200530 & 19 Feb 2020 & -29.541 & 0 / 0.019 & 1582.4 & 249 & 2.09 & 15.33$\pm$0.20 & -55$\pm$28 & 1.7$\pm$0.4 & 2.95 / 4.08 & 221 / 38.6 \\
-80.200386 & 11 Mar 2020 & -29.896 & 4 / 0.115 & 1582.4 & 220 & 2.64 & 15.86$\pm$0.20 & -118$\pm$22 & 0.8$\pm$0.3 & 1.42 / 2.17 & 143 / 14.7 \\
-79.259681 & 12 Mar 2020 & -29.712 & 4 / 0.298 & 1582.4 & 235 & 2.26 & 16.02$\pm$0.20 & -108$\pm$19 & 0.2$\pm$0.3 & 1.54 / 2.59 & 162 / 24.2 \\
\hline
207.891658 & 24 Dec 2020 & -10.968 & 60 / 0.295 & 1604.7 & 316 & 1.82 & 15.75$\pm$0.20 & -57$\pm$21 & -0.3$\pm$0.3 & 2.48 / 4.00 & 333 / 74.4 \\
208.902257 & 25 Dec 2020 & -11.495 & 60 / 0.492 & 1604.7 & 290 & 1.98 & 16.85$\pm$0.20 & -54$\pm$26 & -0.8$\pm$0.3 & 2.88 / 4.50 & 454 / 85.1 \\
211.920555 & 28 Dec 2020 & -13.012 & 61 / 0.080 & 1604.7 & 205 & 2.93 & 15.41$\pm$0.20 & 20$\pm$26 & 0.2$\pm$0.3 & 1.62 / 2.85 & 260 / 48.2 \\
212.902188 & 29 Dec 2020 & -13.441 & 61 / 0.272 & 1604.7 & 235 & 2.51 & 16.46$\pm$0.20 & -44$\pm$25 & -0.8$\pm$0.3 & 1.92 / 3.06 & 306 / 61.3 \\
218.843346 & 04 Jan 2021 & -16.073 & 62 / 0.430 & 1604.7 & 261 & 2.18 & 16.43$\pm$0.20 & -71$\pm$17 & -0.9$\pm$0.3 & 1.32 / 1.70 & 262 / 43.6 \\
219.874312 & 05 Jan 2021 & -16.615 & 62 / 0.631 & 1604.7 & 247 & 2.26 & 16.42$\pm$0.20 & 21$\pm$18 & -0.5$\pm$0.3 & 1.32 / 2.11 & 216 / 30.1 \\
221.862565 & 07 Jan 2021 & -17.472 & 63 / 0.019 & 1604.7 & 232 & 2.42 & 14.65$\pm$0.20 & -8$\pm$26 & 0.4$\pm$0.3 & 2.17 / 3.00 & 253 / 51.7 \\
222.868775 & 08 Jan 2021 & -17.928 & 63 / 0.215 & 1604.7 & 285 & 1.87 & 15.76$\pm$0.20 & -31$\pm$21 & -0.0$\pm$0.3 & 2.35 / 4.92 & 312 / 58.5 \\
\hline
977.892123 & 02 Feb 2023 & -26.588 & 210 / 0.451 & 1627.0 & 296 & 1.83 & 17.05$\pm$0.20 & -32$\pm$23 & -1.4$\pm$0.3 & 2.76 / 4.64 & 276 / 46.7 \\
978.910461 & 03 Feb 2023 & -26.872 & 210 / 0.649 & 1627.0 & 256 & 2.10 & 16.55$\pm$0.20 & 151$\pm$27 & -0.3$\pm$0.3 & 2.80 / 4.76 & 268 / 38.4 \\
981.883376 & 06 Feb 2023 & -27.531 & 211 / 0.229 & 1627.0 & 278 & 1.94 & 16.46$\pm$0.20 & -130$\pm$30 & 0.5$\pm$0.2 & 3.52 / 5.41 & 315 / 61.4 \\
\hline
1329.903770 & 20 Jan 2024 & -22.495 & 279 / 0.096 & 1627.0 & 269 & 2.06 & 16.45$\pm$0.20 & -41$\pm$19 & 0.4$\pm$0.3 & 1.76 / 4.10 & 268 / 44.6 \\
1331.812171 & 22 Jan 2024 & -22.959 & 279 / 0.468 & 1627.0 & 321 & 1.60 & 17.14$\pm$0.20 & 18$\pm$25 & -1.4$\pm$0.5 & 3.62 / 6.14 & 350 / 67.3 \\
1332.871797 & 23 Jan 2024 & -23.461 & 279 / 0.675 & 1627.0 & 286 & 1.82 & 16.27$\pm$0.20 & 95$\pm$20 & 1.0$\pm$0.3 & 2.29 / 4.30 & 329 / 58.5 \\
1337.941736 & 28 Jan 2024 & -25.196 & 280 / 0.663 & 1627.0 & 258 & 2.07 & 16.20$\pm$0.20 & 106$\pm$17 & 1.1$\pm$0.3 & 1.47 / 3.09 & 202 / 32.3 \\
1338.939065 & 29 Jan 2024 & -25.488 & 280 / 0.858 & 1627.0 & 246 & 2.12 & 16.09$\pm$0.20 & 33$\pm$20 & 0.8$\pm$0.3 & 1.86 / 4.24 & 212 / 31.9 \\
1339.928170 & 30 Jan 2024 & -25.758 & 281 / 0.051 & 1627.0 & 284 & 1.85 & 16.54$\pm$0.20 & -7$\pm$19 & -0.4$\pm$0.3 & 2.08 / 4.64 & 201 / 28.1 \\
1340.926524 & 31 Jan 2024 & -26.036 & 281 / 0.245 & 1627.0 & 279 & 1.84 & 16.28$\pm$0.20 & -154$\pm$15 & -0.7$\pm$0.3 & 1.49 / 3.06 & 136 / 21.0 \\
1342.927215 & 02 Feb 2024 & -26.575 & 281 / 0.636 & 1627.0 & 248 & 2.11 & 15.84$\pm$0.20 & 126$\pm$22 & 1.0$\pm$0.3 & 2.10 / 3.78 & 231 / 36.9 \\
1355.833637 & 15 Feb 2024 & -29.084 & 284 / 0.152 & 1627.0 & 301 & 1.71 & 16.81$\pm$0.20 & -105$\pm$19 & 1.0$\pm$0.3 & 2.26 / 4.59 & 218 / 35.6 \\
1358.818568 & 18 Feb 2024 & -29.453 & 284 / 0.735 & 1627.0 & 284 & 1.81 & 16.26$\pm$0.20 & 58$\pm$17 & 1.2$\pm$0.4 & 1.80 / 3.43 & 286 / 46.4 \\
1359.810746 & 19 Feb 2024 & -29.553 & 284 / 0.928 & 1627.0 & 258 & 1.99 & 16.34$\pm$0.20 & 8$\pm$19 & 1.2$\pm$0.3 & 1.89 / 3.44 & 264 / 39.9 \\
1360.821815 & 20 Feb 2024 & -29.682 & 285 / 0.125 & 1627.0 & 269 & 1.97 & 16.15$\pm$0.20 & -133$\pm$15 & -0.1$\pm$0.3 & 1.26 / 2.66 & 231 / 35.7 \\
1361.816345 & 21 Feb 2024 & -29.768 & 285 / 0.319 & 1627.0 & 246 & 2.16 & 16.33$\pm$0.20 & -187$\pm$17 & -0.8$\pm$0.3 & 1.28 / 2.58 & 180 / 30.6 \\
1362.818972 & 22 Feb 2024 & -29.863 & 285 / 0.515 & 1627.0 & 274 & 1.91 & 15.92$\pm$0.20 & 35$\pm$19 & 1.0$\pm$0.3 & 1.89 / 4.16 & 221 / 31.3 \\
1363.752997 & 23 Feb 2024 & -29.798 & 285 / 0.697 & 1627.0 & 280 & 1.84 & 16.16$\pm$0.20 & 77$\pm$16 & 1.1$\pm$0.3 & 1.64 / 3.20 & 219 / 30.6 \\
1364.863166 & 24 Feb 2024 & -30.087 & 285 / 0.913 & 1627.0 & 243 & 2.26 & 16.13$\pm$0.20 & -21$\pm$19 & 0.9$\pm$0.3 & 1.43 / 3.05 & 169 / 18.6 \\
1366.809540 & 26 Feb 2024 & -30.114 & 286 / 0.293 & 1627.0 & 231 & 2.38 & 16.31$\pm$0.20 & -119$\pm$21 & -0.3$\pm$0.2 & 1.61 / 3.54 & 167 / 23.7 \\
1367.794222 & 27 Feb 2024 & -30.130 & 286 / 0.485 & 1627.0 & 263 & 2.03 & 16.31$\pm$0.20 & 14$\pm$21 & -0.9$\pm$0.2 & 2.11 / 4.00 & 254 / 40.1 \\
1368.768013 & 28 Feb 2024 & -30.113 & 286 / 0.675 & 1627.0 & 262 & 2.01 & 16.09$\pm$0.20 & 124$\pm$18 & 0.6$\pm$0.3 & 1.73 / 3.02 & 226 / 38.4 \\
1371.823004 & 02 Mar 2024 & -30.279 & 287 / 0.270 & 1627.0 & 252 & 2.15 & 16.19$\pm$0.20 & -185$\pm$16 & -0.6$\pm$0.3 & 1.27 / 2.24 & 151 / 18.3 \\
\hline
1890.097376 & 02 Aug 2025 & 25.411 & 388 / 0.338 & 1627.0 & 262 & 2.09 & 16.13$\pm$0.20 & -140$\pm$18 & -0.9$\pm$0.3 & 1.49 / 2.65 & 194 / 32.1 \\
1891.134265 & 03 Aug 2025 & 25.614 & 388 / 0.540 & 1627.0 & 241 & 2.36 & 16.21$\pm$0.20 & 114$\pm$20 & -0.3$\pm$0.3 & 1.47 / 2.06 & 144 / 18.0 \\
1892.128748 & 04 Aug 2025 & 25.862 & 388 / 0.734 & 1627.0 & 200 & 2.93 & 15.26$\pm$0.20 & 163$\pm$24 & 0.4$\pm$0.3 & 1.44 / 1.67 & 139 / 18.9 \\
1893.115636 & 05 Aug 2025 & 26.117 & 388 / 0.927 & 1627.0 & 295 & 1.88 & 16.05$\pm$0.20 & -30$\pm$14 & -0.3$\pm$0.3 & 1.26 / 2.11 & 178 / 29.7 \\
1894.119577 & 06 Aug 2025 & 26.338 & 389 / 0.122 & 1627.0 & 268 & 2.06 & 16.58$\pm$0.20 & -132$\pm$15 & -0.3$\pm$0.3 & 1.19 / 1.97 & 107 / 11.8 \\
1918.100650 & 30 Aug 2025 & 29.569 & 393 / 0.799 & 1627.0 & 297 & 1.80 & 15.85$\pm$0.20 & 56$\pm$19 & 0.1$\pm$0.3 & 2.19 / 4.74 & 350 / 66.3 \\
1919.117444 & 31 Aug 2025 & 29.561 & 393 / 0.997 & 1627.0 & 230 & 2.54 & 16.37$\pm$0.20 & -116$\pm$23 & 0.7$\pm$0.3 & 1.68 / 4.12 & 303 / 57.3 \\
1921.060121 & 02 Sep 2025 & 29.708 & 394 / 0.376 & 1627.0 & 269 & 2.06 & 16.32$\pm$0.20 & -94$\pm$17 & -0.3$\pm$0.3 & 1.43 / 2.57 & 164 / 23.8 \\
1922.102829 & 03 Sep 2025 & 29.625 & 394 / 0.579 & 1627.0 & 288 & 1.86 & 16.28$\pm$0.20 & 213$\pm$16 & 0.4$\pm$0.3 & 1.53 / 2.83 & 163 / 21.5 \\
1923.129255 & 04 Sep 2025 & 29.558 & 394 / 0.779 & 1627.0 & 307 & 1.93 & 15.84$\pm$0.20 & 138$\pm$15 & 0.7$\pm$0.3 & 1.38 / 2.47 & 225 / 34.9 \\
1924.093675 & 05 Sep 2025 & 29.626 & 394 / 0.968 & 1627.0 & 277 & 1.97 & 16.29$\pm$0.20 & -3$\pm$14 & 0.2$\pm$0.3 & 1.11 / 1.96 & 160 / 26.3 \\
1925.138891 & 06 Sep 2025 & 29.498 & 395 / 0.171 & 1627.0 & 275 & 2.25 & 16.76$\pm$0.20 & -112$\pm$16 & -0.7$\pm$0.3 & 1.18 / 2.08 & 142 / 21.8 \\
1926.069436 & 07 Sep 2025 & 29.628 & 395 / 0.353 & 1627.0 & 268 & 2.05 & 16.37$\pm$0.20 & -114$\pm$16 & -0.5$\pm$0.3 & 1.29 / 1.89 & 103 / 11.6 \\
1927.132970 & 08 Sep 2025 & 29.445 & 395 / 0.560 & 1627.0 & 280 & 1.98 & 16.18$\pm$0.20 & 188$\pm$16 & -0.2$\pm$0.3 & 1.39 / 2.19 & 197 / 33.9 \\
1928.016337 & 09 Sep 2025 & 29.647 & 395 / 0.733 & 1627.0 & 267 & 2.08 & 15.94$\pm$0.20 & 202$\pm$17 & 0.5$\pm$0.3 & 1.42 / 2.17 & 176 / 36.5 \\
1929.050961 & 10 Sep 2025 & 29.535 & 395 / 0.934 & 1627.0 & 286 & 1.90 & 16.40$\pm$0.20 & 58$\pm$17 & -0.3$\pm$0.3 & 1.64 / 3.09 & 223 / 50.6 \\
1930.120355 & 11 Sep 2025 & 29.315 & 396 / 0.143 & 1627.0 & 286 & 1.90 & 16.95$\pm$0.20 & -170$\pm$16 & -0.5$\pm$0.3 & 1.56 / 2.51 & 205 / 36.2 \\
1931.090549 & 12 Sep 2025 & 29.319 & 396 / 0.332 & 1627.0 & 230 & 2.32 & 16.48$\pm$0.20 & -92$\pm$21 & -0.4$\pm$0.3 & 1.74 / 2.74 & 190 / 30.5 \\
1934.092710 & 15 Sep 2025 & 29.055 & 396 / 0.917 & 1627.0 & 268 & 2.08 & 15.81$\pm$0.20 & 8$\pm$16 & 0.2$\pm$0.3 & 1.34 / 1.93 & 238 / 45.6 \\
1935.094710 & 16 Sep 2025 & 28.947 & 397 / 0.113 & 1627.0 & 266 & 2.10 & 16.81$\pm$0.20 & -100$\pm$18 & -0.7$\pm$0.3 & 1.51 / 2.06 & 215 / 41.6 \\
1953.068354 & 04 Oct 2025 & 25.663 & 400 / 0.618 & 1627.0 & 247 & 2.32 & 15.71$\pm$0.20 & 269$\pm$17 & 0.7$\pm$0.4 & 1.23 / 2.48 & 99 / 8.2 \\
1954.062369 & 05 Oct 2025 & 25.417 & 400 / 0.812 & 1627.0 & 266 & 2.04 & 15.72$\pm$0.20 & 99$\pm$14 & 0.2$\pm$0.3 & 1.05 / 1.84 & 111 / 16.3 \\
1955.087395 & 06 Oct 2025 & 25.079 & 401 / 0.012 & 1627.0 & 264 & 2.10 & 16.10$\pm$0.20 & -71$\pm$15 & -0.2$\pm$0.3 & 1.06 / 1.55 & 157 / 26.3 \\
1956.038567 & 07 Oct 2025 & 24.937 & 401 / 0.197 & 1627.0 & 279 & 1.95 & 15.91$\pm$0.20 & -127$\pm$18 & -0.3$\pm$0.3 & 1.74 / 2.78 & 194 / 27.7 \\
1957.060051 & 08 Oct 2025 & 24.595 & 401 / 0.396 & 1627.0 & 253 & 2.20 & 15.86$\pm$0.20 & 9$\pm$22 & -0.5$\pm$0.3 & 1.95 / 3.10 & 197 / 30.8 \\
1958.135108 & 09 Oct 2025 & 24.103 & 401 / 0.606 & 1627.0 & 266 & 2.06 & 15.58$\pm$0.20 & 261$\pm$19 & 0.6$\pm$0.3 & 1.71 / 2.60 & 178 / 28.3 \\
1959.081662 & 10 Oct 2025 & 23.950 & 401 / 0.790 & 1627.0 & 169 & 4.14 & 15.17$\pm$0.20 & 187$\pm$41 & 0.7$\pm$0.3 & 1.93 / 3.06 & 252 / 46.6 \\
1961.068513 & 12 Oct 2025 & 23.372 & 402 / 0.178 & 1627.0 & 291 & 1.99 & 16.31$\pm$0.20 & -155$\pm$18 & -0.3$\pm$0.3 & 1.68 / 2.69 & 227 / 38.2 \\
1963.134959 & 14 Oct 2025 & 22.545 & 402 / 0.581 & 1627.0 & 305 & 1.78 & 15.45$\pm$0.20 & 276$\pm$20 & 1.1$\pm$0.4 & 2.29 / 3.82 & 245 / 39.3 \\
1980.040457 & 31 Oct 2025 & 16.193 & 405 / 0.878 & 1627.0 & 231 & 2.47 & 15.99$\pm$0.20 & 20$\pm$22 & 0.3$\pm$0.3 & 1.60 / 3.04 & 161 / 27.3 \\
1980.988459 & 01 Nov 2025 & 15.905 & 406 / 0.062 & 1627.0 & 303 & 1.71 & 16.69$\pm$0.20 & -67$\pm$15 & -0.6$\pm$0.3 & 1.56 / 3.26 & 169 / 22.0 \\
1981.936093 & 02 Nov 2025 & 15.605 & 406 / 0.247 & 1627.0 & 267 & 2.00 & 16.90$\pm$0.20 & -137$\pm$19 & -0.9$\pm$0.3 & 1.77 / 3.39 & 173 / 25.8 \\
1983.067733 & 03 Nov 2025 & 14.785 & 406 / 0.468 & 1627.0 & 165 & 3.92 & 16.49$\pm$0.20 & 64$\pm$35 & -0.5$\pm$0.3 & 1.65 / 2.31 & 110 / 10.8 \\
1983.998058 & 04 Nov 2025 & 14.533 & 406 / 0.649 & 1627.0 & 235 & 2.50 & 15.88$\pm$0.20 & 258$\pm$23 & 1.3$\pm$0.3 & 1.67 / 2.85 & 129 / 16.5 \\
1984.908775 & 05 Nov 2025 & 14.331 & 406 / 0.827 & 1627.0 & 172 & 3.47 & 15.64$\pm$0.20 & 124$\pm$34 & 0.4$\pm$0.4 & 1.94 / 3.73 & 190 / 35.6 \\
1985.896779 & 06 Nov 2025 & 13.913 & 407 / 0.020 & 1627.0 & 154 & 3.91 & 16.27$\pm$0.20 & -31$\pm$34 & 1.0$\pm$0.4 & 1.58 / 3.34 & 234 / 41.4 \\
1987.104494 & 07 Nov 2025 & 12.850 & 407 / 0.255 & 1627.0 & 260 & 2.19 & 16.41$\pm$0.20 & -193$\pm$19 & 1.1$\pm$0.3 & 1.53 / 2.46 & 205 / 32.3 \\
1988.101042 & 08 Nov 2025 & 12.389 & 407 / 0.450 & 1627.0 & 246 & 2.24 & 16.61$\pm$0.20 & 26$\pm$18 & -1.1$\pm$0.3 & 1.36 / 2.34 & 123 / 12.8 \\
1989.047462 & 09 Nov 2025 & 12.059 & 407 / 0.634 & 1627.0 & 211 & 2.66 & 16.02$\pm$0.20 & 226$\pm$21 & 1.1$\pm$0.3 & 1.36 / 2.38 & 133 / 15.0 \\
1990.153093 & 10 Nov 2025 & 11.327 & 407 / 0.850 & 1627.0 & 245 & 2.25 & 16.10$\pm$0.20 & 108$\pm$18 & 0.2$\pm$0.3 & 1.36 / 2.63 & 173 / 23.9 \\
1991.018669 & 11 Nov 2025 & 11.176 & 408 / 0.018 & 1627.0 & 244 & 2.25 & 16.42$\pm$0.20 & -1$\pm$15 & 0.4$\pm$0.3 & 1.04 / 1.88 & 92 / 7.2 \\
1992.097601 & 12 Nov 2025 & 10.476 & 408 / 0.229 & 1627.0 & 233 & 2.29 & 16.90$\pm$0.20 & -143$\pm$19 & -1.2$\pm$0.3 & 1.42 / 2.37 & 189 / 26.3 \\
1993.079506 & 13 Nov 2025 & 10.029 & 408 / 0.420 & 1627.0 & 200 & 2.66 & 16.28$\pm$0.20 & 7$\pm$22 & -1.0$\pm$0.3 & 1.45 / 2.31 & 138 / 16.6 \\
\end{tabular}
}
\tablefoot{For each visit, we list the barycentric Julian date BJD, the UT date, the barycentric Earth RV (BERV), the rotation cycle c and phase $\phi$ (computed 
as indicated in Sec.~\ref{sec:obs}), the total observing time t$_{\rm exp}$, the peak S/N in the spectrum (in the $H$ band) per 2.3~\kms\ pixel, the noise level 
in the LSD Stokes $V$ profile, the LBL RVs and error bar, the estimated \Bl\ and BIS from LSD profiles of atomic lines (with error bars), the measured veiling in 
the $JH$ and $K$ bands, $r_{JH}$ and $r_K$, and finally the measured \pab\ and \brg\ EWs (uncorrected for veiling). }
\label{tab:log}
\end{table*}

\section{ZDI modeling: Additional material}
\label{sec:appC}

\begin{figure*}[ht!]
\centerline{\large\bf 2021\raisebox{0.3\totalheight}{\includegraphics[scale=0.18,angle=-90]{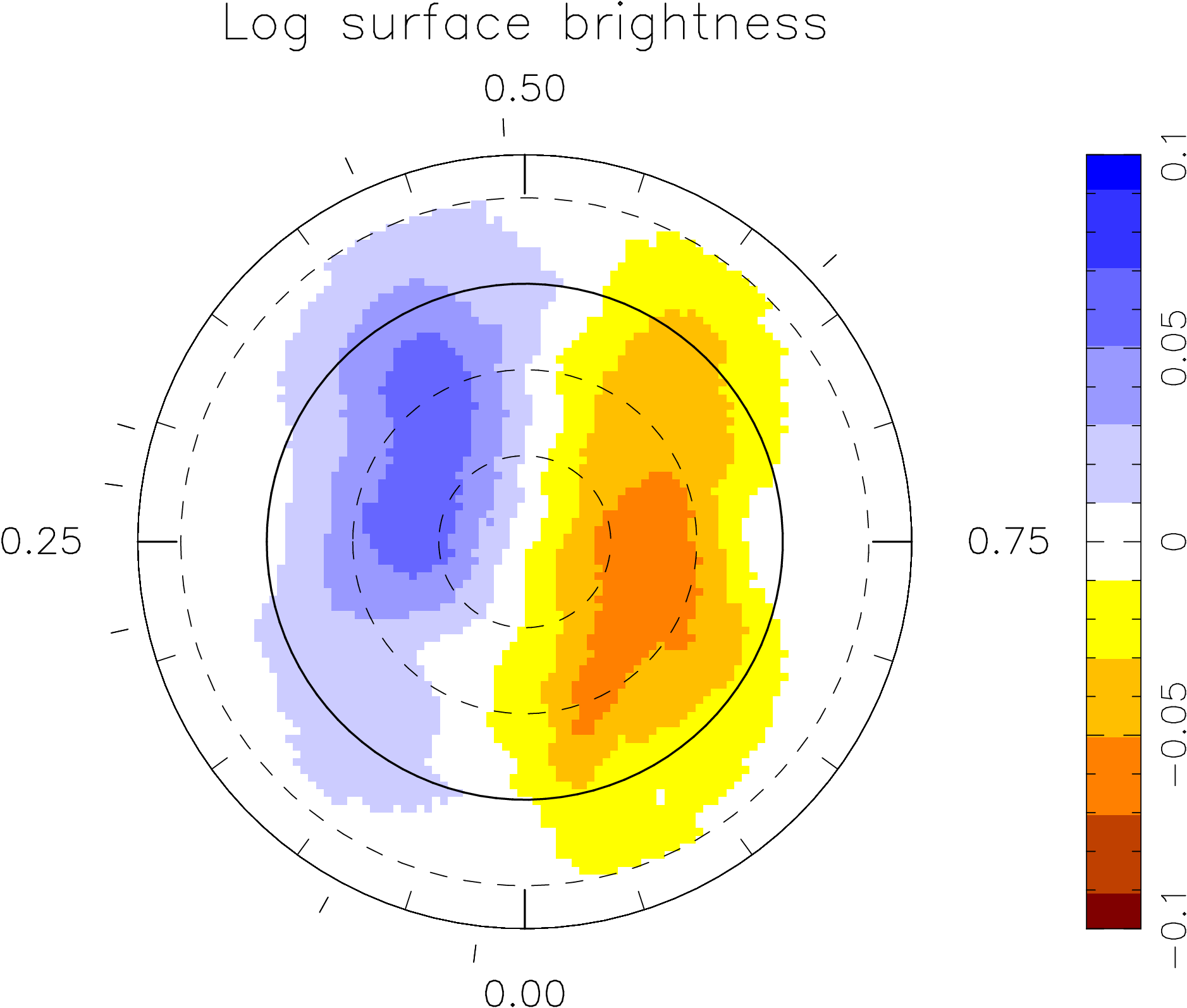}\hspace{9mm}\includegraphics[scale=0.32,angle=-90]{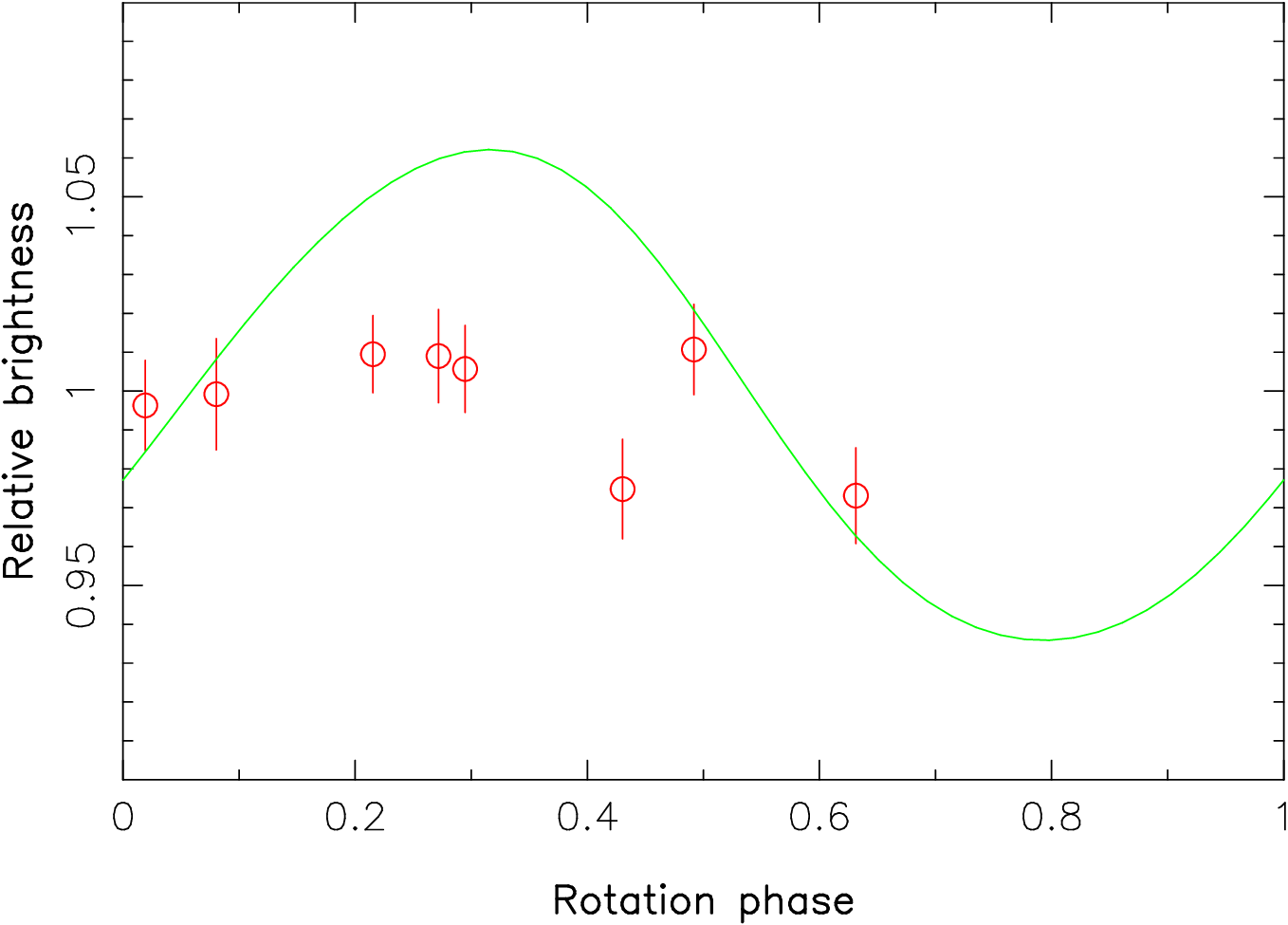}}\vspace{3mm}}
\centerline{\large\bf 2024\raisebox{0.3\totalheight}{\includegraphics[scale=0.18,angle=-90]{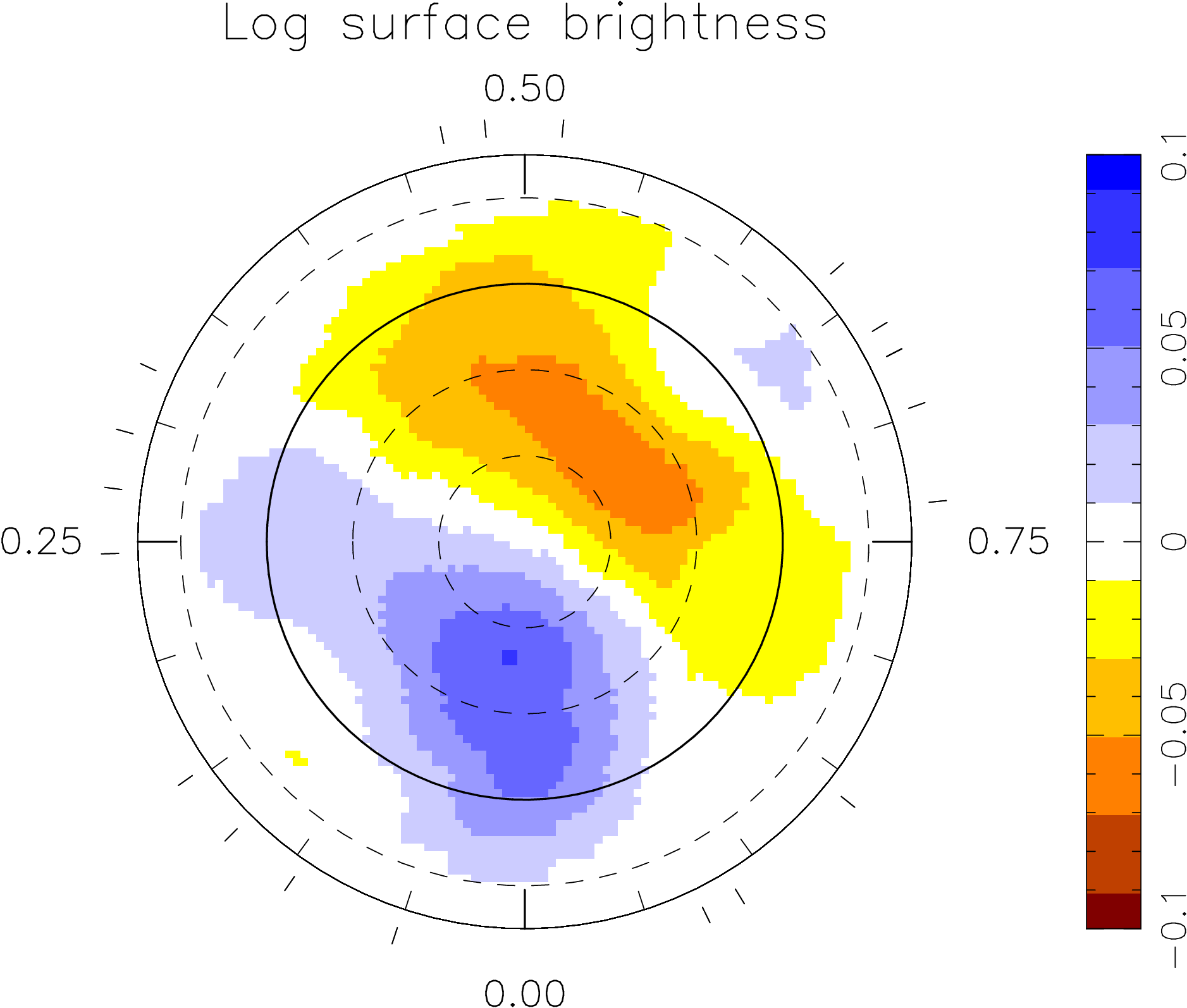}\hspace{9mm}\includegraphics[scale=0.32,angle=-90]{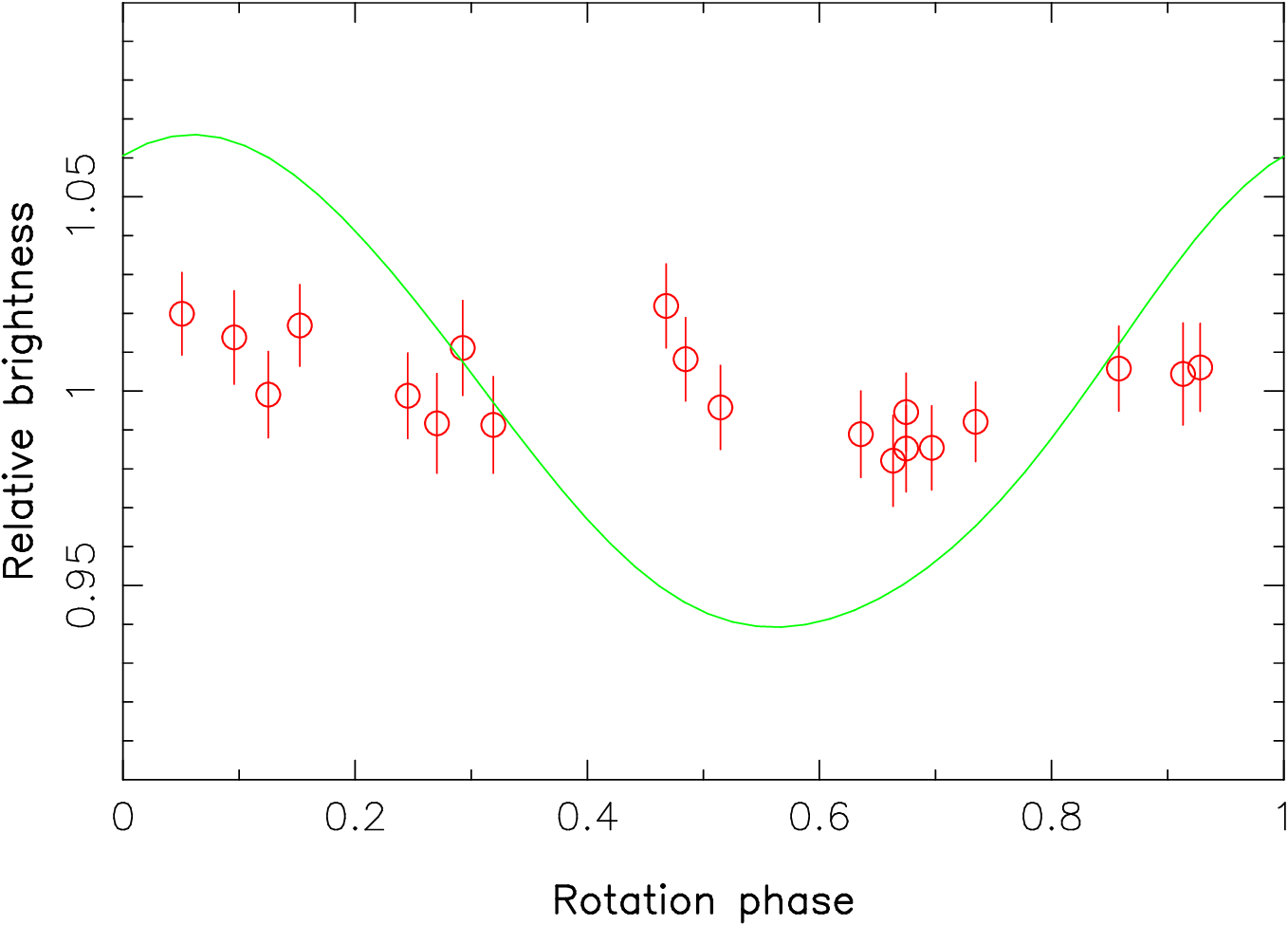}}\vspace{3mm}}
\centerline{\large\bf 2025\raisebox{0.3\totalheight}{\includegraphics[scale=0.18,angle=-90]{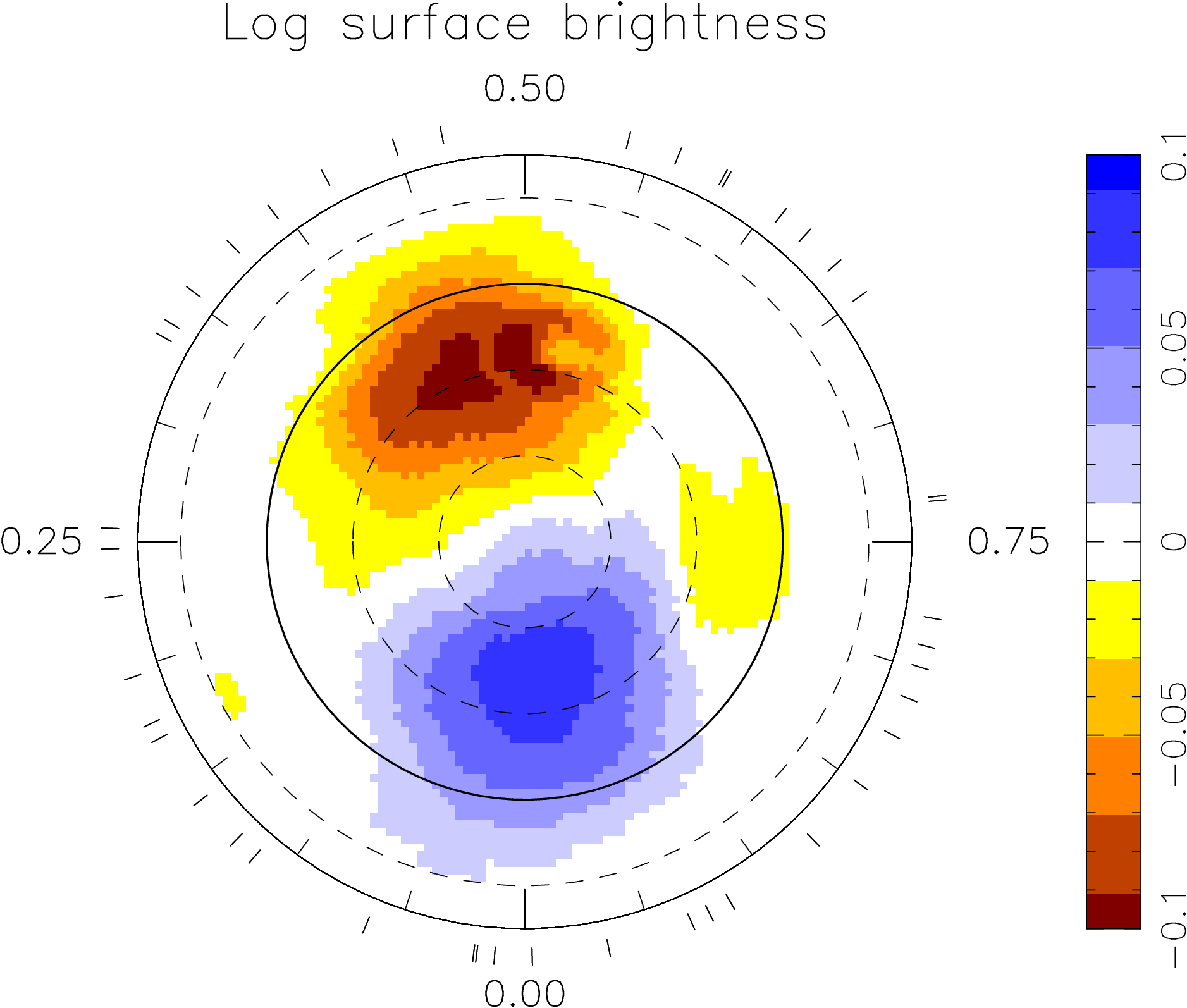}\hspace{9mm}\includegraphics[scale=0.32,angle=-90]{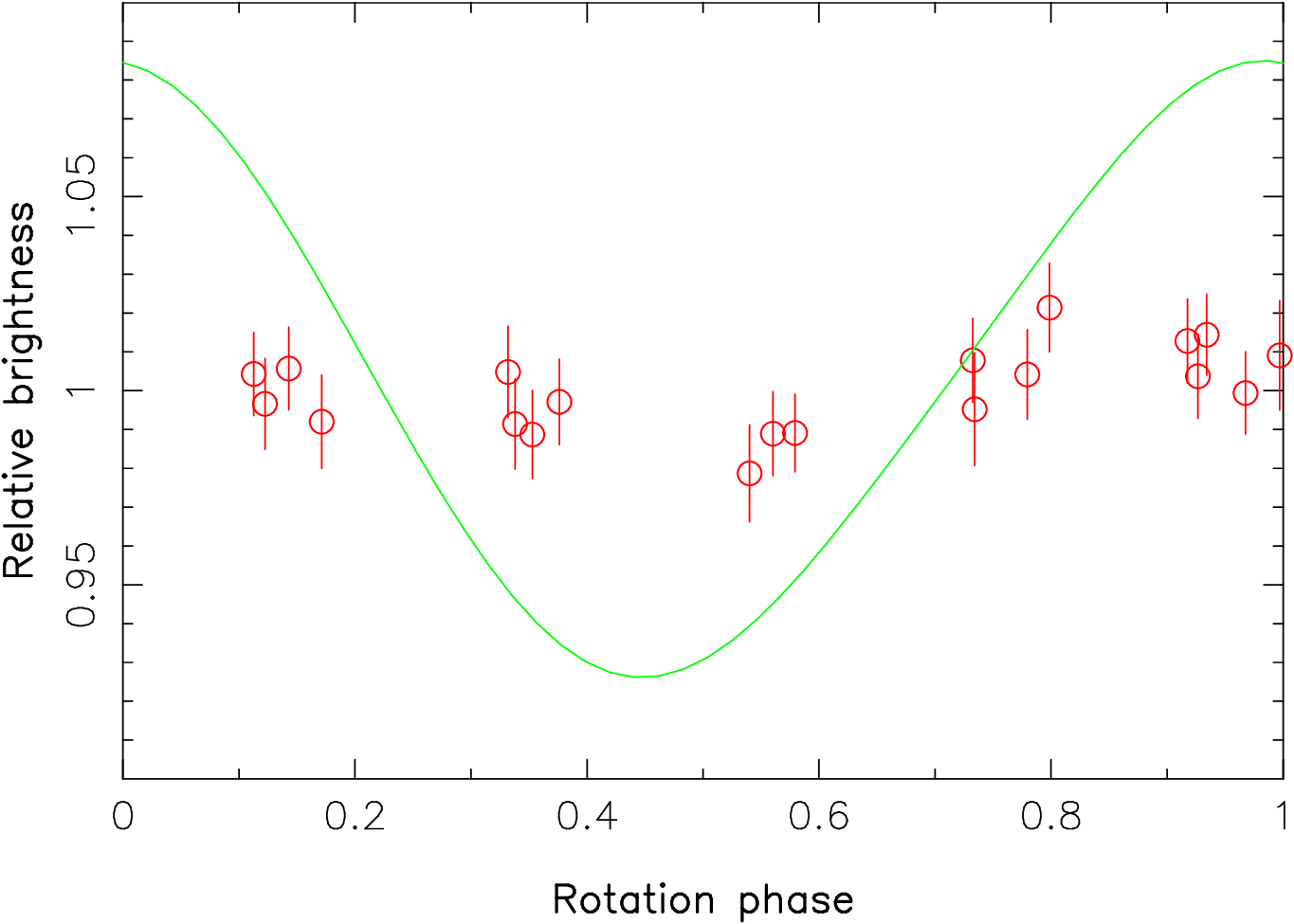}}}
\caption[]{Maps of the logarithmic relative surface brightness (with respect to the quiet photosphere, left panels) reconstructed simultaneously with the magnetic maps of Fig.~\ref{fig:map} 
with ZDI from the LSD Stokes $I$ and $V$ profiles of Fig.~\ref{fig:fit}, and corresponding photometric light curves (right panels, green curves) along with estimates inferred from $dT$ 
measurements (assuming a blackbody at photospheric temperature, red circles).  In the brightness maps, yellow and blue depict regions darker and brighter than the quiet photosphere, 
respectively.  }
\label{fig:pho}
\end{figure*}

Figure~\ref{fig:pho} shows, for each season, the brightness maps derived simultaneously with the magnetic maps of Fig.~\ref{fig:map} and the corresponding 
photometric light curves compared to those derived from $dT$.  

\section{RV modeling: Additional material}
\label{sec:appD}

\begin{figure}[ht!]
\centerline{\includegraphics[scale=0.38,angle=-90]{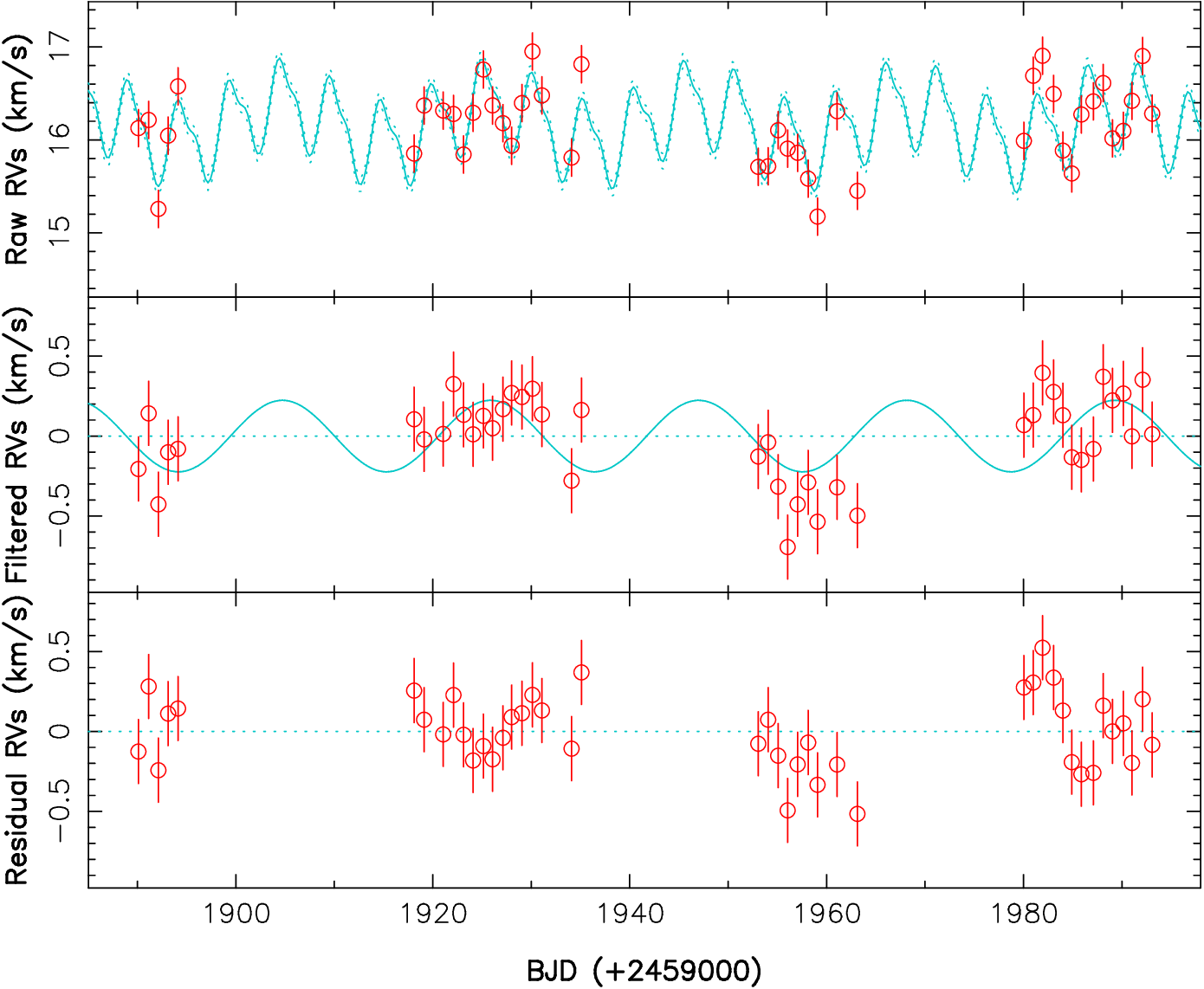}}
\caption[]{Same as Fig.~\ref{fig:rv}, zooming-in on the 2025 data}  
\label{fig:rv2}
\end{figure}

\begin{figure}[ht!]
\centerline{\includegraphics[scale=0.38,angle=-90]{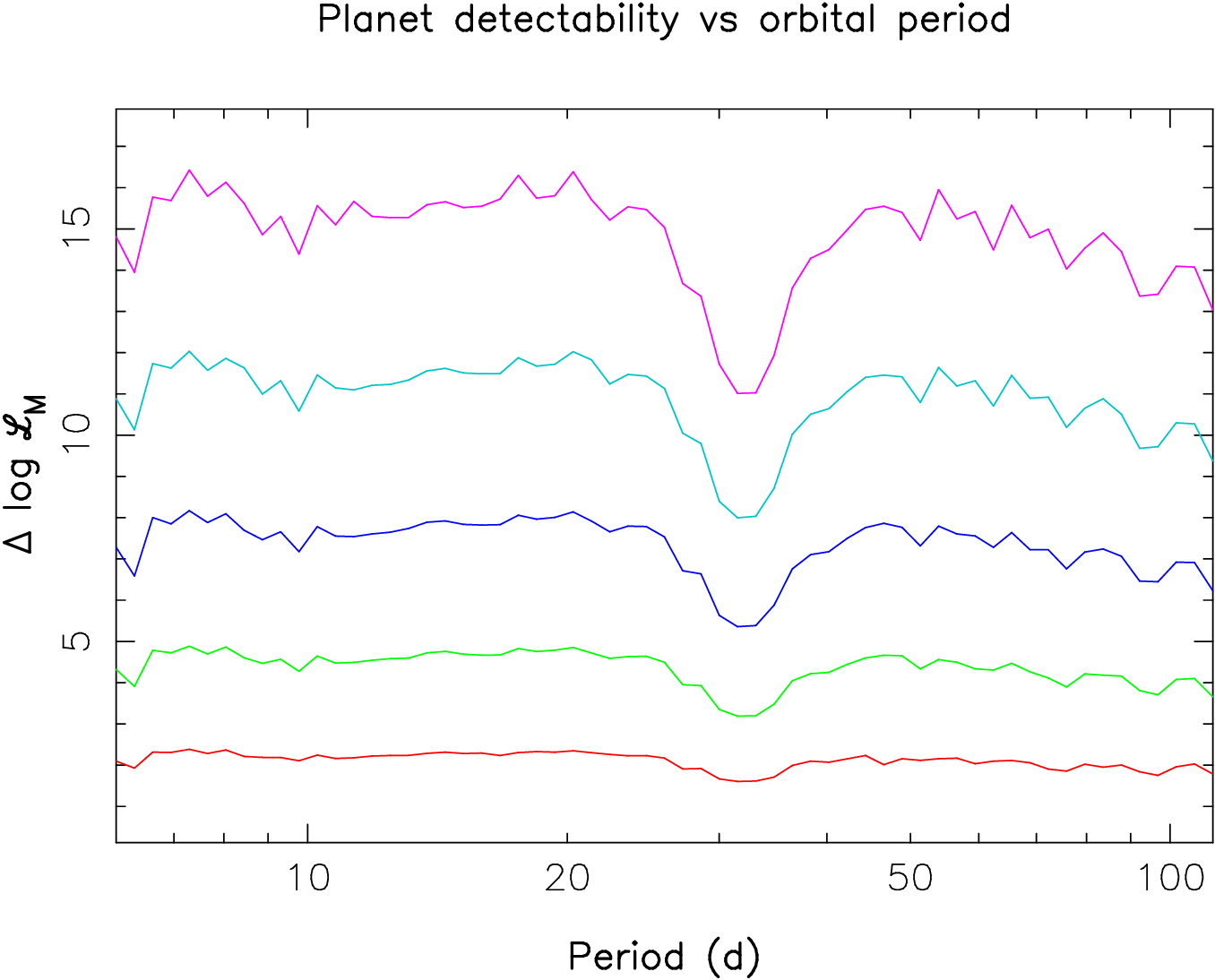}}
\caption[]{Logarithmic Bayes factor $\Delta \log \mathcal{L}_M$ as a function of orbital period for planet RV signatures of semi-amplitudes 0.10 (red), 0.15 (green), 0.20 (blue), 
0.25 (cyan), and 0.30~\kms (purple), once averaged over all phases.  Note the drop in $\Delta \log \mathcal{L}_M$ at orbital periods close to the synodic period of the Moon. }  
\label{fig:det}
\end{figure}

Figure~\ref{fig:rv2} shows the RV curve and MCMC + GPR fit, zooming-in on the 2025 data.  
Figure~\ref{fig:det} shows the detectability of RV signatures as a function of orbital period for our DO~Tau RV data.  

\section{Accretion and activity: Additional material}
\label{sec:appE}

\begin{figure*}[ht!]
\centerline{\hspace{-2mm}\includegraphics[scale=0.32,angle=-90]{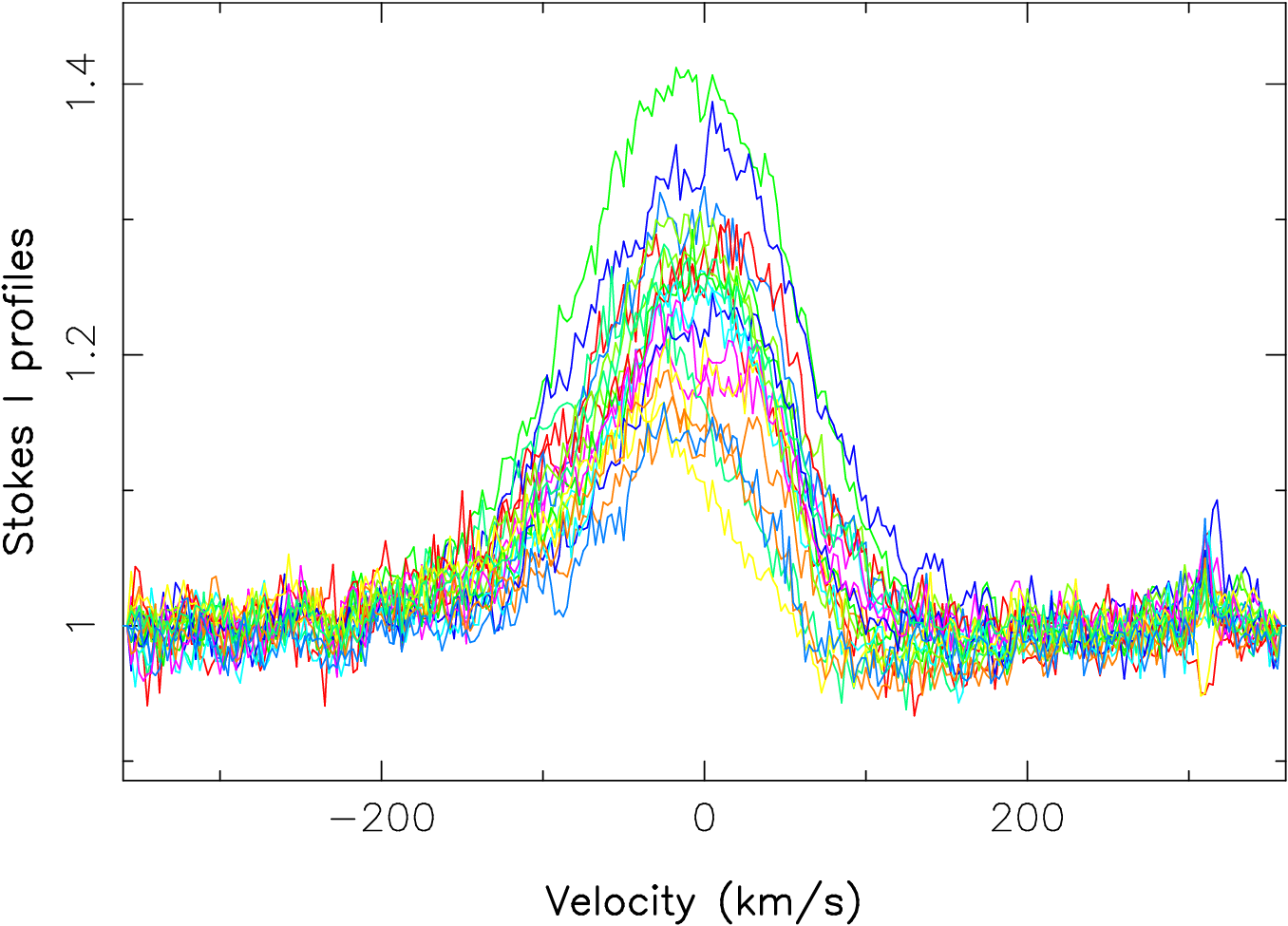}\hspace{14.5mm}\includegraphics[scale=0.32,angle=-90]{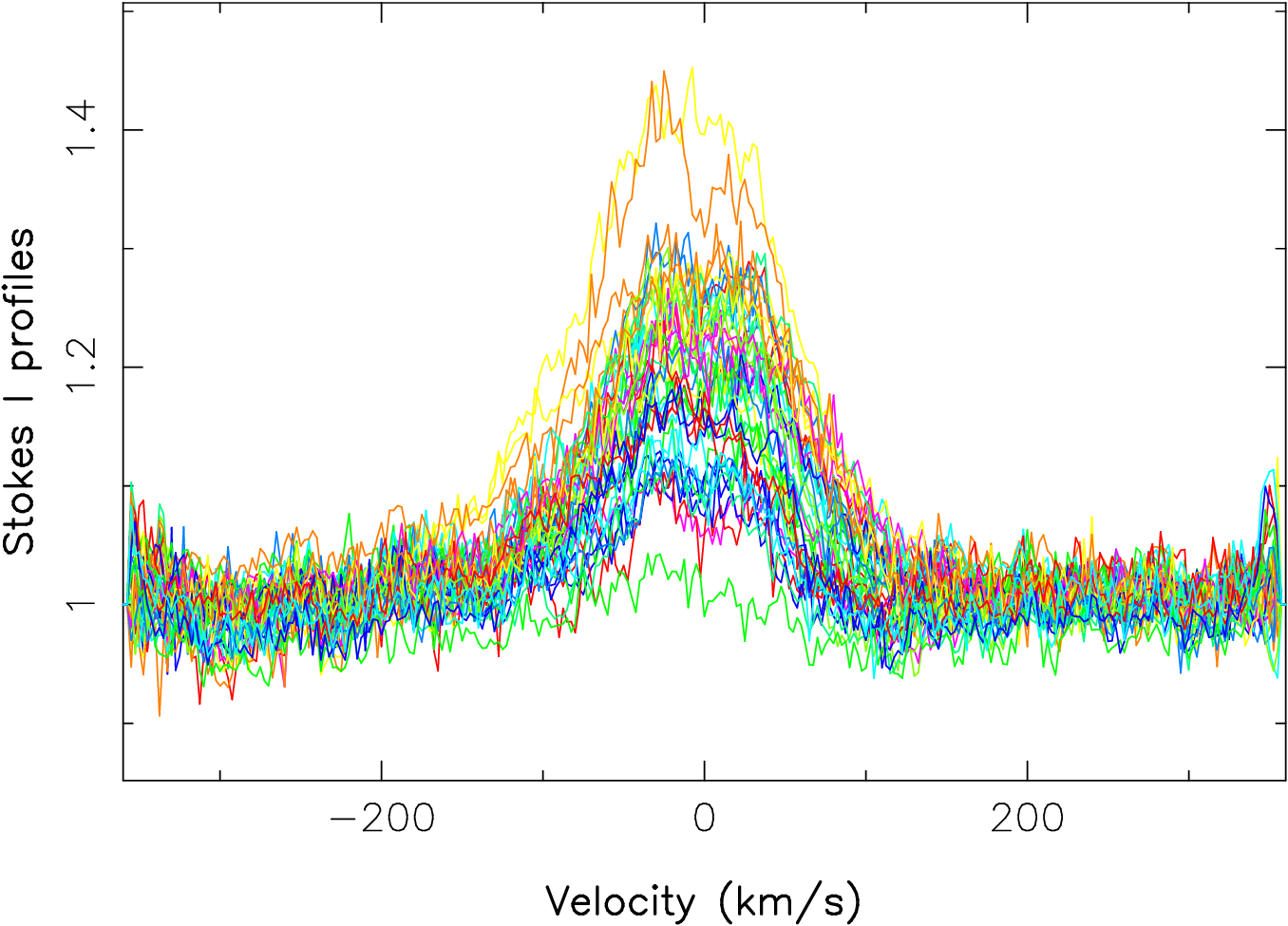}\vspace{2mm}}
\centerline{\includegraphics[scale=0.55,angle=-90]{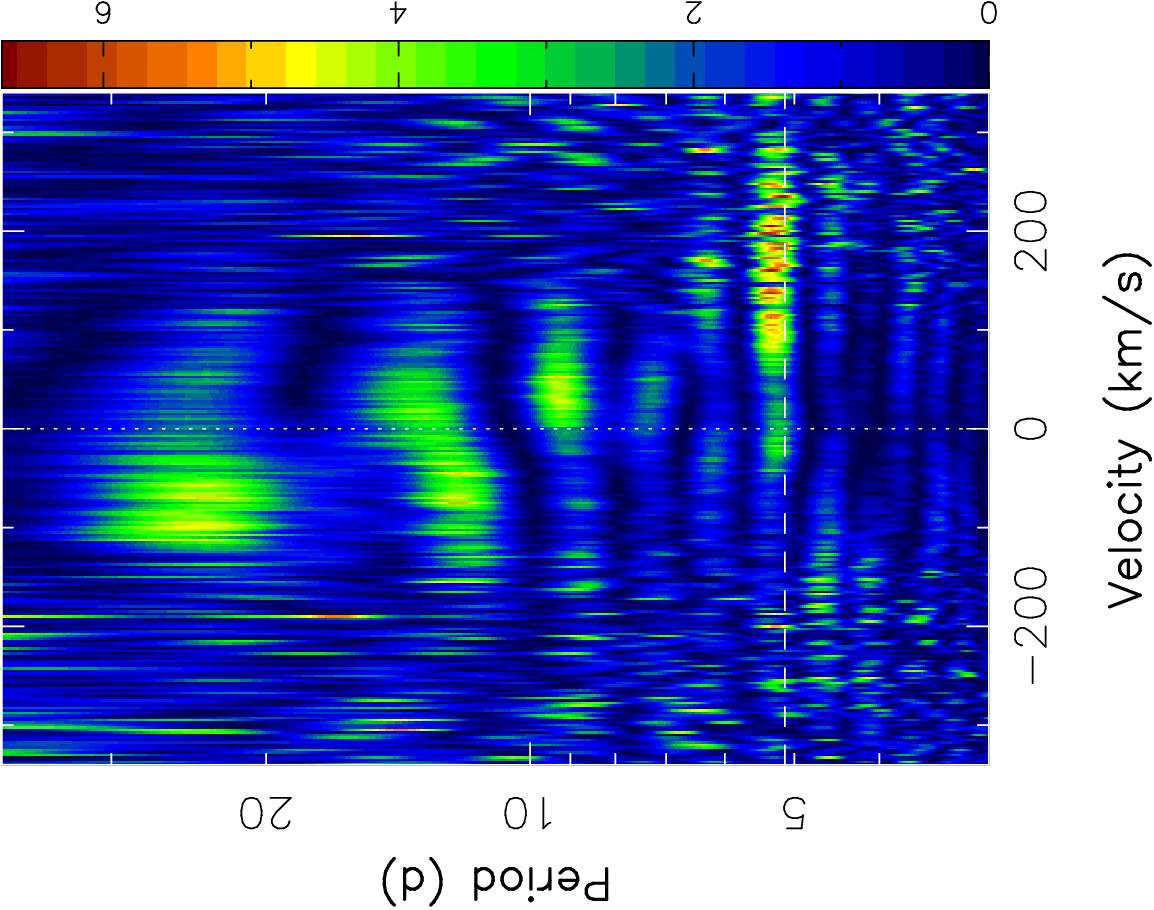}\hspace{3mm}\includegraphics[scale=0.55,angle=-90]{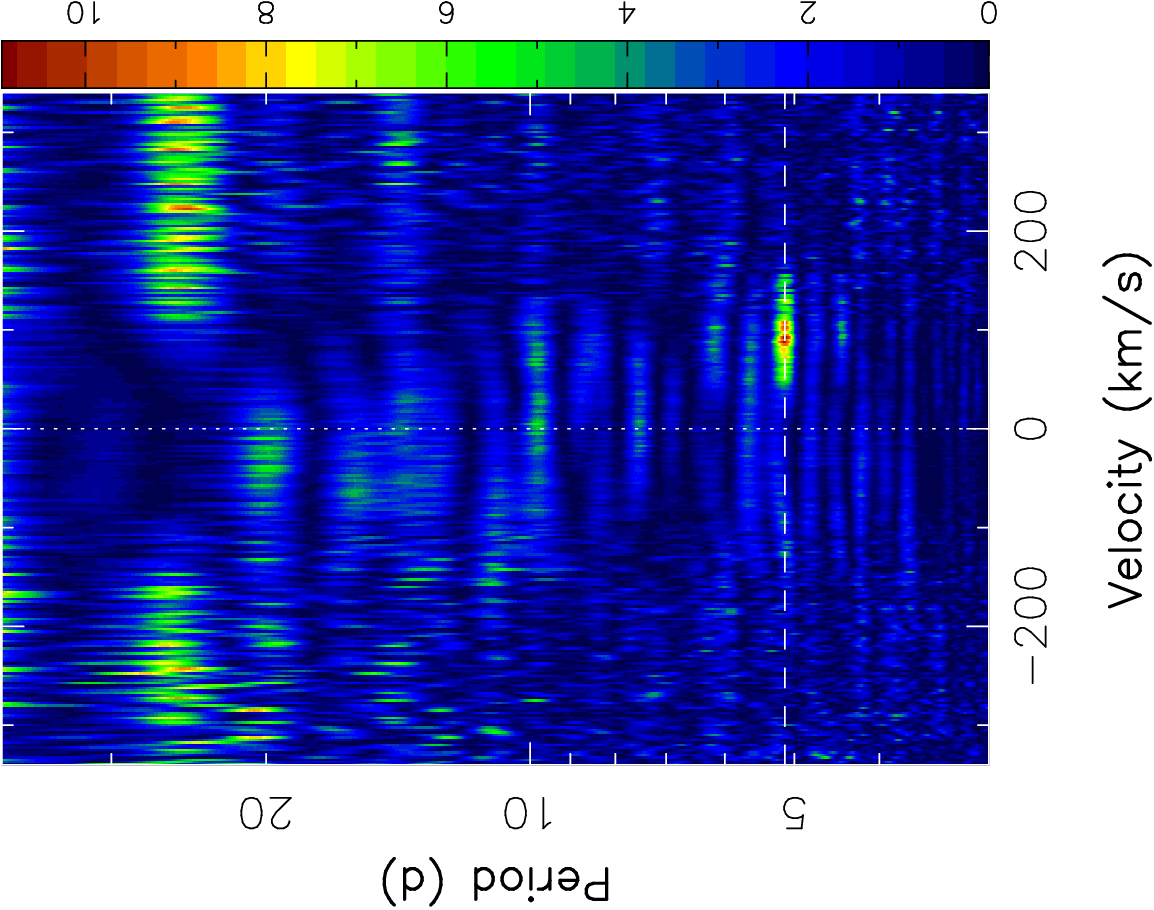}}
\caption[]{
\emr Same as Fig.~\ref{fig:hei} for the 2268~nm \brg\ line (left: 2024, right: 2025).}                                       
\label{fig:brg}
\end{figure*}

\begin{figure*}[ht!]
\centerline{\hspace{-2mm}\includegraphics[scale=0.32,angle=-90]{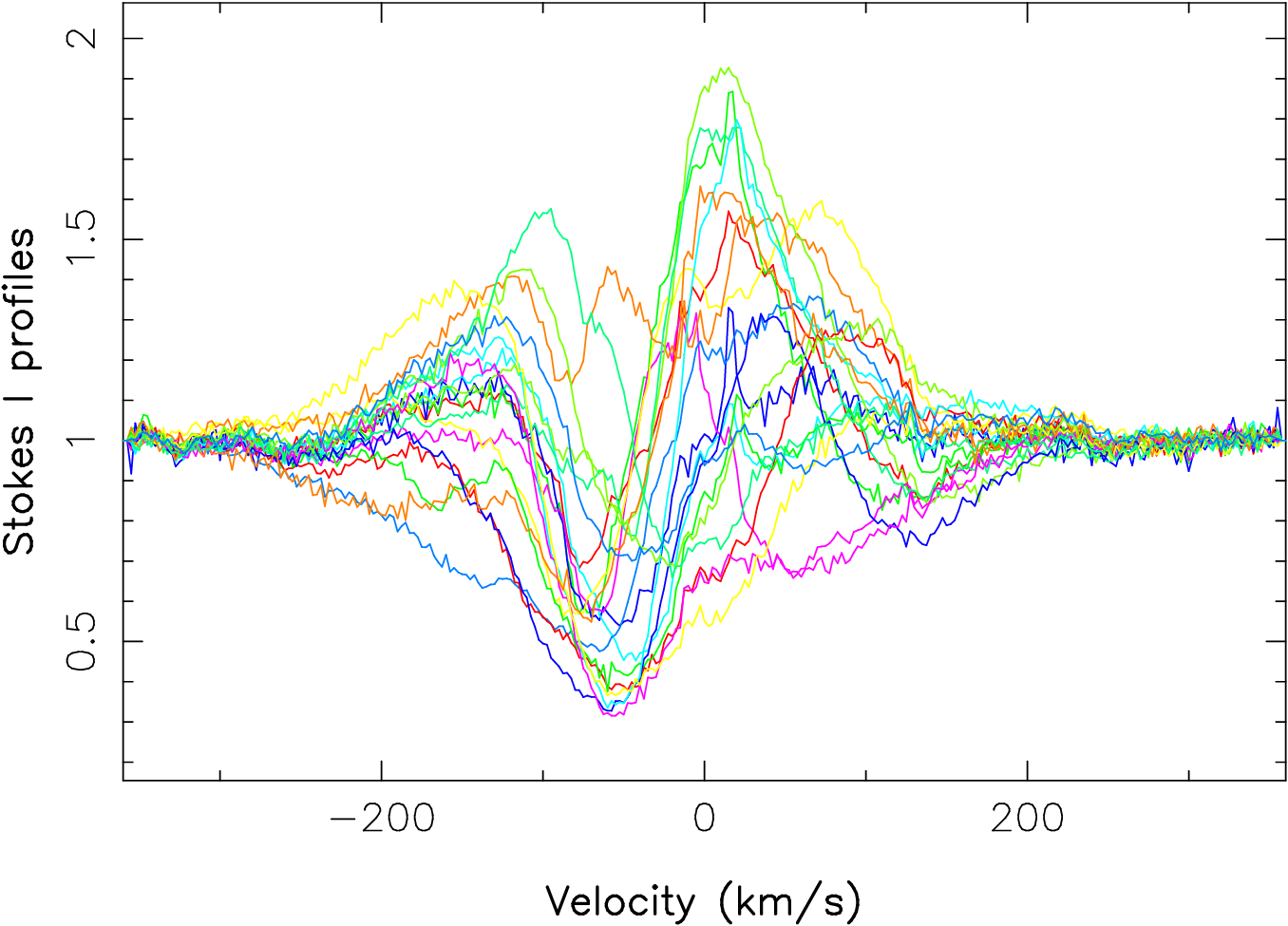}\hspace{14.5mm}\includegraphics[scale=0.32,angle=-90]{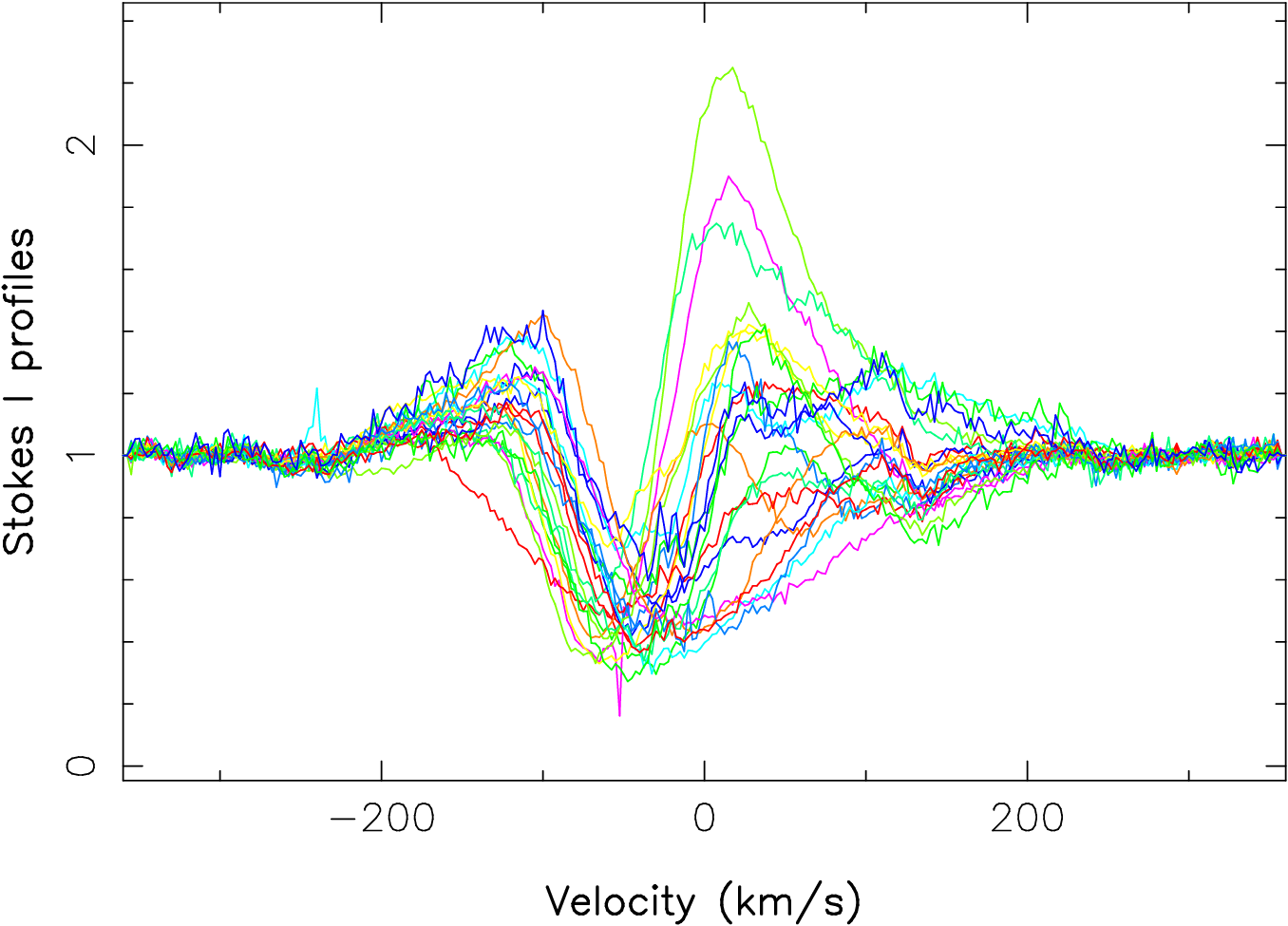}\vspace{2mm}}
\centerline{\includegraphics[scale=0.55,angle=-90]{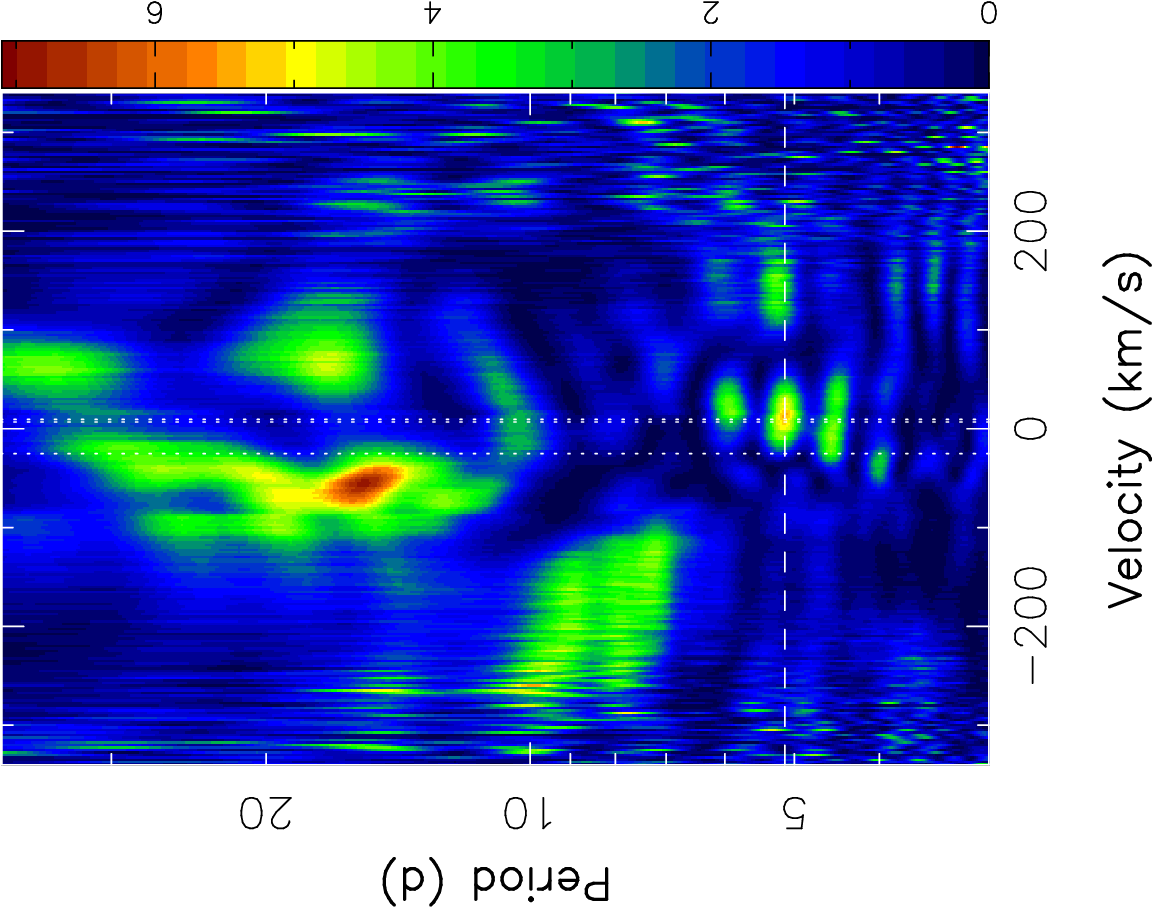}\hspace{3mm}\includegraphics[scale=0.55,angle=-90]{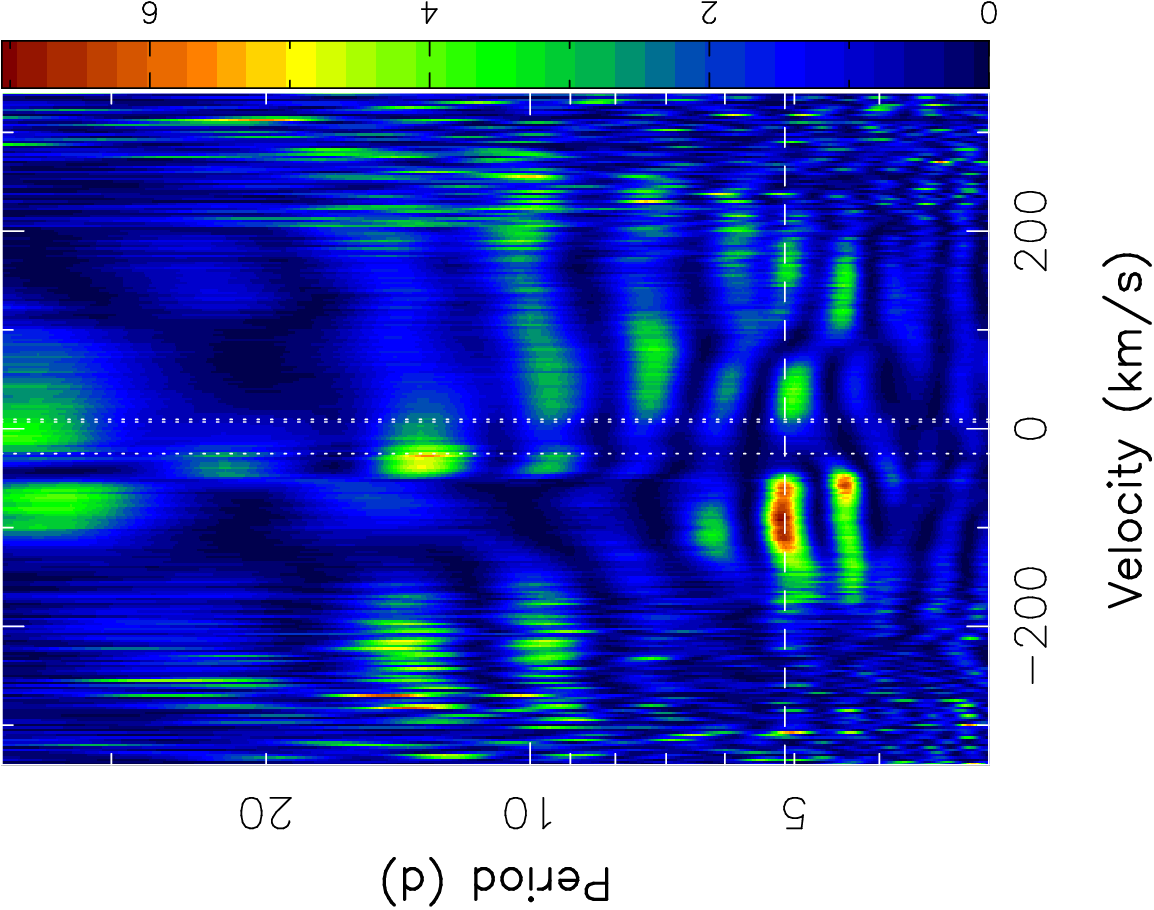}}
\caption[]{
Same as Fig.~\ref{fig:hei} for the first (left panel) and second (right panel) half of the 2025 season data set.} 
\label{fig:hei2}
\end{figure*}

\begin{figure*}[ht!]
\centerline{\hspace{-2mm}\includegraphics[scale=0.32,angle=-90]{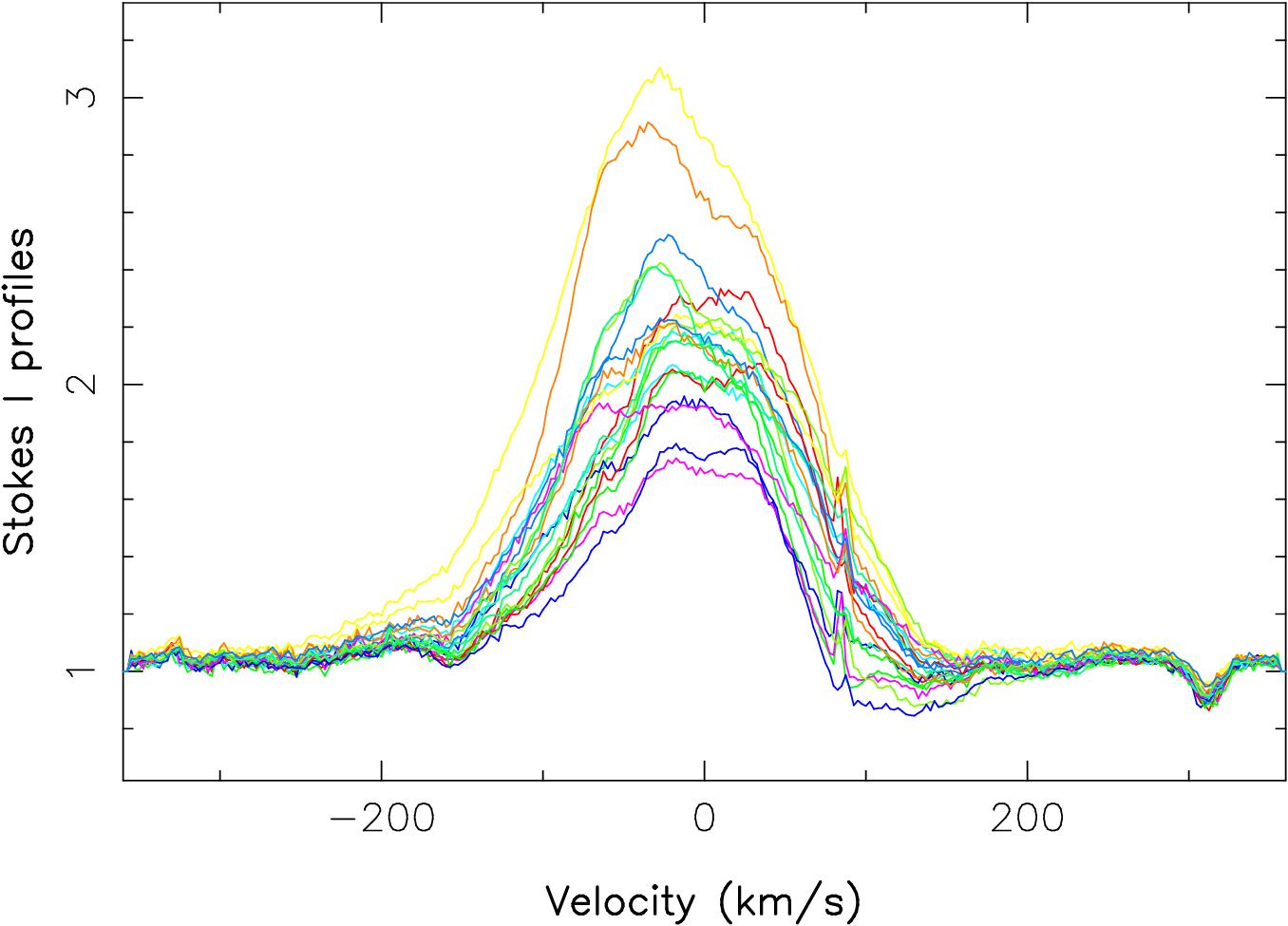}\hspace{14.5mm}\includegraphics[scale=0.32,angle=-90]{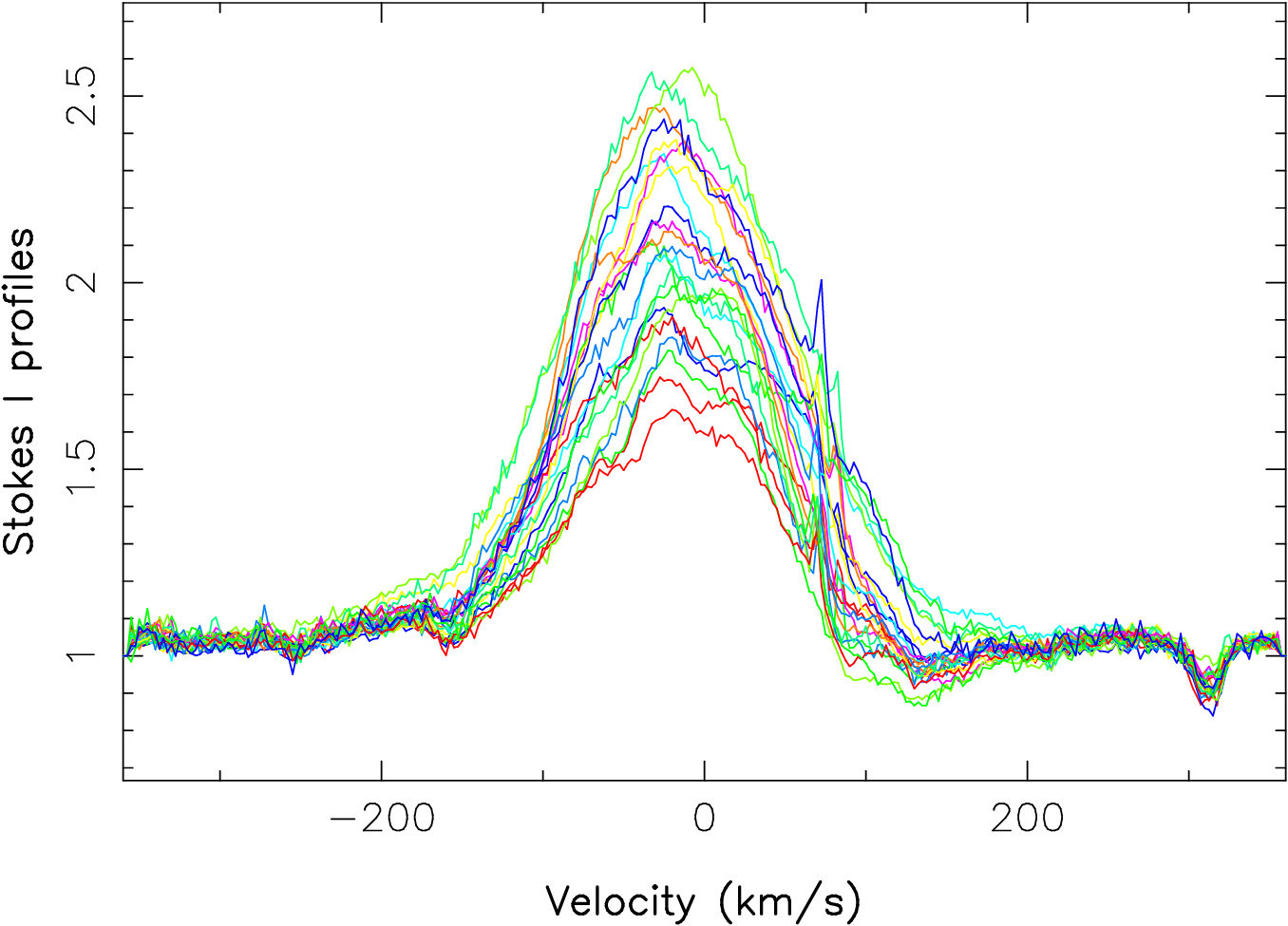}\vspace{2mm}}
\centerline{\includegraphics[scale=0.55,angle=-90]{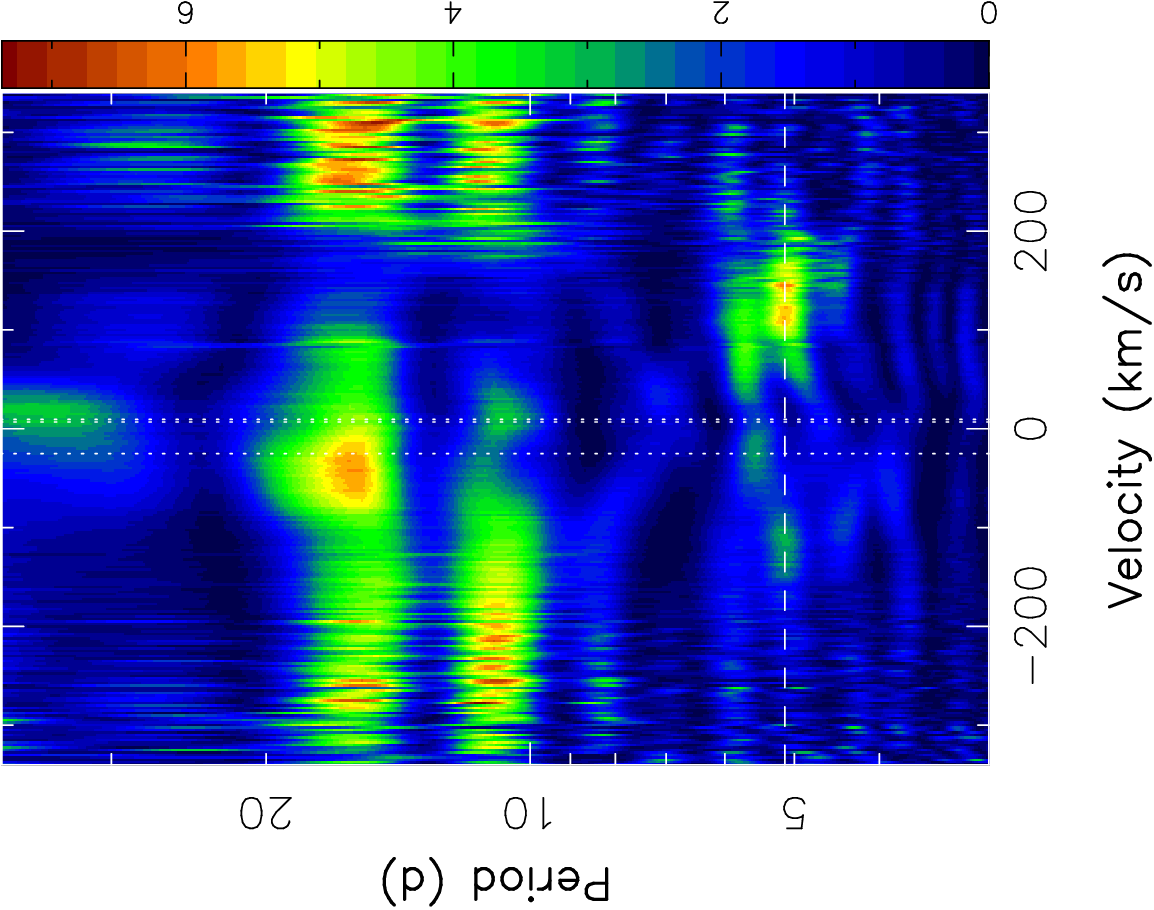}\hspace{3mm}\includegraphics[scale=0.55,angle=-90]{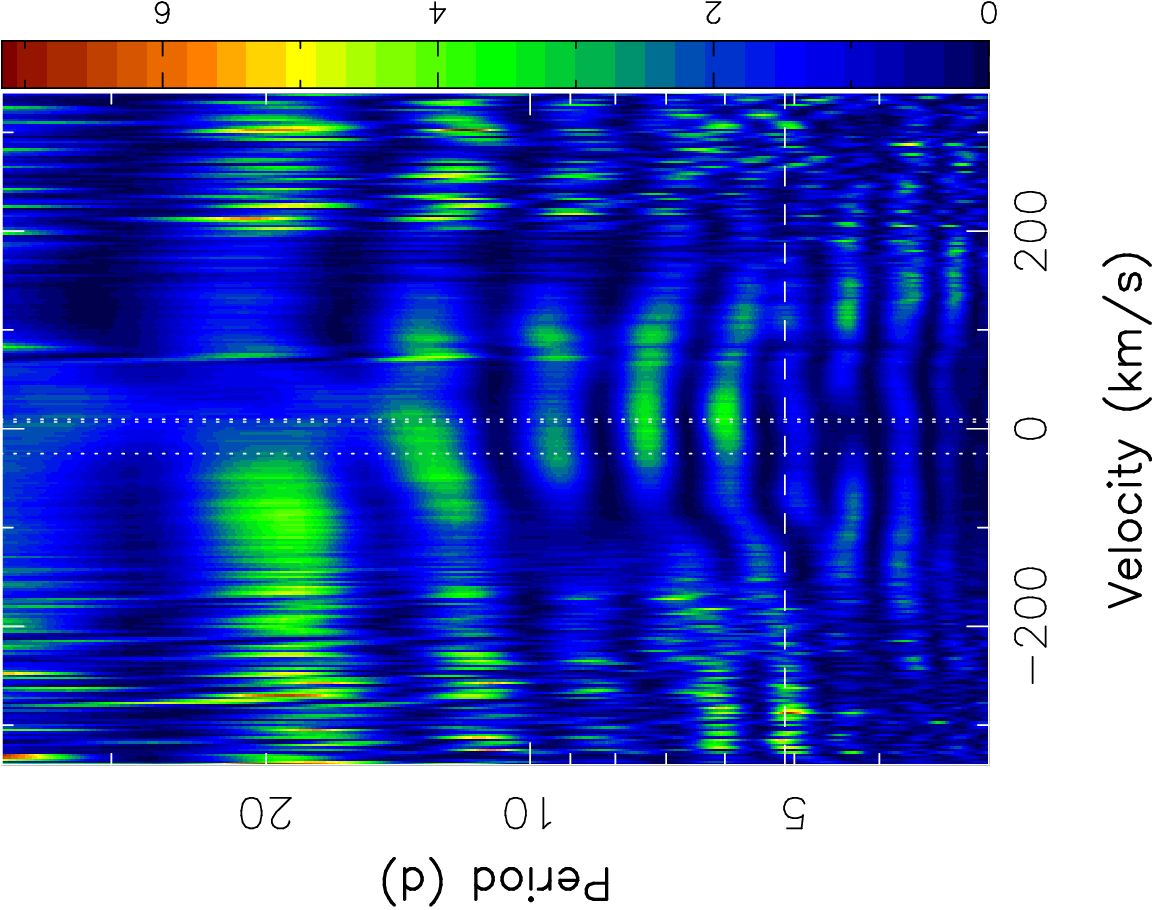}}
\caption[]{
\emr Same as Fig.~\ref{fig:hei2} for the 1282~nm \pab\ line (left: first half of 2025, right: second half).} 
\label{fig:pab2}
\end{figure*}

\begin{figure*}[ht!]
\centerline{\hspace{-2mm}\includegraphics[scale=0.32,angle=-90]{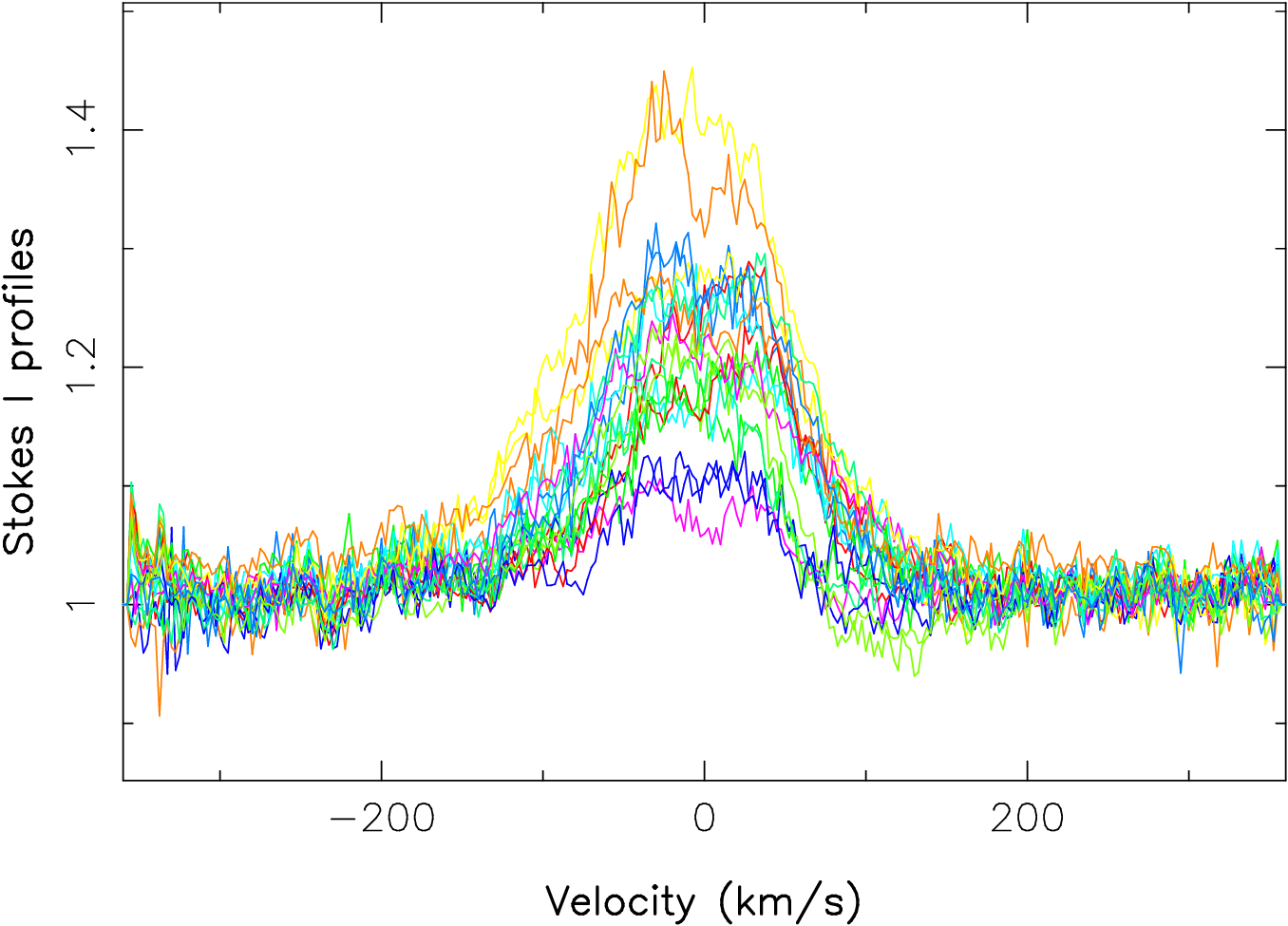}\hspace{14.5mm}\includegraphics[scale=0.32,angle=-90]{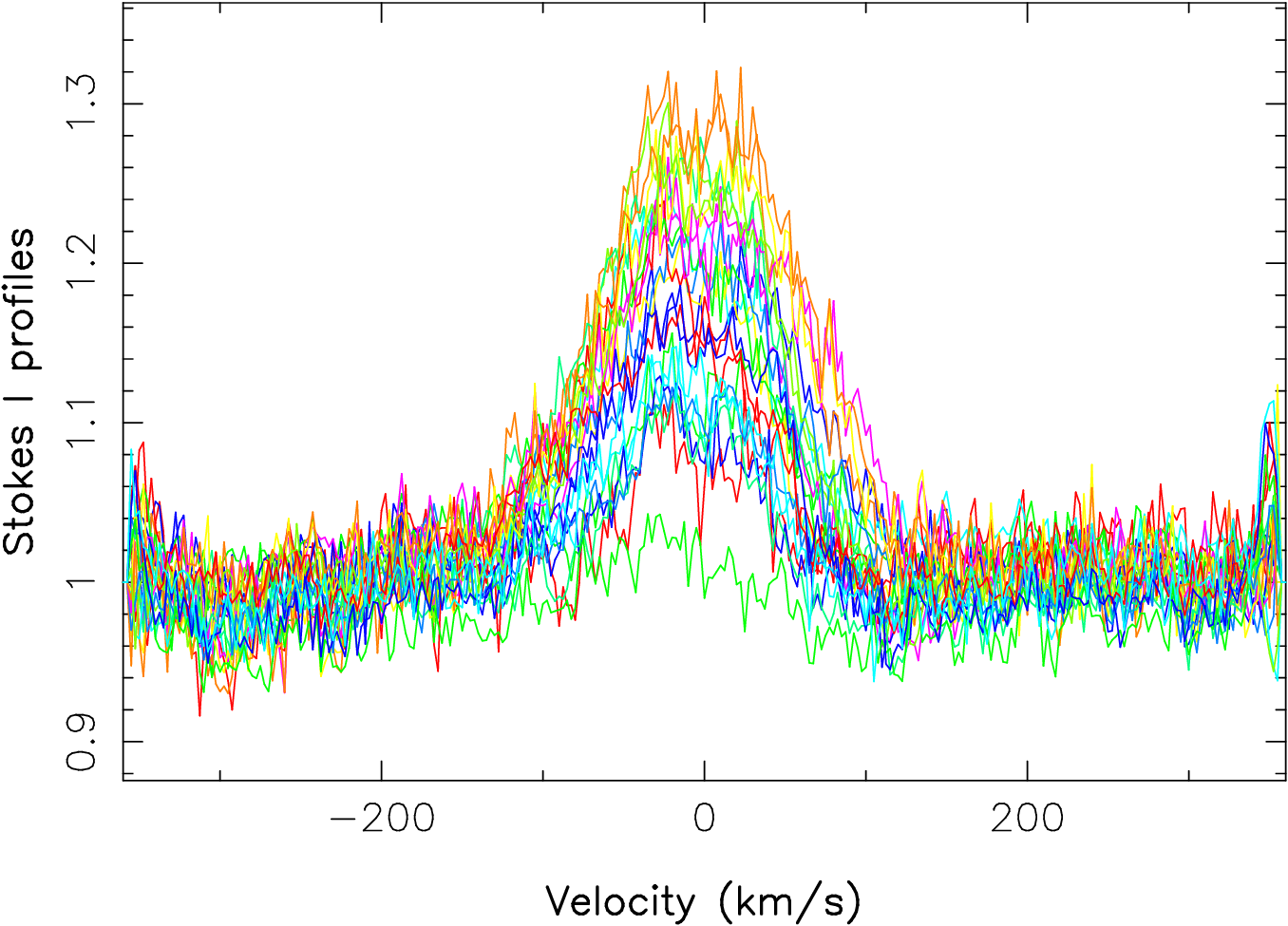}\vspace{2mm}}
\centerline{\includegraphics[scale=0.55,angle=-90]{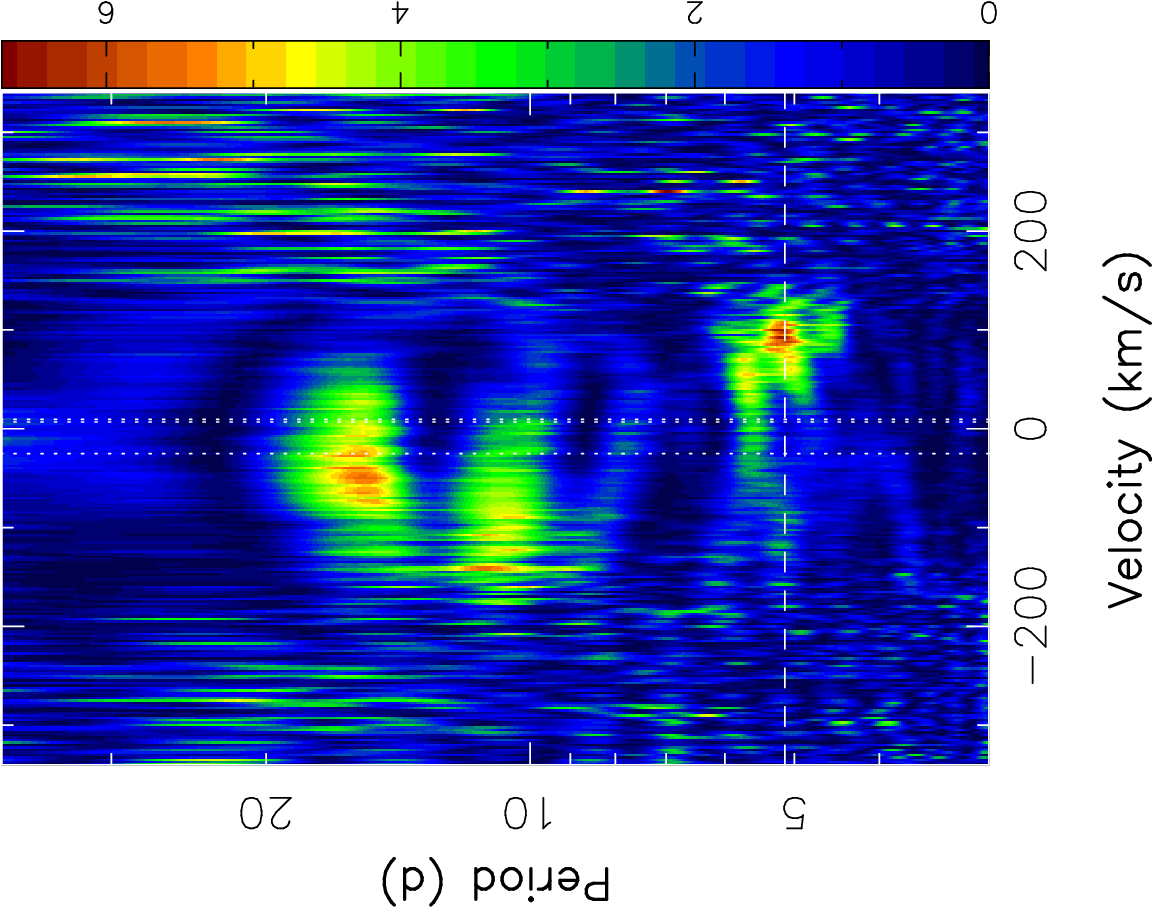}\hspace{3mm}\includegraphics[scale=0.55,angle=-90]{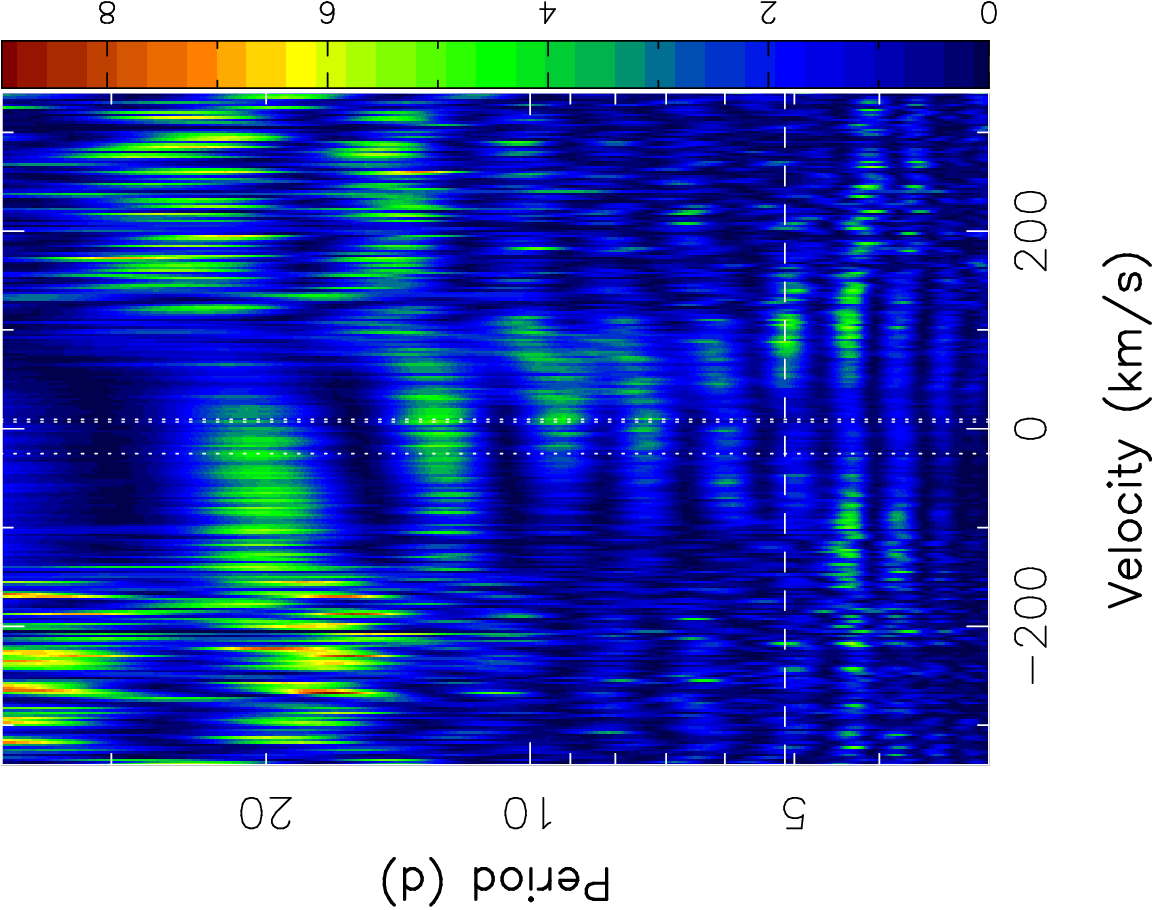}}
\caption[]{
\emr Same as Fig.~\ref{fig:hei2} for the 2268~nm \brg\ line (left: first half of 2025, right: second half).} 
\label{fig:brg2}
\end{figure*}

Figure~\ref{fig:brg} shows the 2D periodogram of the \brg\ line in season 2024 and 2025.  Figures~\ref{fig:hei2}, \ref{fig:pab2}, and \ref{fig:brg2} show similar plots for the first 
and second half of the 2025 season, for the \hei, \pab, and \brg\ lines.  

\FloatBarrier %\usepackage{placeins}
%\clearpage

\end{appendix}
\end{document}